\pdfoutput=1

\documentclass[final,authoryear,5p,times,twocolumn]{elsarticle}
\usepackage{natbib,geometry,pifont}
\usepackage[section] {placeins}
\usepackage{amsmath,amssymb,amstext,amscd,bm}
\usepackage{rotating}
\usepackage{pdflscape}
\usepackage{txfonts}
\usepackage[latin1]{inputenx}
\usepackage{dcolumn}
\usepackage{multirow}
\usepackage{lscape}
\usepackage{amsmath}
\usepackage{multicol}
\usepackage{graphics}
\usepackage[maxfloats=100]{morefloats}
\usepackage{amsthm}
\usepackage{lineno}
\usepackage[toc,page]{appendix}

\journal{New Astronomy}
\def\astrobj#1{#1}

\input{aas_macros.sty}

\begin{document}

\begin{frontmatter}

  \title{The OPD Photometric Survey of Open Clusters\\
    II. robust determination of the fundamental parameters of 24 open clusters 
    \tnoteref{mytitlenote}} 
\tnotetext[mytitlenote]{Based on
    observations made at Pico dos Dias Observatory - LNA/MCTI}

  \author[UNIFEI]{H. Monteiro} \author[UNIFEI,IAG]{W. S. Dias}
  \author[UNIFEI]{G. R. Hickel} \author[UNIFEI,IAG]{T. C. Caetano}
 \fntext[*]{E-mail: hmonteiro@unifei.edu.br}
 \cortext[cor1]{Corresponding author}

\address[UNIFEI]{Universidade Federal de Itajub\'a, Instituto de F\'isica e Qu\'imica, Itajub\'a-MG, Brazil}
\address[IAG]{Universidade de S\~ao Paulo, Instituto de Astronomia, Geof\'isica e Ci\^encias Atmosf\'ericas, S\~ao Paulo - SP, Brazil}

\begin{abstract}

  In the second paper of the series we continue the investigation of
  open cluster fundamental parameters using a robust global
  optimization method to fit model isochrones to photometric data. We
  present optical UBVRI CCD photometry (Johnsons-Cousins system)
  observations for 24 neglected open clusters, of which 14 have high
  quality data in the visible obtained for the first time, as a part
  of our ongoing survey being carried out in the 0.6m telescope of the
  Pico dos Dias Observatory in Brazil. All objects were then analyzed
  with a global optimization tool developed by our group which
  estimates the membership likelihood of the observed stars and fits
  an isochrone from which a distance, age, reddening, total to
  selective extinction ratio $R_{V}$ (included in this work as a new
  free parameter) and metallicity are estimated. Based on those
  estimates and their associated errors we analyzed the status of each
  object as real clusters or not, finding that two are likely to be
  asterisms. We also identify important discrepancies between our
  results and previous ones obtained in the literature which were
  determined using 2MASS photometry.

\end{abstract}

\end{frontmatter}
\newpage

\section{Introduction}

Open clusters have long been recognized as key objects to investigate
the kinematics of star formation regions, aspects of the Galactic
spiral structure, or even the chemical abundance gradients in the disk
of our Galaxy.

As examples of galactic parameters derived from such studies we can
mention the measurement of the rotational velocity of the spiral
pattern of the Galaxy by \citet{Dias2005}, which allowed for the
determination of the location of the co-rotation radius, and a lower
limit of the age for the spiral pattern.  The connection of the
co-rotation radius and the gradient of [Fe] abundance in the Galaxy
was also obtained using open clusters as discussed in
\citet{Lepine2011} and \citet{Barros2013}.
 
The efforts of our group have focused on the study of these objects
throughout the last years to determine, in a systematic and consistent
manner, fundamental parameters such as age, distance, reddening,
metallicity and kinematical information for a large sample as
possible. The main long term goal is to achieve robust statistical
significance of the results obtained for open clusters.  In spite of
all studies that have been made, examining the DAML02 open clusters
catalog\footnote{Available on-line at
  http://www.wilton.unifei.edu.br/ocdb} \citep{Dias2002}, it is clear
that much work remains to be done. Although most of the objects have
estimates of the distance and age, about 50$\%$ have no reliable
determination based on good quality optical photometry. The majority
of clusters had parameters estimated using only near infra-red data
from The Two Micron All Sky Survey\footnote{Available on-line at
  http://www.ipac.caltech.edu/2mass/} (2MASS) catalog photometry. The
main problem with this scenario is that 2MASS data is good for
clusters which are well defined and clearly differentiated from the
surrounding field which may not be always the case. Indeed as shown in
\citet{Netopil2015}, where the authors present a thorough analysis of
the major large scale open cluster homogeneous parameters databases,
there are trends or constant offsets which are significant. The
authors also show that in some cases the databases can have 20\% or
more problematic objects, which limits the usability of these
databases.

In that context, optical photometric data is still the one of the best
choices to study open clusters. In this work we present 24 objects
studied in our photometric Survey of Open Clusters. The survey has the
goal to increase the open clusters homogeneous database with good
quality observational optical data, based on UBVRI photometry
performed at Pico Dos Dias Observatory - LNA/BRAZIL. The survey is
described in detail in \citet{Caetano2015} (hereafter paper I). Of the
24 clusters presented, 14 have had their UBVRI data obtained for the
first time. 

Along with good quality observational data, the determination of
fundamental parameters such as age, distance and reddening is also a
key aspect of open cluster studies. In an attempt to maintain a good
level of homogeneity in the determination of these parameters, we have
in last years developed a tool based on global optimization
techniques that removes most of the subjectivity involved in
isochrone fitting while also allowing it to be reproducible within
the uncertainties. The procedure allows the user to take into account
important factors such as the binary fraction, initial mass function
and observational uncertainties. Following the procedure outlined in
detail in our previous works, the fundamental parameters for the 24
open clusters presented in this work were robustly determined from
isochrone fits to the data.

Having achieved reasonable stability, we also make the code freely
available to the community through a web-site dedicated to the results
of this project
\footnote{http://www.wilton.unifei.edu.br/OPDSurvey.html}. The code
produces a wide range of output information, as stellar membership
probabilities, stellar density maps, among others, which may prove to be
suitable for other studies of open clusters.

The paper is organized as follows: In Sec. 2 we summarize the
observational procedure, the data reduction technique as well as the
transformation of the instrumental magnitudes to the standard system.
In Sec. 3. we give an overview of the code and how it works and
mention the main changes introduced since the last version used in
Paper I. In Appendix A the details of the changes are presented.  In
Sec. 4 we present the color-color and color-magnitude diagrams
analysis for each object and discuss the results.  In Sec. 5 we give
our final conclusions.

\section{Observations and data reduction}

All clusters targeted in the survey were observed using the same
instrumental setup which is described in detailed in paper I. We used
the CCD 106, a SITe SI003AB $1024 \times 1024$ CCD back-illuminated
camera with Johnson-Cousins UBVRI filters attached to Boller \&
Chivens 0.6 m telescope.

For the 14 new clusters presented in this work, images were collected
in 24 through 26 of June 2010, during the same observing run which ended
on the 28th June 2010. The observational strategy adopted is exactly
the same as the one presented in detail in paper I, as well as the
data reduction, which was performed using the softwares and techniques
briefly summarized below. 

All the images were pre-processed in a standard way, e.g. were
trimmed, had the bias subtracted, were corrected for shutter timing
effects and for sky flat field.

The instrumental magnitudes and the position of the stars in each
frame were derived by the point spread function (PSF) method.  We used
the software STARFINDER \citep{Diolaiti2000} which was developed for
crowed stellar field
analysis\footnote{http://www.bo.astro.it/\~{}giangi/StarFinder/index.htm},
adapted to be executed automatically, as described in detail in paper
I.

The equatorial coordinates for each detected star were computed using
the coordinates expressed in the detector reference system given by
STARFINDER and the UCAC4 \citep{Zacharias2013}. The transformation
between the CCD reference system and equatorial system was made
through linear equations since the field is small.

\subsection{Transformation to the standard system}

In each night we performed observations of at least five Landolt
standard fields from \citet{Clem2013} at different air masses. The
standard fields were used to calibrate the images to the standard
Johnson system considering the calibration equations given below:

\begin{eqnarray}
u = U + u1 + u2  X + u3 (U-B)\\
b = B + b1 + b2  X + b3 (B-V)\\
v = V + v1 + v2  X + v3 (B-V)\\
r = R + r1 + r2  X + r3 (V-R)\\ 
i = I + i1 + i2  X + i3 (V-I)
\end{eqnarray}
where upper case letters represent the magnitudes and colors in the
standard system and lower case letters were adopted for the
instrumental ones and X is the air mass. The coefficient values are
reported in an online Table available at the web-site of the
project\footnote{http://www.wilton.unifei.edu.br/OPDSurvey.html}.

The best fit was obtained by a global optimization procedure that
minimized the differences between the magnitudes of the observed
standard stars calculated in the standard system with those catalogued
values from \citet{Clem2013}. The final values of the coefficients
are given by the mean of the results of one hundred runs of the
fitting procedure and the uncertainty was estimated by the standard
deviation of the solution sample. The global optimization procedure
requires the definition of a parameter space which will be searched
and for this we adopted typical coefficients values for the OPD site
plus a range of $10\%$ to accommodate unknown variations and errors.

We present all the transformation coefficients, rms values,
interesting plots as the residual of the fit to the standard stars and
the errors as a function of the magnitude of all observed stars, as
well as the data, in the web-site dedicated to the technical
information and results of this project, and we refer the reader to it
for further details.

The quality of the nights presented in this paper was checked
comparing the data of the standard stars with those obtained in the
night published in the paper I.  The mean differences (not systematic)
are typically about 0.02 mag with standard deviation lower than 0.05
mag in each filter.

\section{Determining Fundamental parameters}

To obtain the fundamental parameters for open clusters (age, distance,
E(B-V) and metalicity) we have developed a code that employs a global
optimization technique known as Cross-Entropy to perform the fitting of
theoretical isochrones to photometric data. The basic procedure
  of the global optimization method can be summarized as follows:

\begin{enumerate}[(i)]

\item a sample containing the initial values of the parameters to be
  optimized is randomly generated, according to predetermined
  criteria;

\item a synthetic cluster is generated for each set of parameters
  based on a given initial mass function and binary fraction and the
  theoretical isochrones of \citet{Girardi2000} and
  \citet{Marigo2008};
 
\item the synthetic cluster is compared with the observational data
  and a likelihood is calculated;

\item the full set of solutions is ranked based on the likelihood and a
  pre-defined percentage of those is selected;

\item a new sample of solutions is randomly generated, based on the
  distribution of the best ranked solutions of the item (iv). 

\item the optimization process in items (ii) to (v) is then repeated until
  a stopping criteria (for example, number of iterations) is
  satisfied.

\end{enumerate}

The procedure above is repeated for a user defined number of times
performing a bootstrap procedure to re-sample with replacement the
observational data. The objective is to obtain the final distribution
of solutions from where we obtain the most probable one, which we
adopt as the mode of the distribution. The uncertainty in the solution
is the robust standard deviation of the solution sample.

The objective function used to obtain the likelihood of the solution
in item (iii) is weighted by a non-parametric membership probability
computed for the stars, taking into consideration their position
relative to the cluster center, the star density in the particular
position of the field and the multidimensional photometric data, in
addition to the photometric errors. The main objective of this
procedure is to minimize the subjectivity in the selection of stars
and maximize the contrast of cluster features in relation to the field
stars in the CMDs, allowing for a more robust fit to the available
data.

The in depth code details and other relevant information are described
in \citet{Oliveira2013}, \citet{Dias2012}, \citet{Monteiro2011} and
\citet{Monteiro2010}. Also in the context of the results of the OPD
Survey, we have also made a summarized discussion in
\citet{Caetano2015}, paper I in this series.

Since the publication of paper I we have also made several
improvements to the code. We performed a general reorganization,
introduced a new free parameter, the total to selective extinction
ratio $R_V$ and implemented a way of defining the cluster region not
as circle but as an iso-density region from the density map. Other
minor improvements were also made aiming at making the tool publicly
available. We discuss the modifications in detail in Appendix A.

For consistency, given the changes introduced in the code, we
  checked previously determined solutions for clusters in paper
  I. The comparison of our results obtained with the present version
of the code to those given in the paper I shows no significant
difference.  The average and standard deviation of the differences of
our results to those of paper I are $-0.1\pm0.1$ mag in E(B-V),
$-1\pm332$ pc in distance, $-0.6\pm0.2$ yr in logt, $0,005\pm0,008$ in
metallicity.

The code is written in IDL (Interactive Data Language) and has been
optimized to run on multi-core on OS LINUX. A WINDOWS version is
available for one core only.  Basically the execution time depends on
the number of the stars, number of filters, the number of the runs and
the number of cores.  As an example to illustrate the performance, for
one hundred member stars with UBVRI data, using five i7 cores the code
takes about ten hours to finalize 50 runs. As mentioned before, the
code is freely available to the comunity through a web-site dedicated
to the results of this project
\footnote{http://www.wilton.unifei.edu.br/OPDSurvey.html}

\section{Results and discussions\label{analysis}}

In this work we applied the method described previously to our UBVRI
data with respective errors to determine the fundamental parameters of
the studied objects.  As in paper I and references therein, to make
the CE more efficient in finding the global solution we adopt band $=
0.15$ and $\alpha = 0.4 $ for the tuning parameters of the method. In
practice, in each iteration $15\%$ of the best solutions were used to
generate new parameter value distributions and the lower $\alpha$
value is used to slow the convergence rate of the algorithm to keep it
from converging to a local minimum.

In Table 1 the code parameters used in the fitting procedure for all
studied clusters are given.  For all clusters $100\%$ of the
  stars were considered as binary. As discussed in
  \citep{Monteiro2010}, in the most extreme case the choice of 100\%
  binary fraction introduces a small systematic error in the
  determined distance and reddening (usually within the parameter
  uncertainties of the fit) and does not affect the age. Ideally the
  binary fraction should be determined through other means and then
  incorporated in fitting procedure, but adopting a binary fraction of
  100\% essentially gives us upper limits for distance and reddening.

The tabulated isochrones used in the code were taken from
\citet{Girardi2000} and \citet{Marigo2008}, the same ones used in
Paper I, and the parameter space was chosen as follows:
\begin{itemize}
\item age: from log(age) =6.60 to log(age) =10.15;
\item distance: from 1 to 10 000 parsecs;
\item $E(B-V)$: from 0.0 to 3.0; 
\item Metallicity (Z): from 0.001 to 0.30 dex with steps of $Z = 0.05$ dex.
\item $R_V$: from 2.0 to 4.0
\end{itemize}

The final value of the metallicity parameter was transformed to
[Fe/H], adopting the same approximation considered in the Padova
database of stellar evolutionary tracks and isochrones:
$[Fe/H]=logZ/Z\odot$ with $Z\odot=0.019$. The errors were obtained by
the usual propagation formula.

To determine parameter estimate errors we perform the fit for
  each data set 50 times, each time re-sampling from the original data
  set with replacement to perform a bootstrap procedure. Through this
  Monte-Carlo procedure new synthetic clusters are also generated each
  time from the theoretical isochrones using the adopted IMF. In
  practice this allows us to incorporate in the error estimate the
  effect of variations in the realizations of the synthetic
  clusters. The final uncertainties in each parameter are then
obtained by the standard deviation of the runs.

The current version of the code can provide a series of outputs,
including an estimation of the photometric membership. It also gives
the user several plots that allow for a general inspection before
running the full CE isochrone fit.  As examples we mention important
diagnostic plots such as the density map, sky chart with the proper
motion of the stars, CMDs with membership, CMDs with vector proper
motion overploted. Many secondary plots are also given such as the
CMDs of stars inside and outside the chosen cluster region, CMDs of
stars in different radii (or density selected), among others.  The The
Digitized Sky Survey (DSS)
\footnote{http://archive.stsci.edu/dss/index.html} images may also be
useful to provide independent estimates such as the radius $R_{c}$ to
be used in tuning parameters. All of the plots mentioned above were
used extensively in obtaining the results for the clusters in this
work. In the interest of brevity we do not include all these plots
here but we refer the reader to the survey web-site for the complete
graphs for all studied clusters.

\begin{figure*}
\centering
\includegraphics[height = 6.0cm, width = 6.0cm]{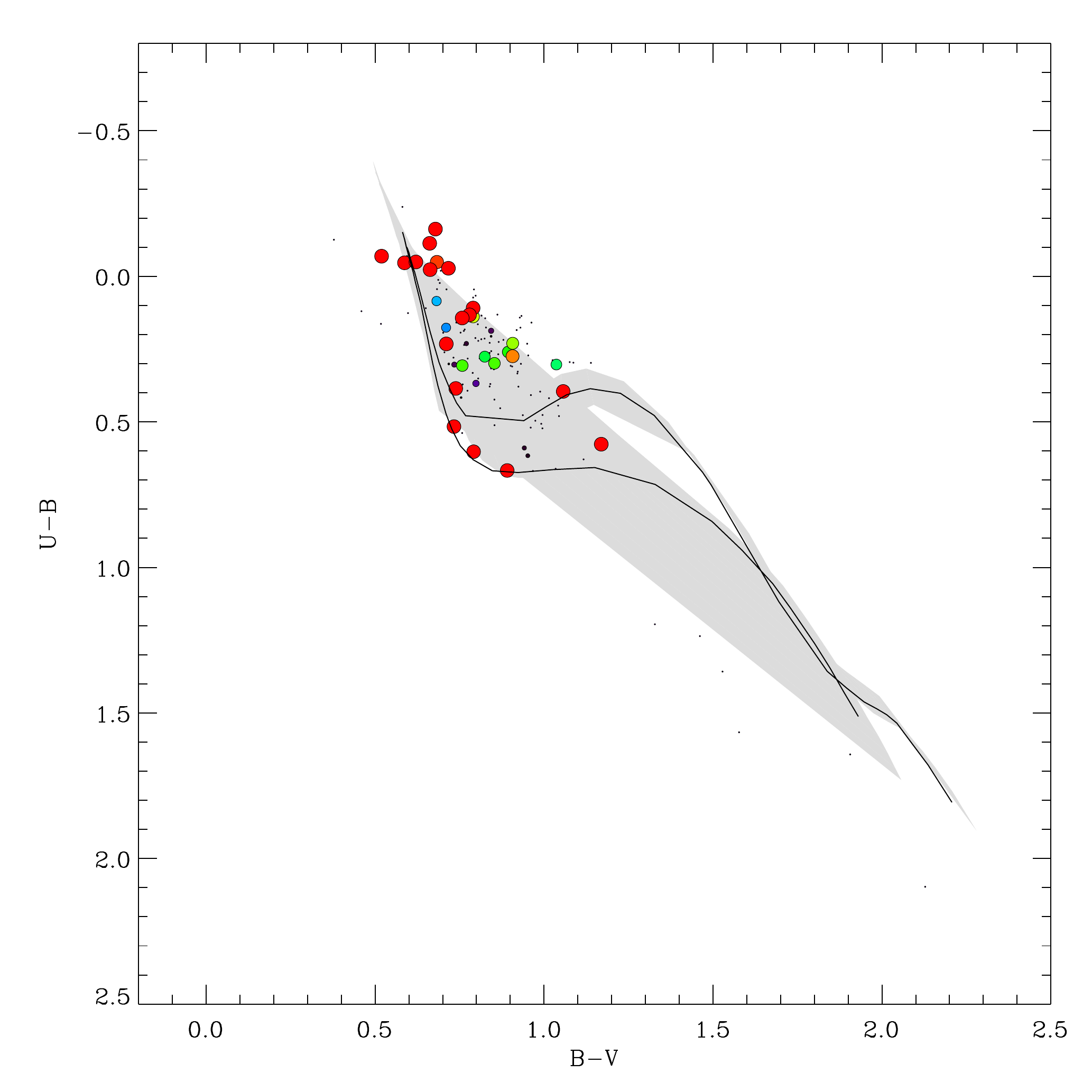}
\includegraphics[height = 6.0cm, width = 6.0cm]{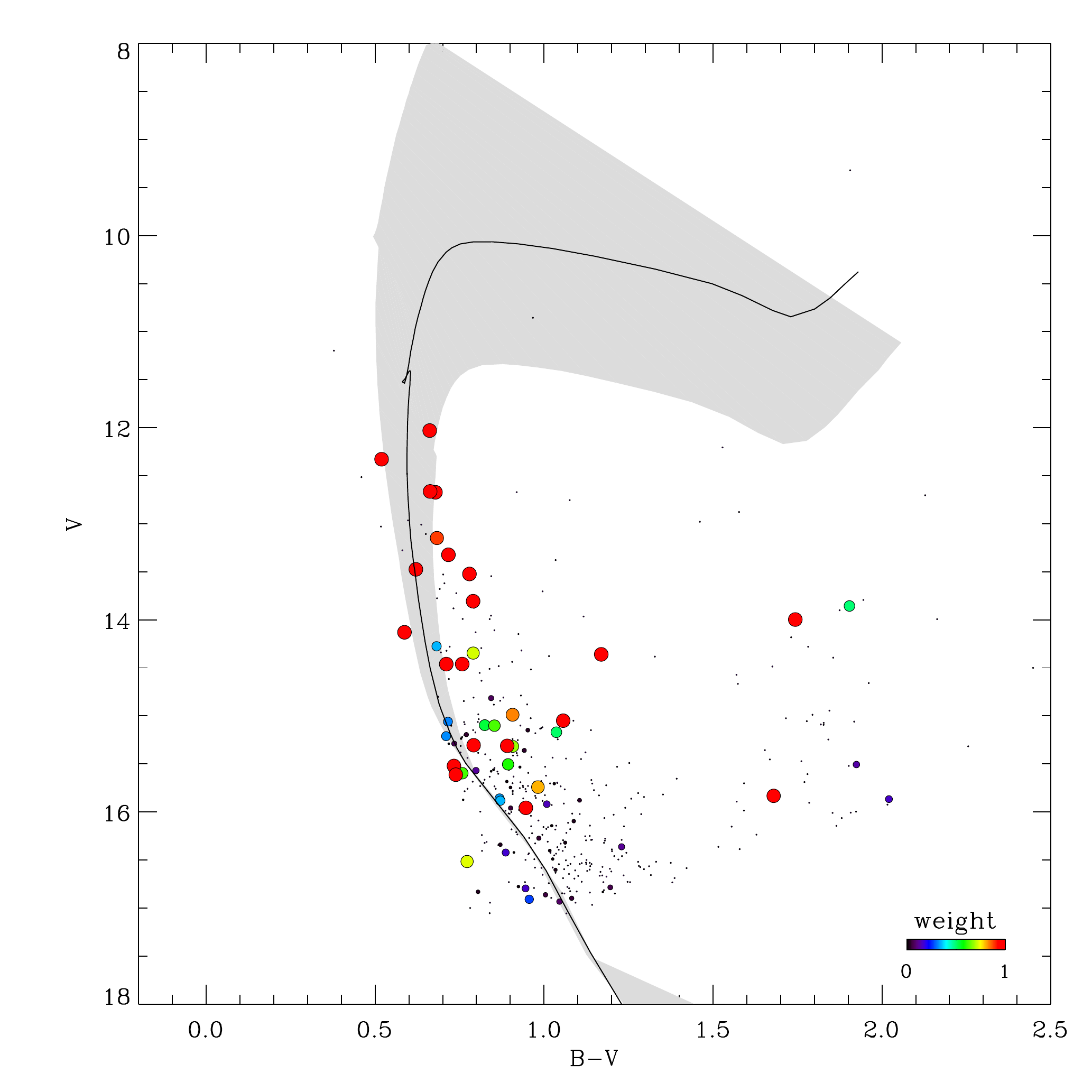}
\includegraphics[height = 6.0cm, width = 6.0cm]{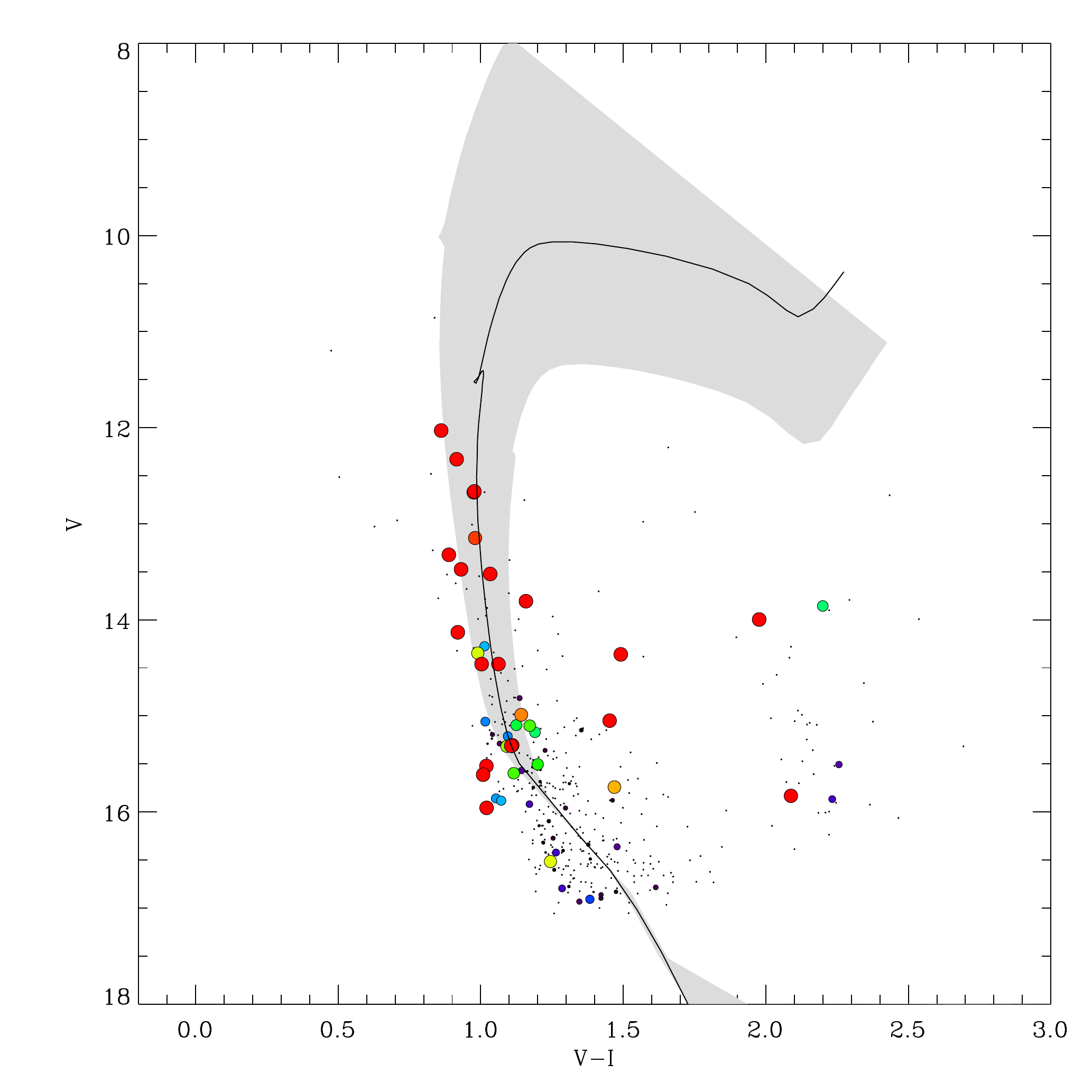}
\caption{The fit results obtained by the cross-entropy method applied
  to the UBVRI data for the cluster BH 200. The symbol size and
    color are proportional to membership of each star and are meant as
    a guideline only. The individual values of membership are given in
    the survey web-site along with other useful data. The line
  represents the fit of the isochrones from \citet{Girardi2000}. The
  light-shaded region shown represents the estimated $1\sigma$
  uncertainty in the fit.}
              \label{bh200}%
\end{figure*}

\begin{figure*}
\centering
\includegraphics[height = 6.0cm, width = 6.0cm]{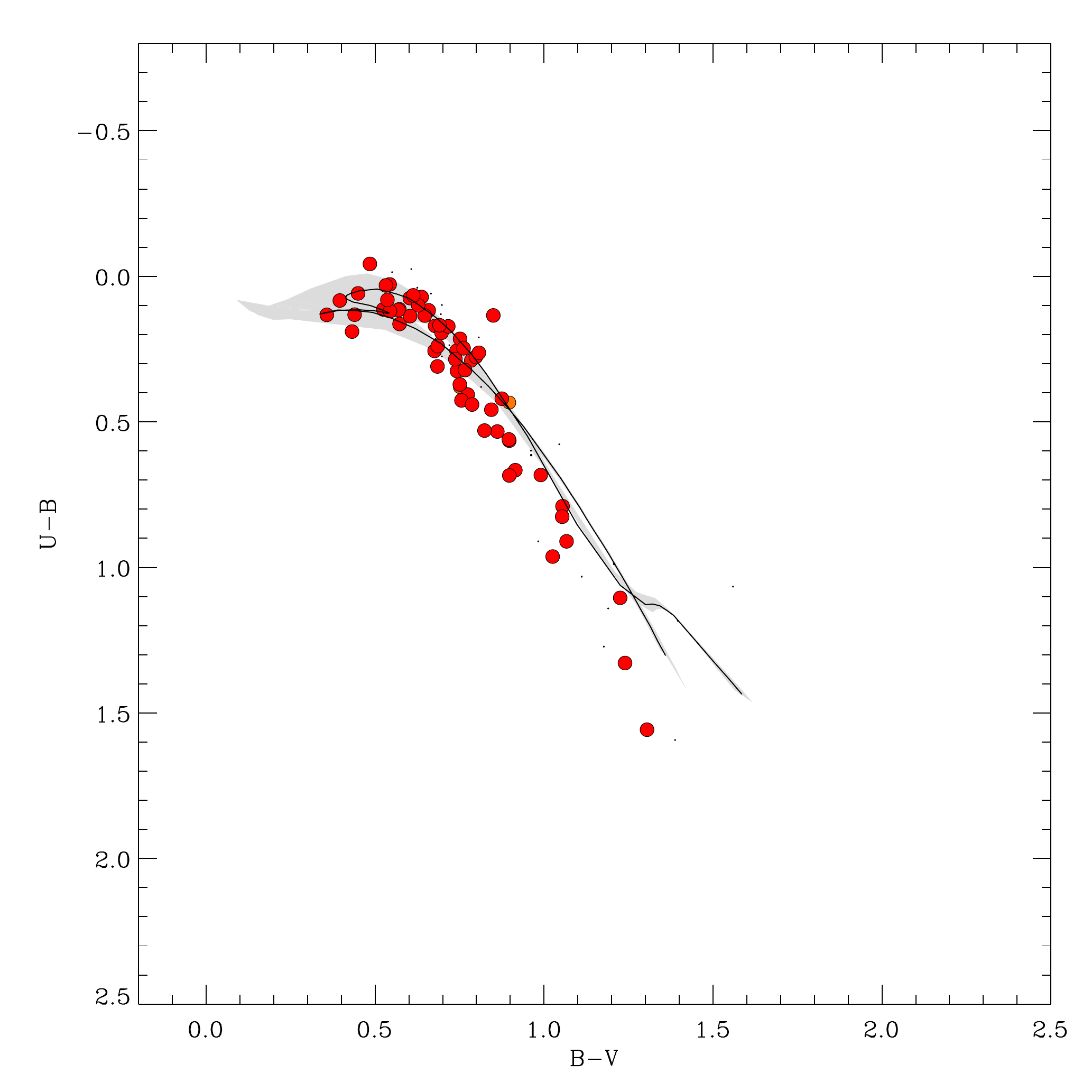}
\includegraphics[height = 6.0cm, width = 6.0cm]{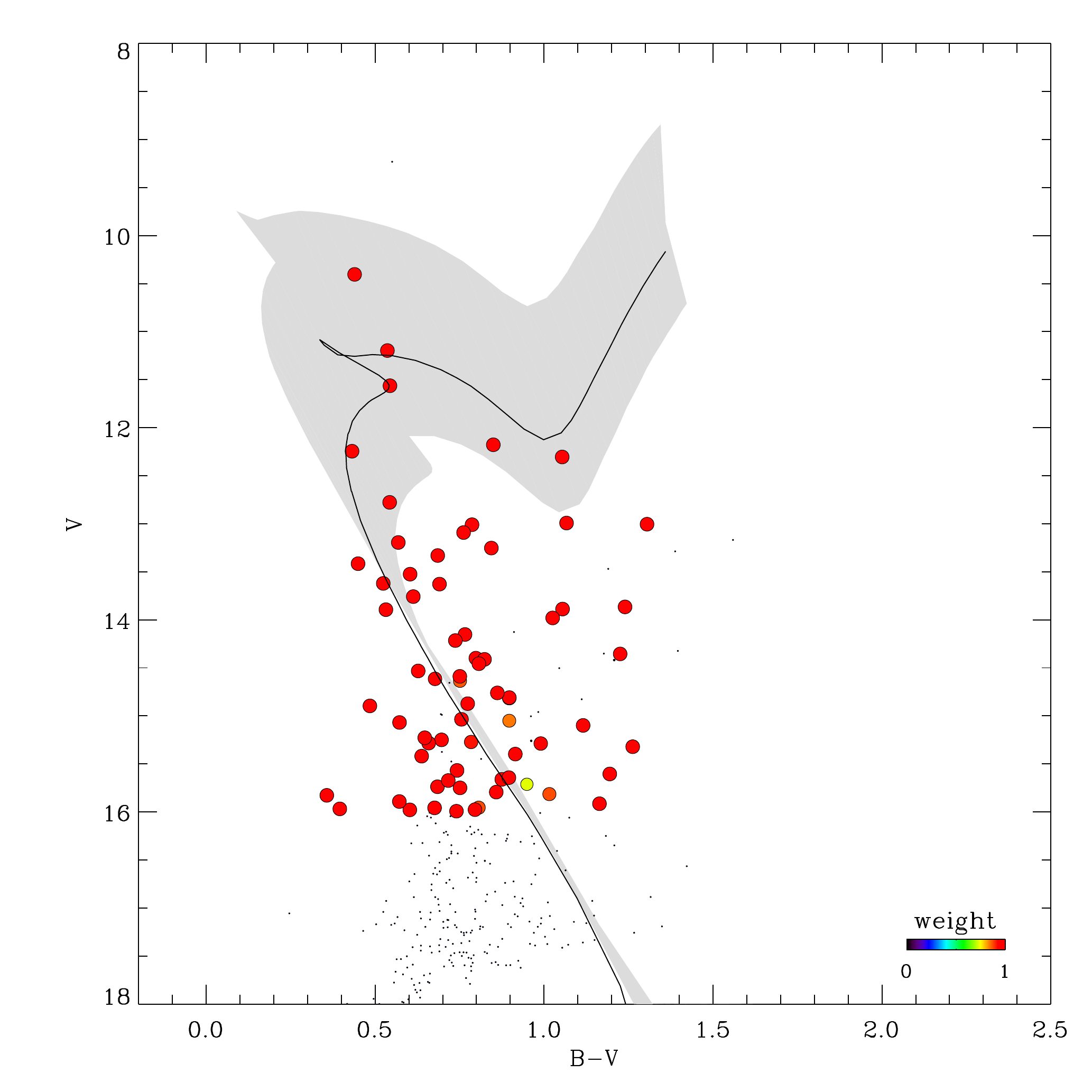}
\includegraphics[height = 6.0cm, width = 6.0cm]{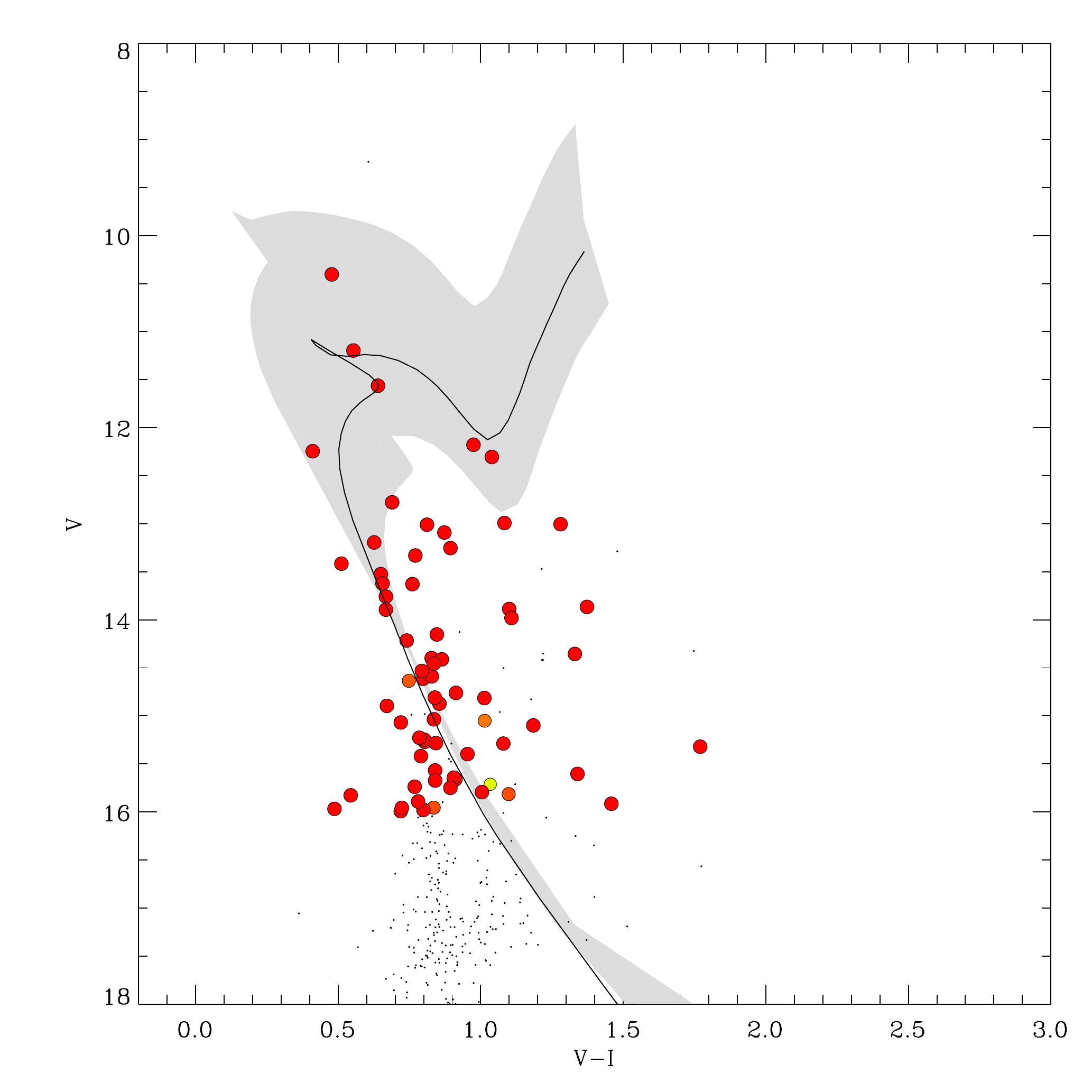}
   \caption{Same as Figure \ref{bh200}, but for ESO139-54.}
              \label{ESO139-54}%
\end{figure*}

\begin{figure*}
\centering
\includegraphics[height = 6.0cm, width = 6.0cm]{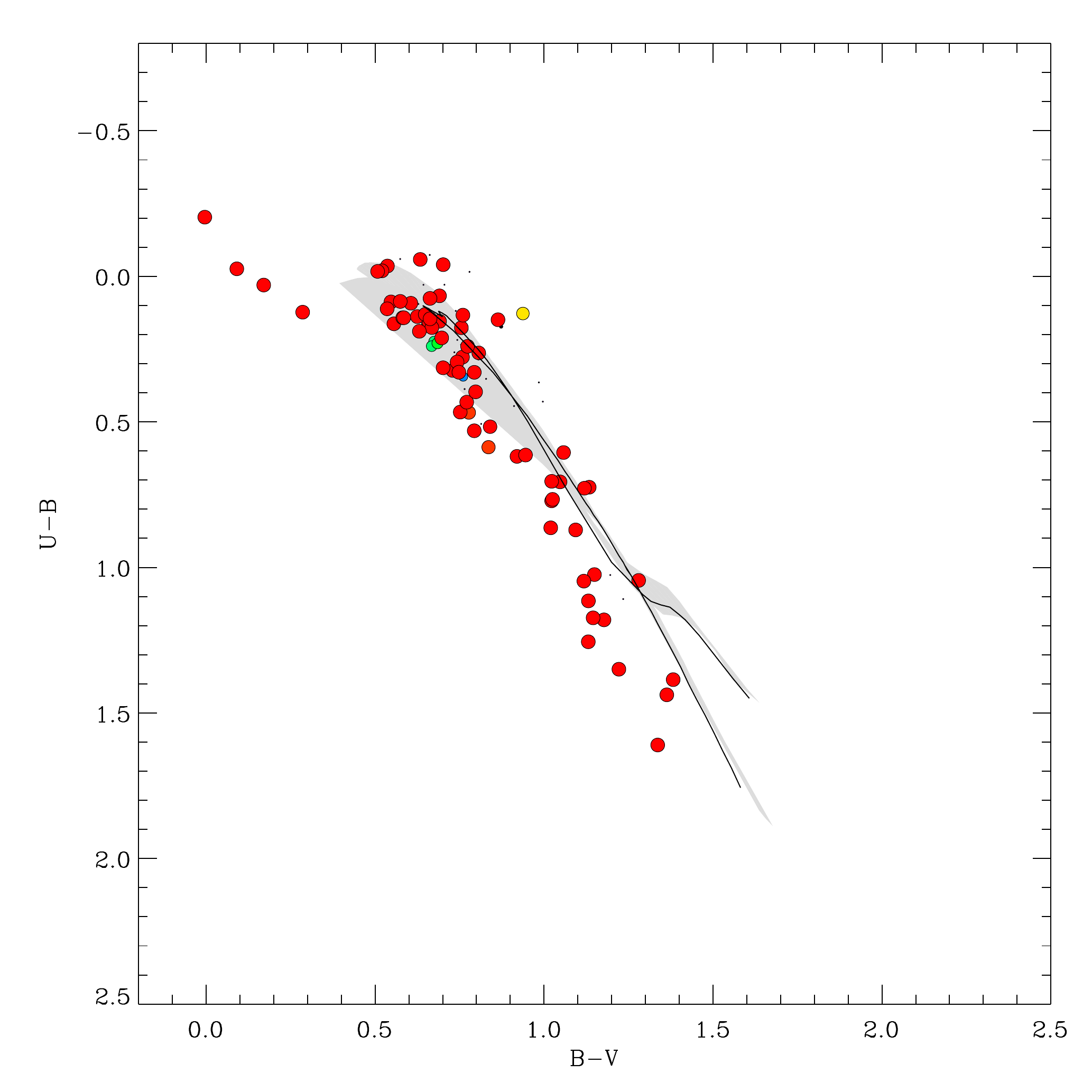}
\includegraphics[height = 6.0cm, width = 6.0cm]{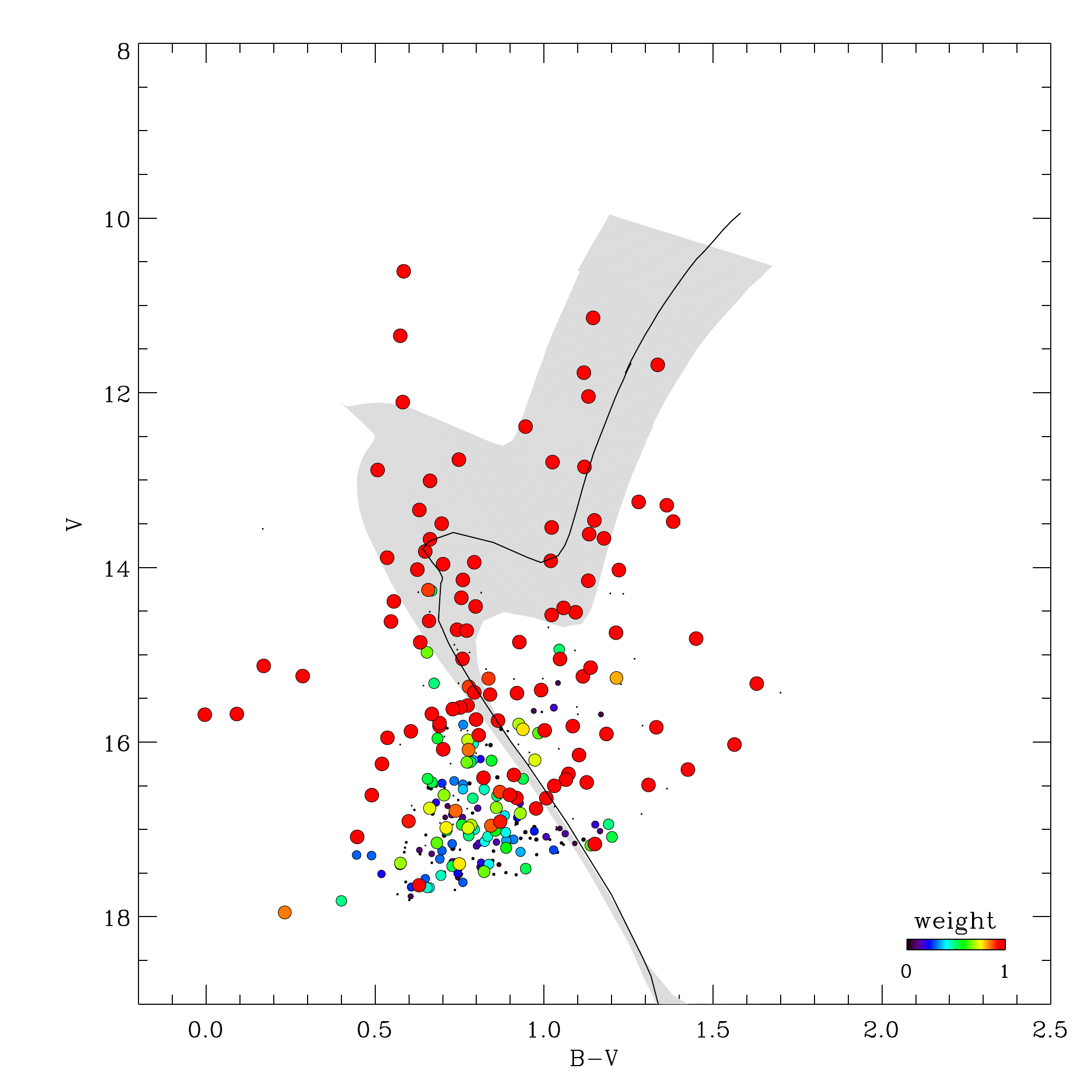}
\includegraphics[height = 6.0cm, width = 6.0cm]{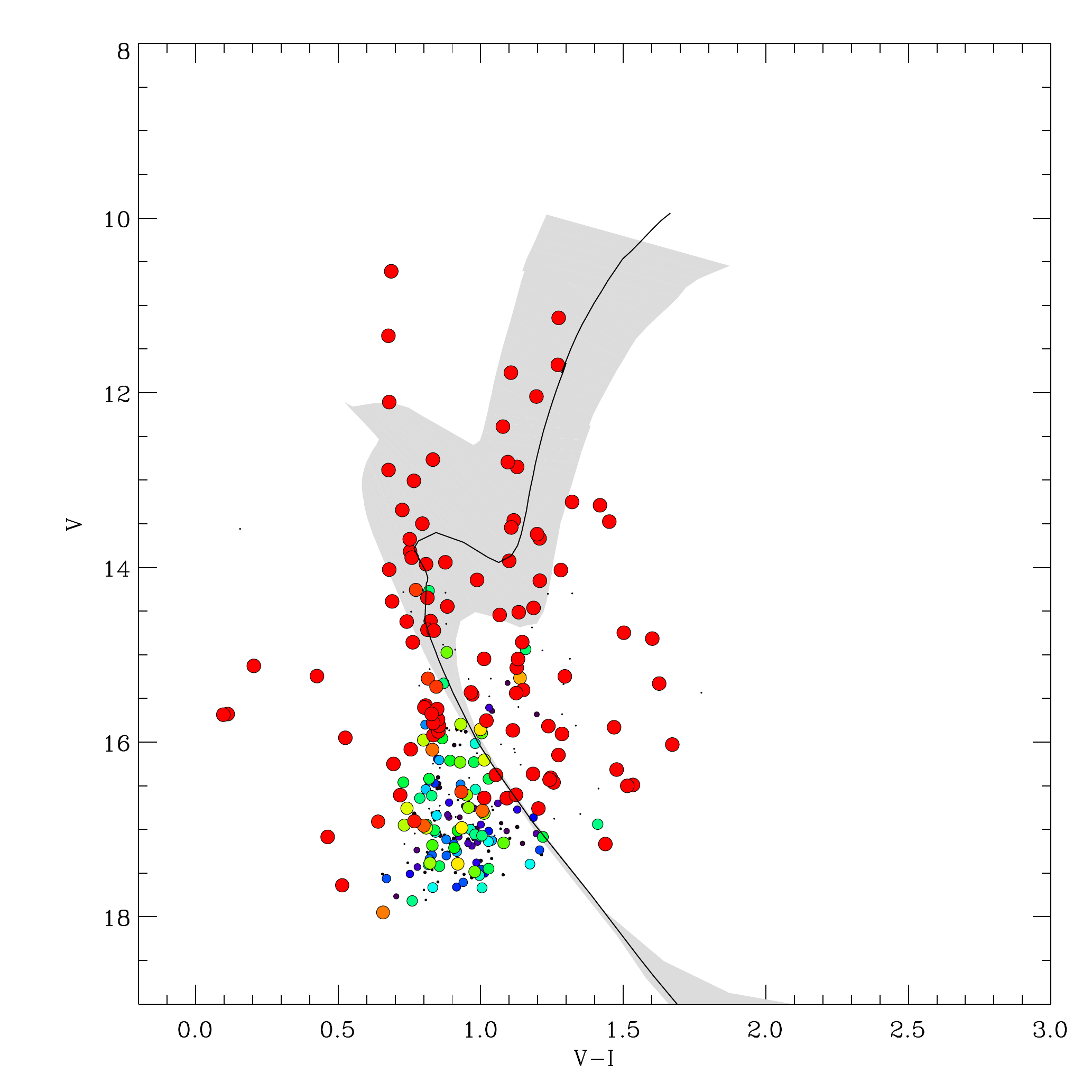}
   \caption{Same as Figure \ref{bh200}, but for ESO397-01.}
              \label{ESO397-01}%
\end{figure*}

\begin{figure*}
\centering
\includegraphics[height = 6.0cm, width = 6.0cm]{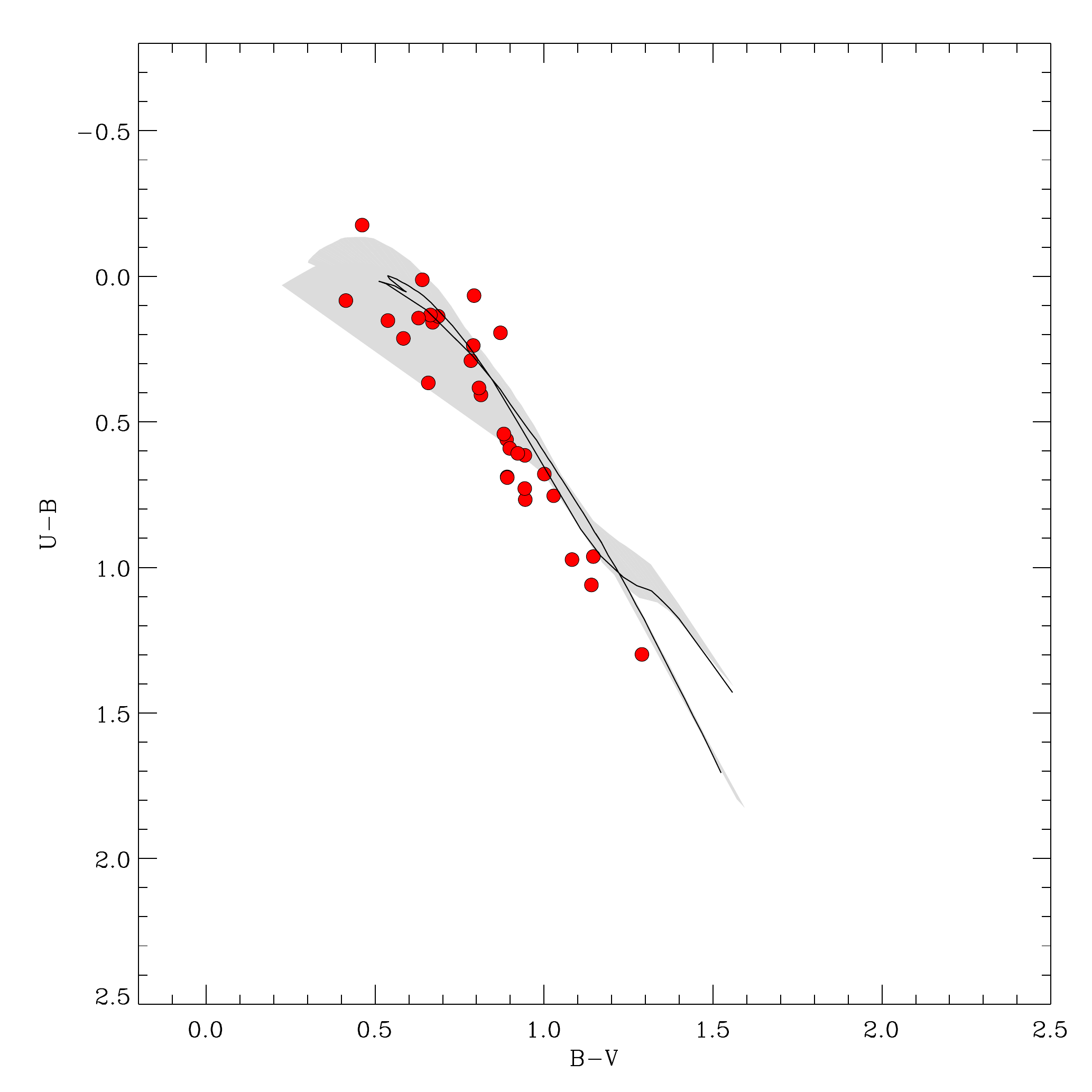}
\includegraphics[height = 6.0cm, width = 6.0cm]{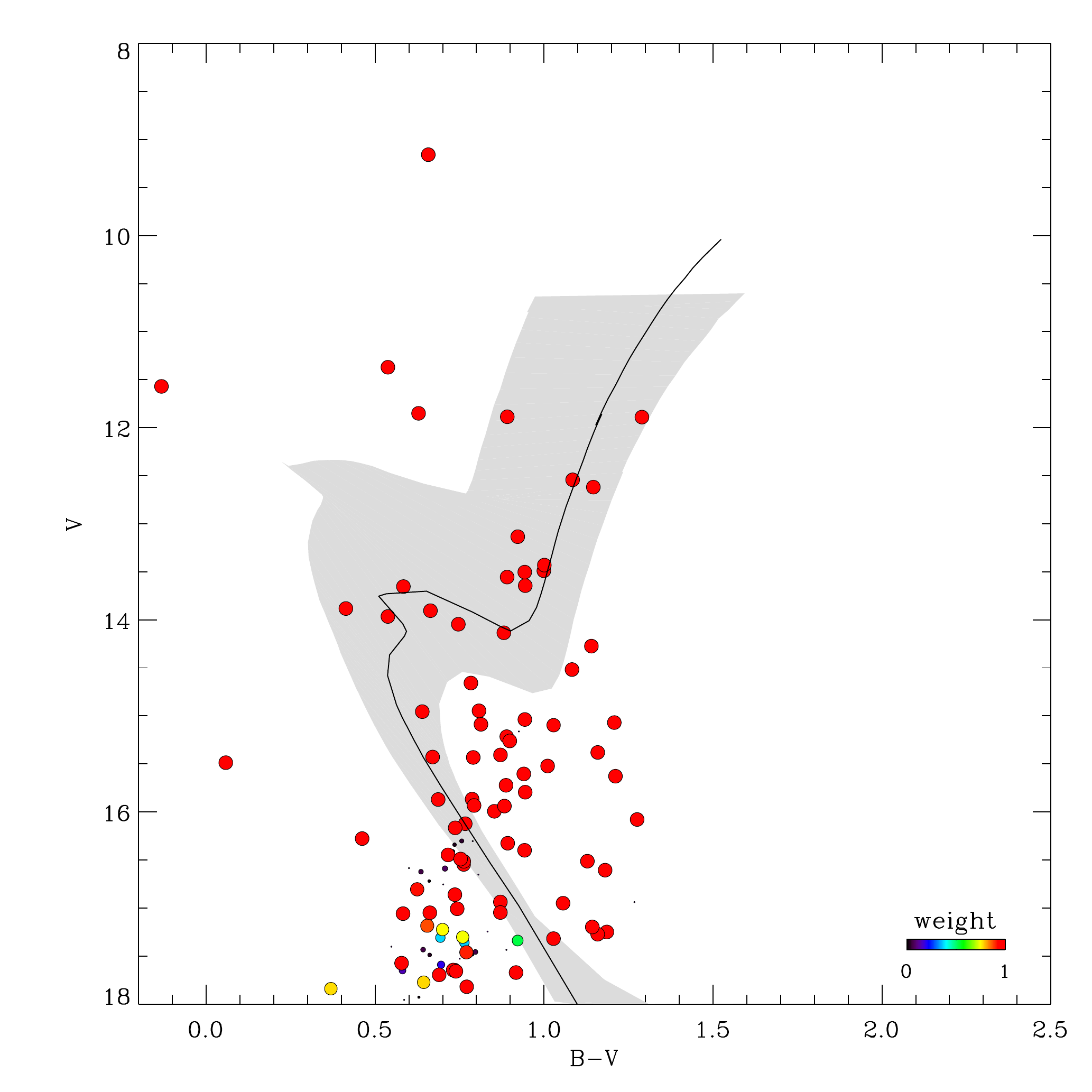}
\includegraphics[height = 6.0cm, width = 6.0cm]{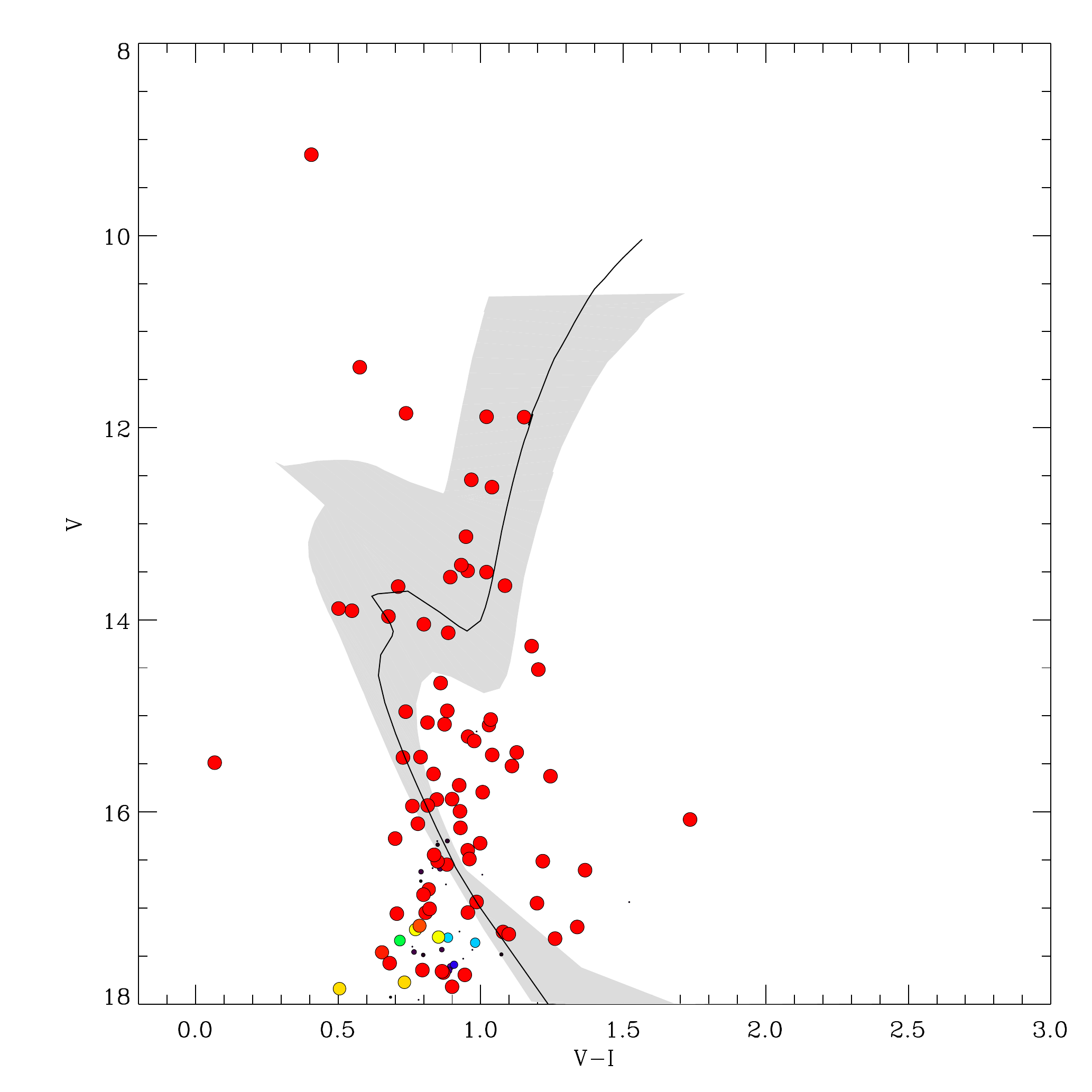}
   \caption{Same as Figure \ref{bh200}, but for ESO447-29.}
              \label{ESO447-29}%
\end{figure*}

\begin{figure*}
\centering
\includegraphics[height = 6.0cm, width = 6.0cm]{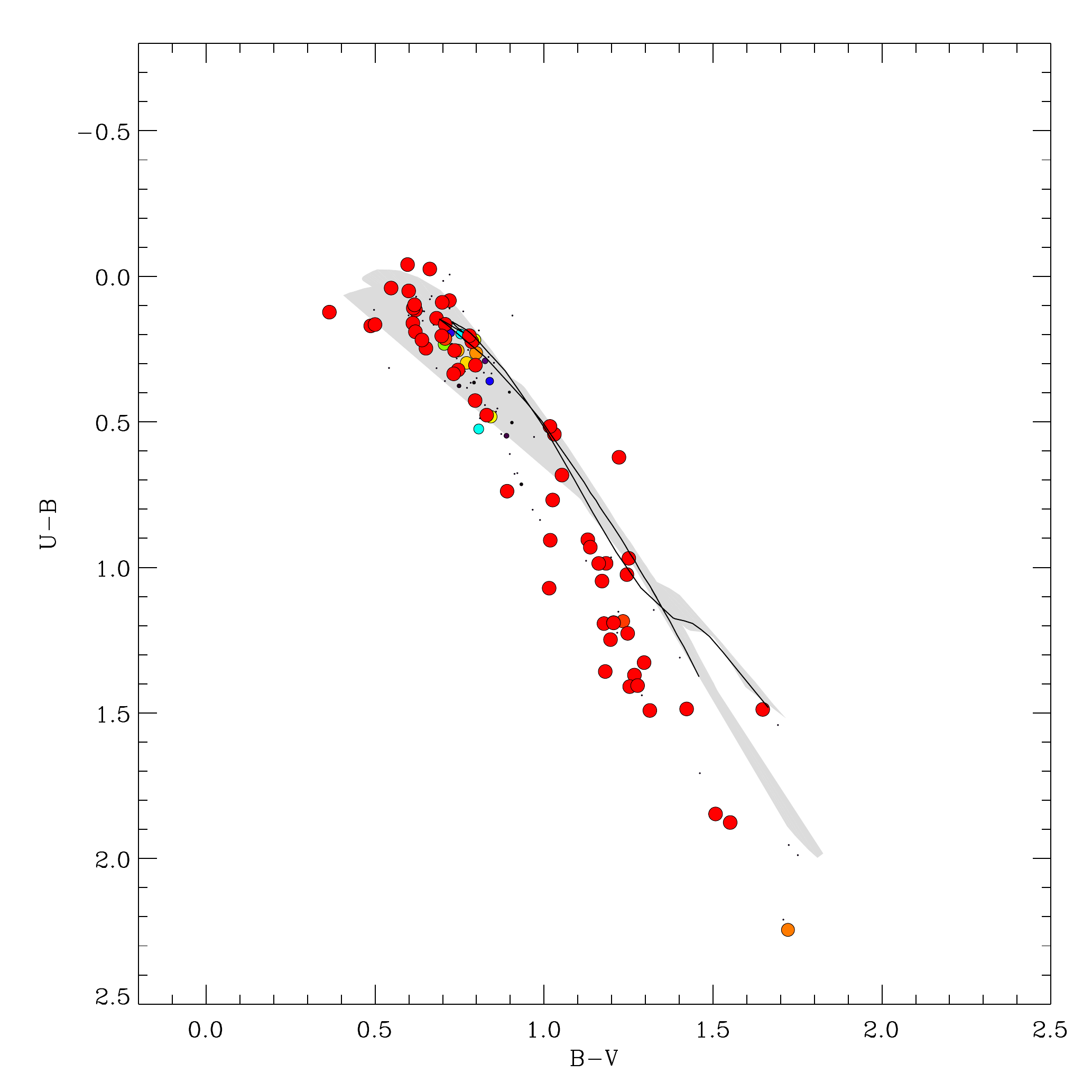}
\includegraphics[height = 6.0cm, width = 6.0cm]{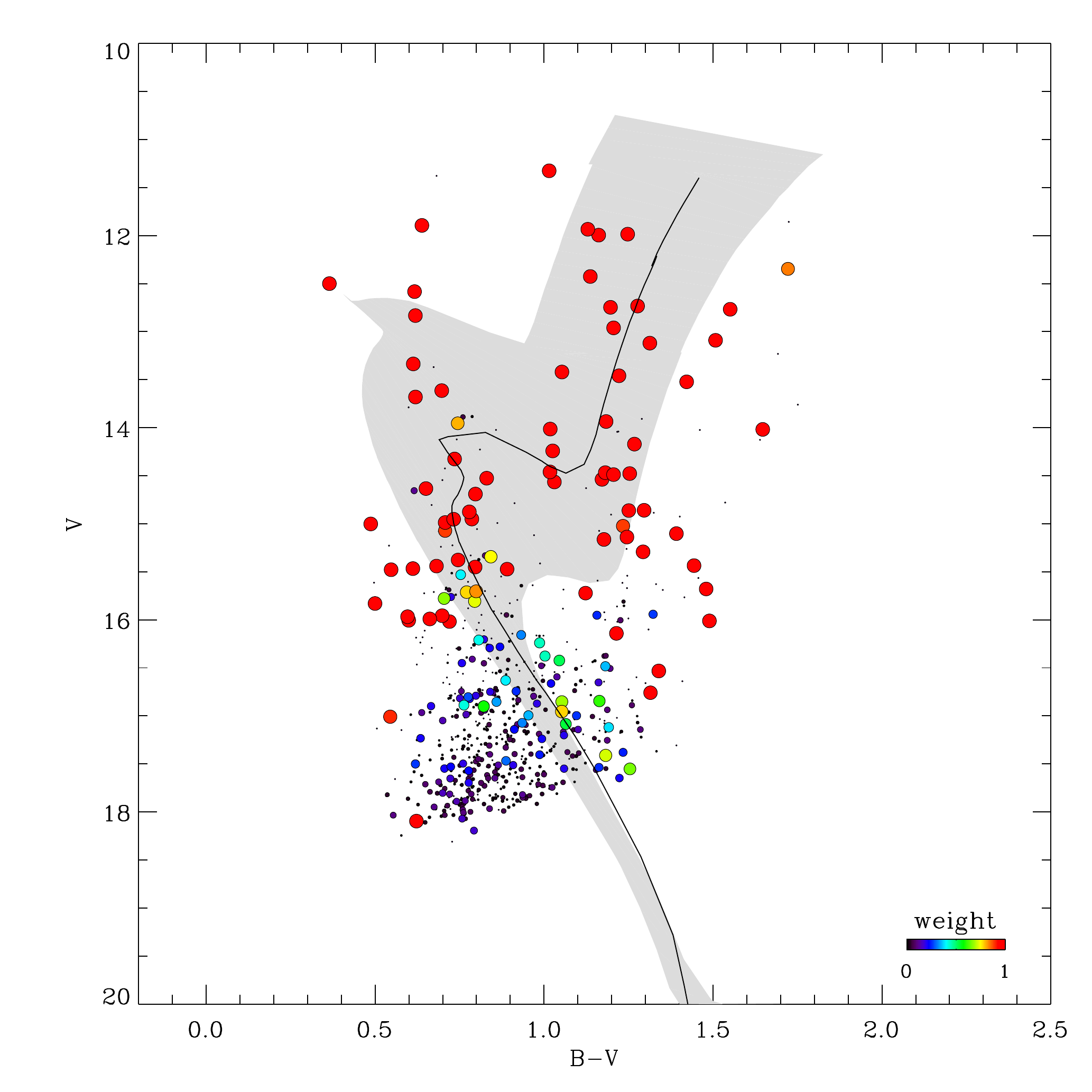}
\includegraphics[height = 6.0cm, width = 6.0cm]{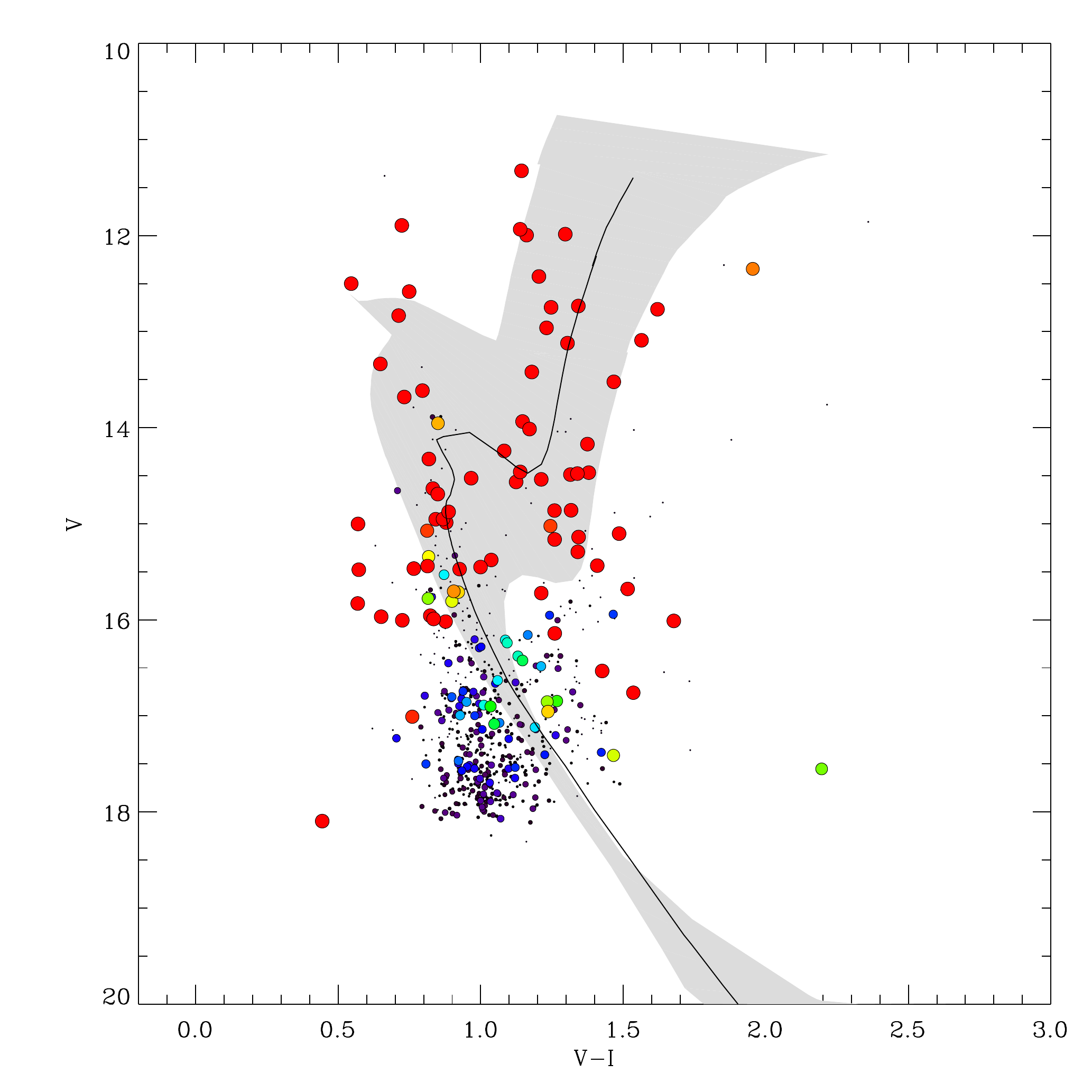}
   \caption{Same as Figure \ref{bh200}, but for NGC6840.}
              \label{NGC6840}%
\end{figure*}

\begin{figure*}
\centering
\includegraphics[height = 6.0cm, width = 6.0cm]{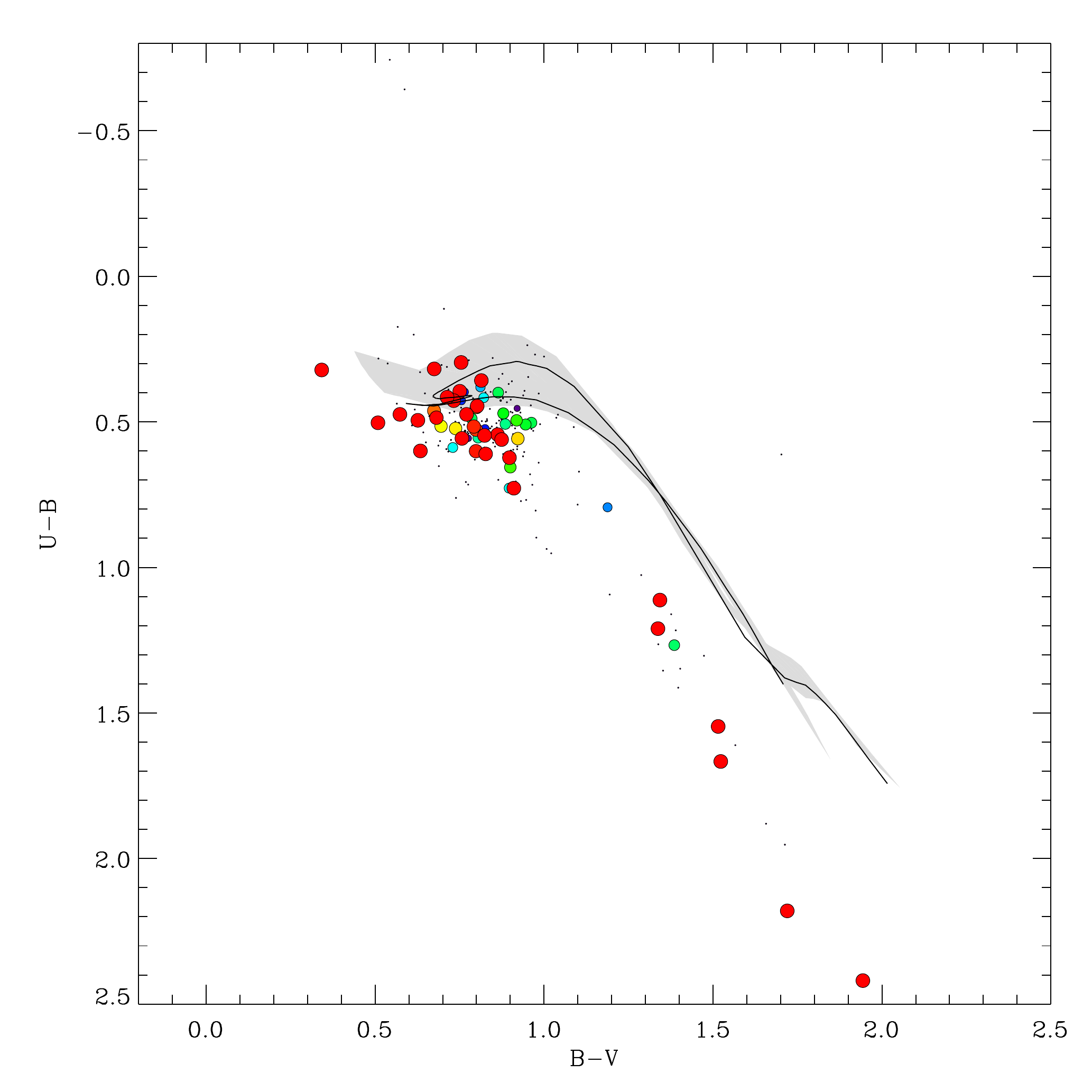}
\includegraphics[height = 6.0cm, width = 6.0cm]{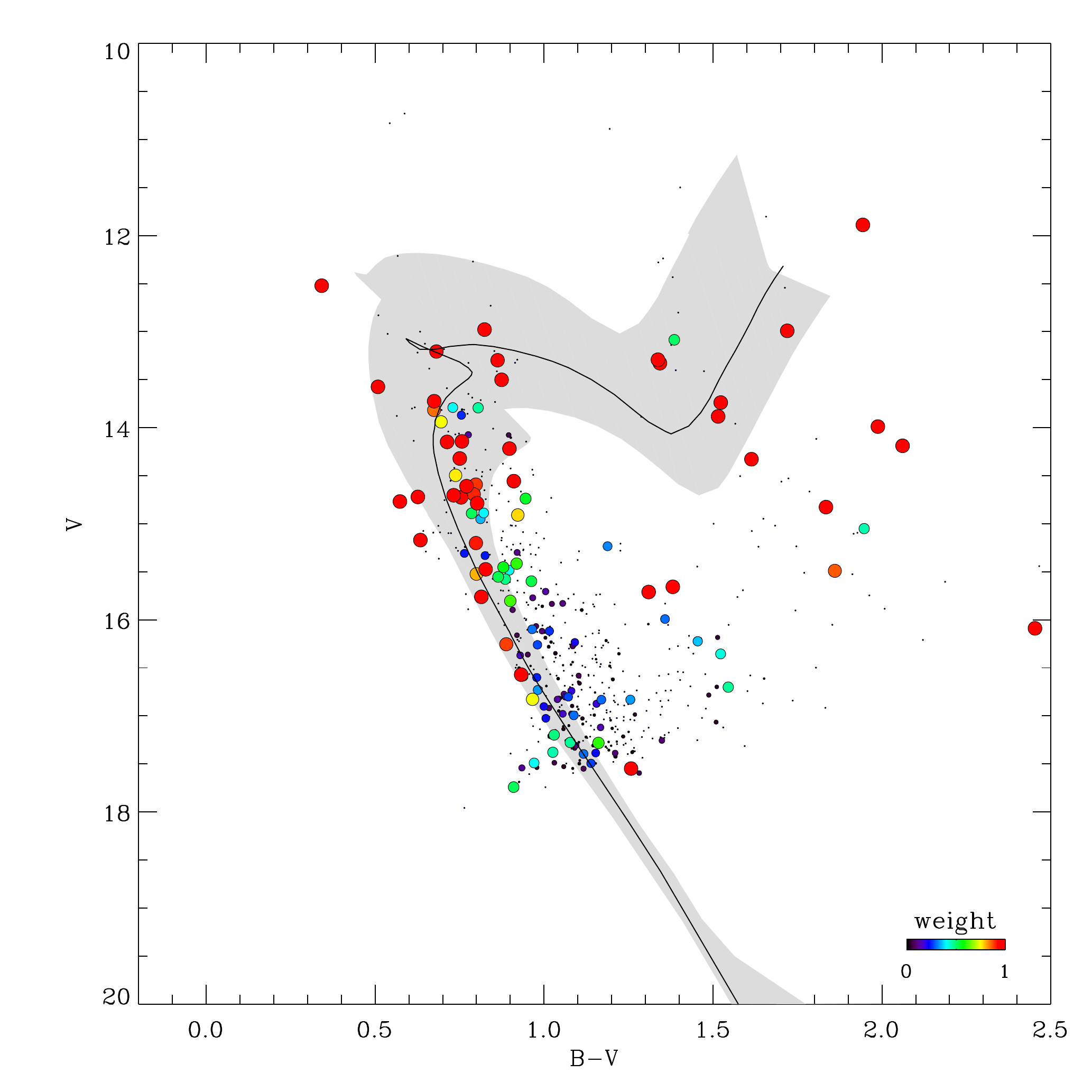}
\includegraphics[height = 6.0cm, width = 6.0cm]{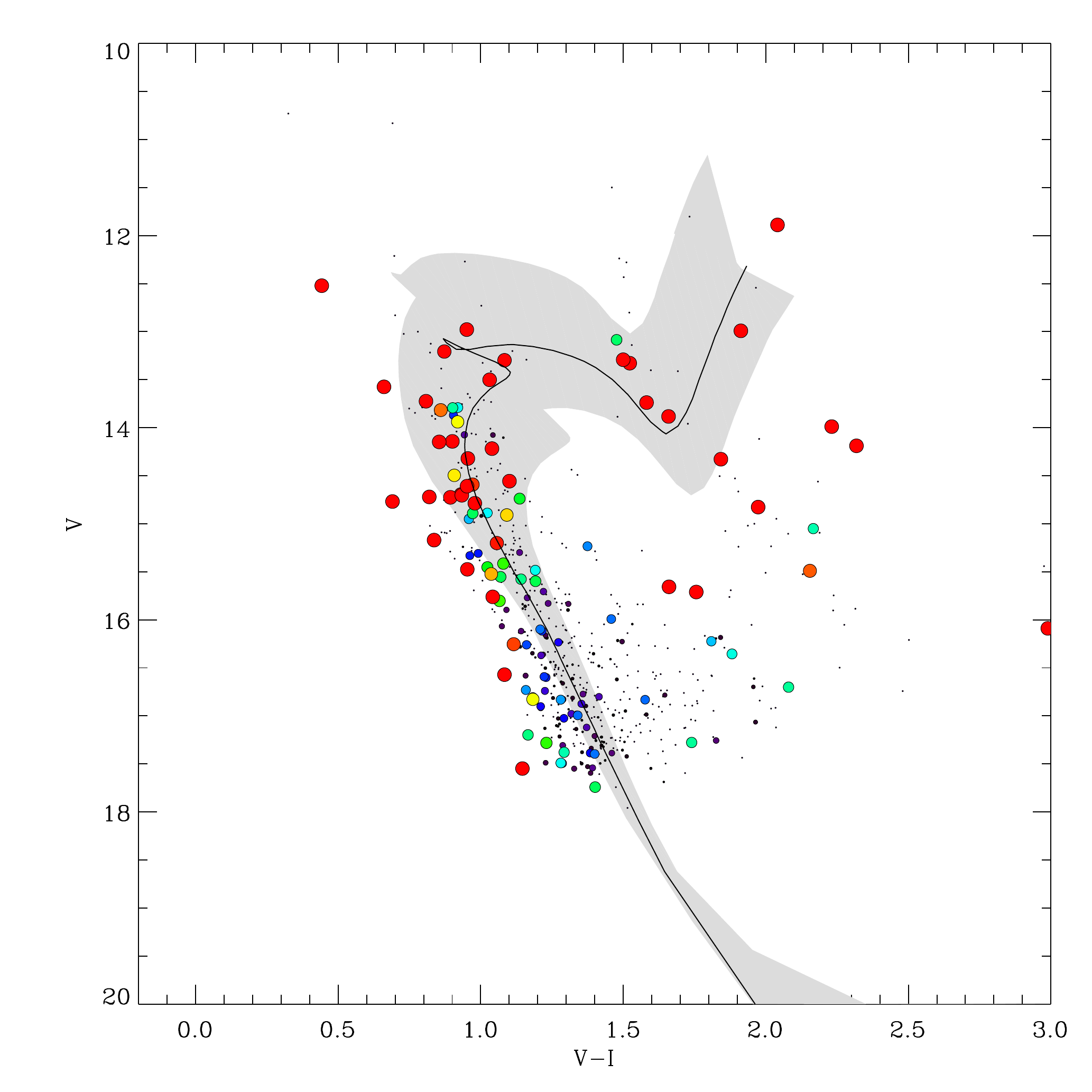}
   \caption{Same as Figure \ref{bh200}, but for Ruprecht111.}
              \label{Rup111}%
\end{figure*}

In Table 2 we present the final results for the 24 clusters
investigated in this work.  The isochrones fitted to the data for each
cluster are shown in Figures \ref{bh200} through \ref{Rup121} as well
as Figures \ref{dol33} and \ref{Rup100} and details are discussed
below. We point out that the studied clusters in this work have not
had UBVRI photometry obtained previously and fundamental parameters
values available in the DAML02 catalog up to now were mostly from
other authors using 2MASS data. For the 14 new clusters in this
  work, this is the first determination of reddening, distance, age
  and metallicity based on high quality CCD UBVRI photometry.

Below a detailed discussion of some of the most interesting clusters
studied is presented.

\subsection{Asterisms or real clusters?} 

As reported in paper I several candidates observed in this project may
not be a real open clusters. Some cases have unclear CMDs which are
difficult to analyze, even with automated statistical
tools. Therefore, as in paper I, we opted to also perform a visual
inspection of multiple data sources and diagnostics to provide better
constrains to the code parameters used in the fitting procedure. Other
more subjective criteria were also used in the final decision to
indicate if the object is or not a real cluster. 

One of the first aspects we evaluate, even before performing the
  fit, is the CMD of the object, which we plot with the estimated
  photometric membership likelihood for the stars (as in Figures
  \ref{bh200} through \ref{Rup121}). This allows us to gauge the level
  of contamination present in the data and make decisions on weather
  restrictions in the definition of cluster area (smaller radius or
  higher iso-density region) will be necessary or not. It also helps
  in deciding if additional cuts based on color or magnitude may
  improve the fitting.

  The second important aspect to be evaluated is the density map of
  the field (see for example Figure \ref{Dol33-den-vpd}). With this
  plot we can evaluate if the over-density, which is used to
  automatically determine the cluster center, is clearly present. It
  may be the case that the over-densities cannot be separated from
  random density fluctuations of the non cluster stars spatial
  distribution. In such cases, although the code finds a fit, we tend
  to take lack of clear over-density as indication that the object may
  not be a real cluster. The density plot also lets the user decide if
  a radial definition of cluster region is adequate or if a
  iso-density would be more appropriate such as in cases where the
  typical King profile distribution is not present.

  A third aspect which we have also incorporated in our quality
  control of the results is the evaluation of all the available proper
  motion information by visual inspection of proper motion charts
  combined with the CMD plot such as displayed on Figure
  \ref{Dol33-den-vpd}. The code also produces a plot of the sky chart
  of the field with the proper motion vectors over-plotted. In both
  cases we evaluate visually if there is any coherence in the proper
  motion vectors, taking into account the known errors in the UCAC4
  catalog. We are working in ways to implement an automation of this
  procedure into the code. Also related to the proper motions is the
  evaluation of results from kinematic analysis of \citet{Dias2014}
  when available where we compare proper motions to cluster proper
  motions obtained in that work.

A forth consideration is also made based on visual inspection of
available catalog images such as those from 2MASS and most
frequently the ones from DSS. In those images we are mainly searching
for patterns that could be the result of stellar over-densities.

Finally, since the code produces probability distributions for the
estimated parameters, we can also use this information, in particular
the shape of the distributions to infer some properties of the
solution. For well defined clusters the distributions are usually
gaussian in shape. Distributions that tend to be flat are the result
of degeneracy of that particular parameter. Large deviations from the
mode are also indication of poor fits. By evaluating these properties
we can subjectively decide if the procedure is fitting a real cluster
or not.  

When the analysis above gives contradicting results and the cluster
being studied is flagged d (dubious) or nf (not found) in the DAML02
catalog, we understand that this is a strong evidence that the
candidate is not a real cluster.  This was the case for the objects
Dolidze 33, ESO 447-29 and ESO 392-13 as discussed below.

\subsection{Dolidze 33} 

The candidate Dolidze 33 had originally a diameter of 9 arcmin and
flag d (dubious) in DAML02 catalog.  In Figure \ref{Dol33-den-vpd} we
present the density map and the CMD with stellar vector proper motion
over-plotted.  Although the density map shows a region with a density
2.5 times greater than the field, the number of stars is small
($5 stars/arcmin^2$). No initial selection criterion, either by radius
or density, produced a clear signature in the CMD.  There are only
nine stars with $V \leq 14$ of which seven had membership greater than
$51\%$. However these stars have incoherent proper motions. Finally
the isochrone fit obtained by the CE (see Figure \ref{dol33}) is not
satisfactory for the blue filters (B-V) and red filters (V-I) and does
not present a Gaussian distribution of solutions for the age.

Considering the criteria mentioned above we decided to classify the
Dolidze 33 as an asterism and not a real cluster.

\subsection{ESO 447-29} 

The candidate ESO 447-29 seems to be made up of the brightest stars in
the observed field of about 10 arcmin.  However there is no clear open
cluster feature in the density map or the CMD as one can see in the
Figures \ref{ESO447-29} and \ref{dens-VPD-ESO447-29}. In the density
diagram the density peak is only two times greater than background as
can be seen in Fig. \ref{dens-VPD-ESO447-29}. The CMD with VPDs also
shows that the brightest stars of the field ($V \leq 14$) have no
similar proper motion as can be seen in Fig. \ref{dens-VPD-ESO447-29}.

One issue with this cluster is that the data seem to indicate that it
has a radius larger than the field of the observations. Perhaps high
quality observations with a very wide field would allow us to define
better the CLUSTER and FIELD populations. As it is with our available
data and the considerations above we decide to classify this object as
an asterism.

\subsection{ESO 392-13}			

The candidate ESO 392-13 received the flag nf in the DAML02 catalog,
indicating that it was not found in the visual inspection of the
visual image of the DSS.  In fact only one hundred stars have been
detected in our observations and the density map indicates low
contrast with respect to the background.

However, one interpretation of the CMD is a signature of a possible open
cluster with a main sequence between $V\approx12$ and $V\approx19$, a
turn-off $V\approx12$ and possible giants at $(B-V)\approx1.4$.

Inspecting the color-color diagram we see a group of stars with
$(U-B) \leq 0.25$ and $(B-V) \leq 0.75$ that are clearly not members
if we take the solution found to be true (see the Figure
\ref{ESO392-13}). However, eliminating those stars would leave a large
gap in the main sequence that would be inconsistent with a real open
cluster. This is corroborated by the inspection of the vector proper
motion of that particular group showing no clear signature of group
motion. Based on these considerations we classify this object as not a
cluster.

\subsection{Ruprecht~100}  

The open cluster Ruprecht 100 is an interesting case where the CMD
shows two apparent populations of different ages and located about the
same distance, which occur because the cluster is located toward the
Sagittarius-Carina spiral arm ($l=297.7210^{\rm o}$ and $b =
-0.2067^{\rm o}$). In this case, the addition of the possibility to
delineate cluster regions with iso-density contours helps us to obtain
a correct interpretation of the CMD and consequently a better fit to
the data, as discussed below.
 
With the density map we can choose different regions to extract the
stars initially considered as members of the cluster trying to
maximize the contrast of cluster features in relation to the field
stars in the CMDs.  In the case of Ruprecht 100, the density map
clearly shows two distinct peaks as can be seen in Fig. 16. In the fit
done in Paper I, the region selected was centered on the higher
density peak (middle panel in the Figure \ref{Rup100-panel}), with a
radius of 3 arcmin and the CMD displays a turn-off at $V\approx14$.
However, we see that the secondary peak (lower panel in the Figure
\ref{Rup100-panel}), distant about 3.5 arcmin in declination from the
center, shows a population with little difference from that presented
from the peak at the center of the image.  In the CMD it coincides
with a younger population.  In the Figure \ref{Rup100-panel} we also
present a third region taken in the field out of the density peak
which shows an expected field population.

These similarities in the two density peaks led us to the hypothesis
that they are actually all part of the same cluster which does not
have the classical King profile from a spherically symmetric
distribution of stars. This would make the turn-off at $V\approx14$ a
contamination from the content of the Sagittarius-Carina arm, mainly
constituted of a young population.  In the CMD there are giants stars
seen at different distances and reddening values, indicated also by
the somewhat random distribution of stars with $B-V > 1.5$. The final
fit for this cluster presented in the Figure \ref{Rup100} was then done with the
cluster region being defined by an iso-density region taken at 32
$star/arcmin^2$.

The scenario seen here is similar to the one reported by
\citet{Carraro2006} for two open cluster candidates in their work,
which they concluded to be just visual effects, accumulations of stars
produced by the patchy nature of the inter- stellar absorption toward
the Galactic bulge. Both clusters are in the Sagittarius-Carina arm
region. Also according to Carraro, spatial over-densities in this
direction can be effects of variable extinction. We take this
discussion as evidence that indeed the cluster is a non circular
structure with a slightly younger age as previously determined.

\subsection{Trumpler 25}
 
\begin{figure*}
\centering
\includegraphics[height = 6.0cm, width = 6.0cm]{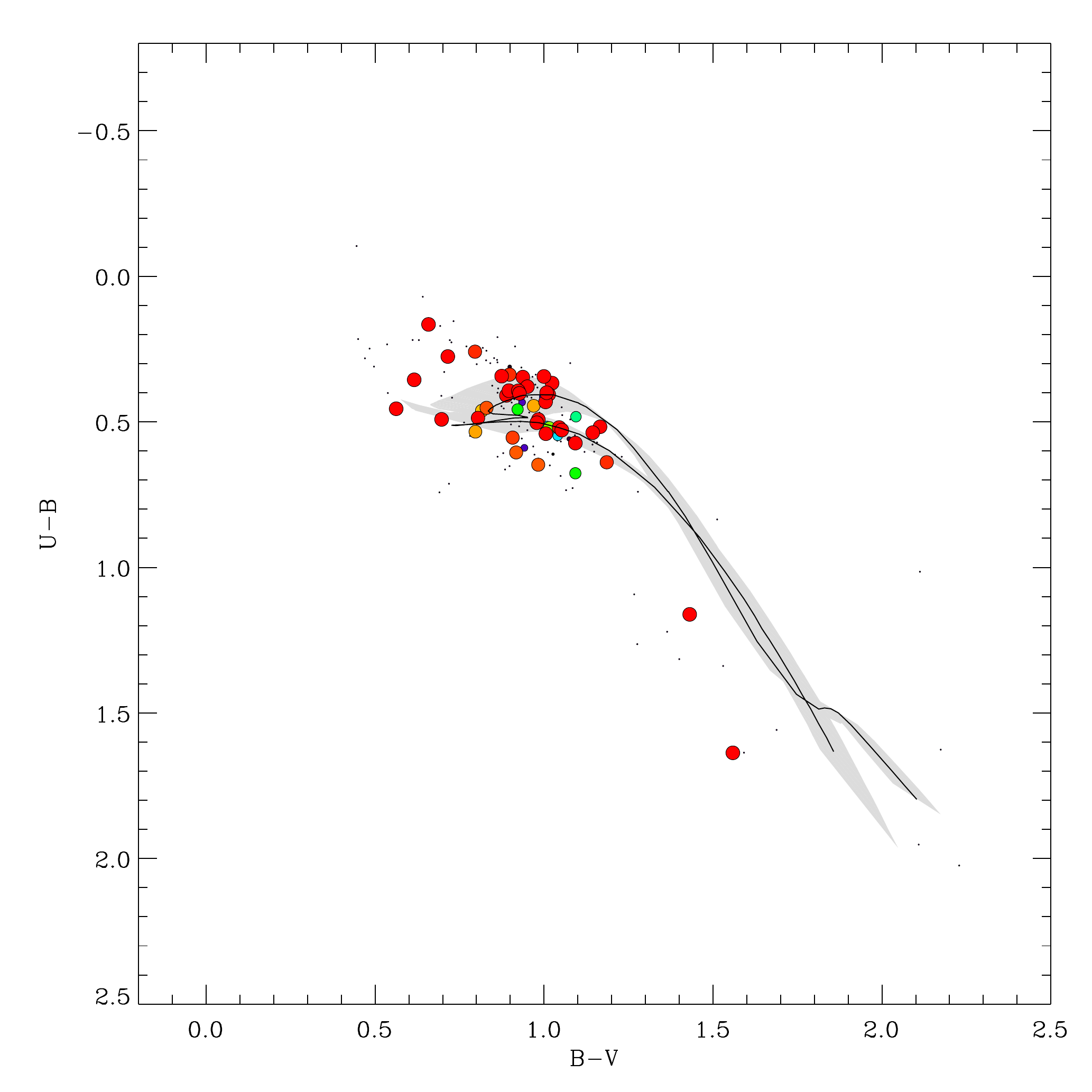}
\includegraphics[height = 6.0cm, width = 6.0cm]{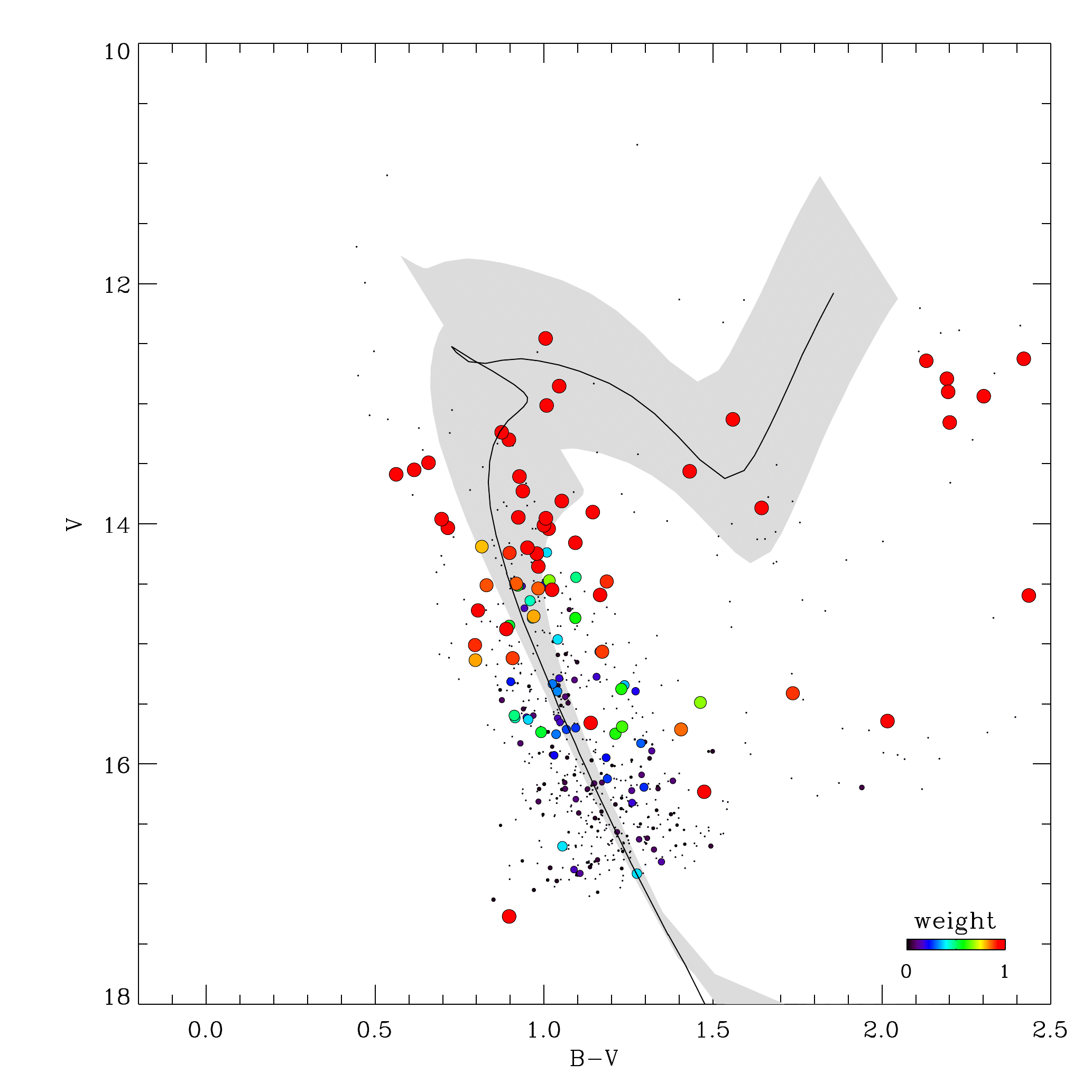}
\includegraphics[height = 6.0cm, width = 6.0cm]{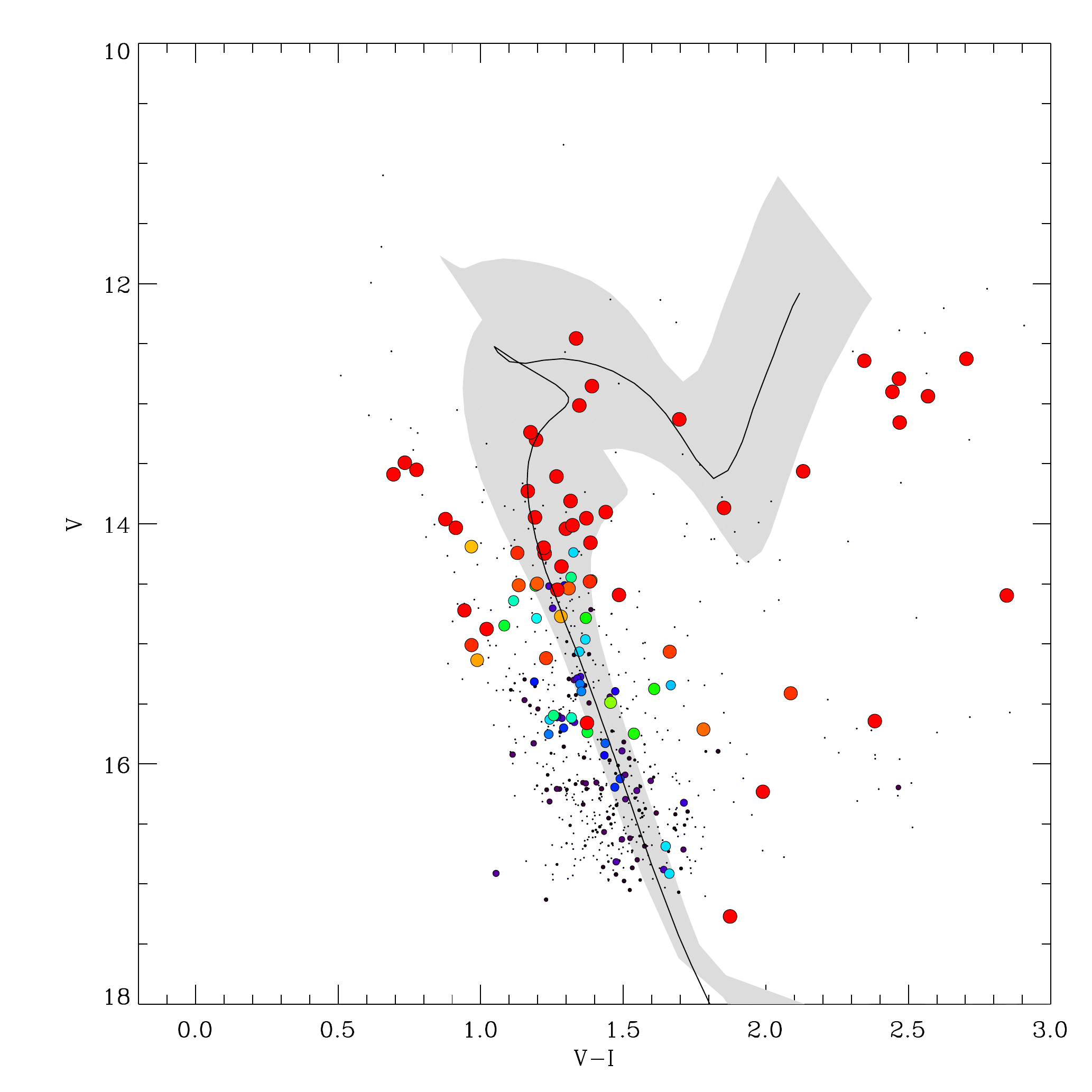}
   \caption{Same as Figure \ref{bh200}, but for Trumpler25.}
              \label{Trumpler25}%
\end{figure*}

\begin{figure*}
\centering
\includegraphics[height = 6.0cm, width = 6.0cm]{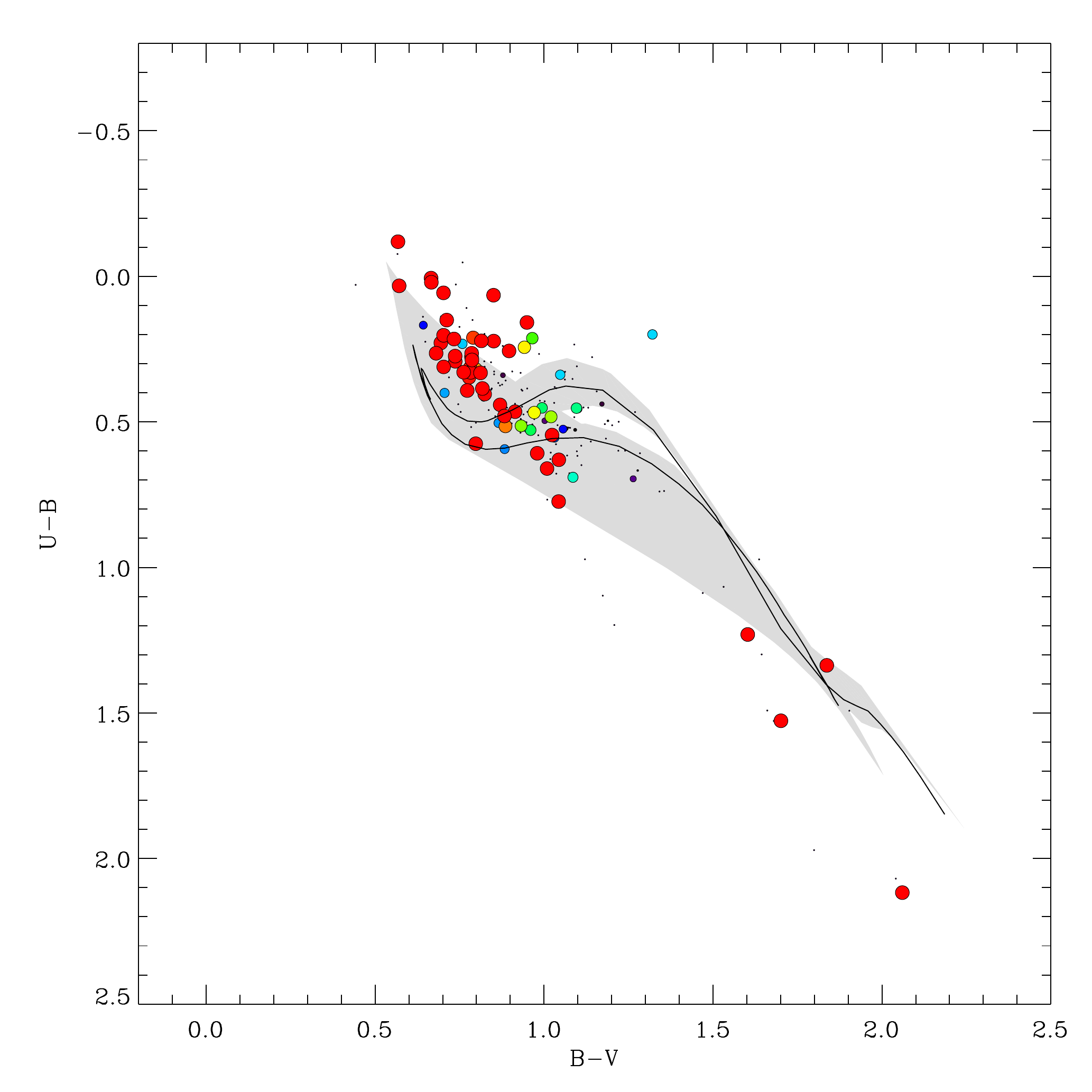}
\includegraphics[height = 6.0cm, width = 6.0cm]{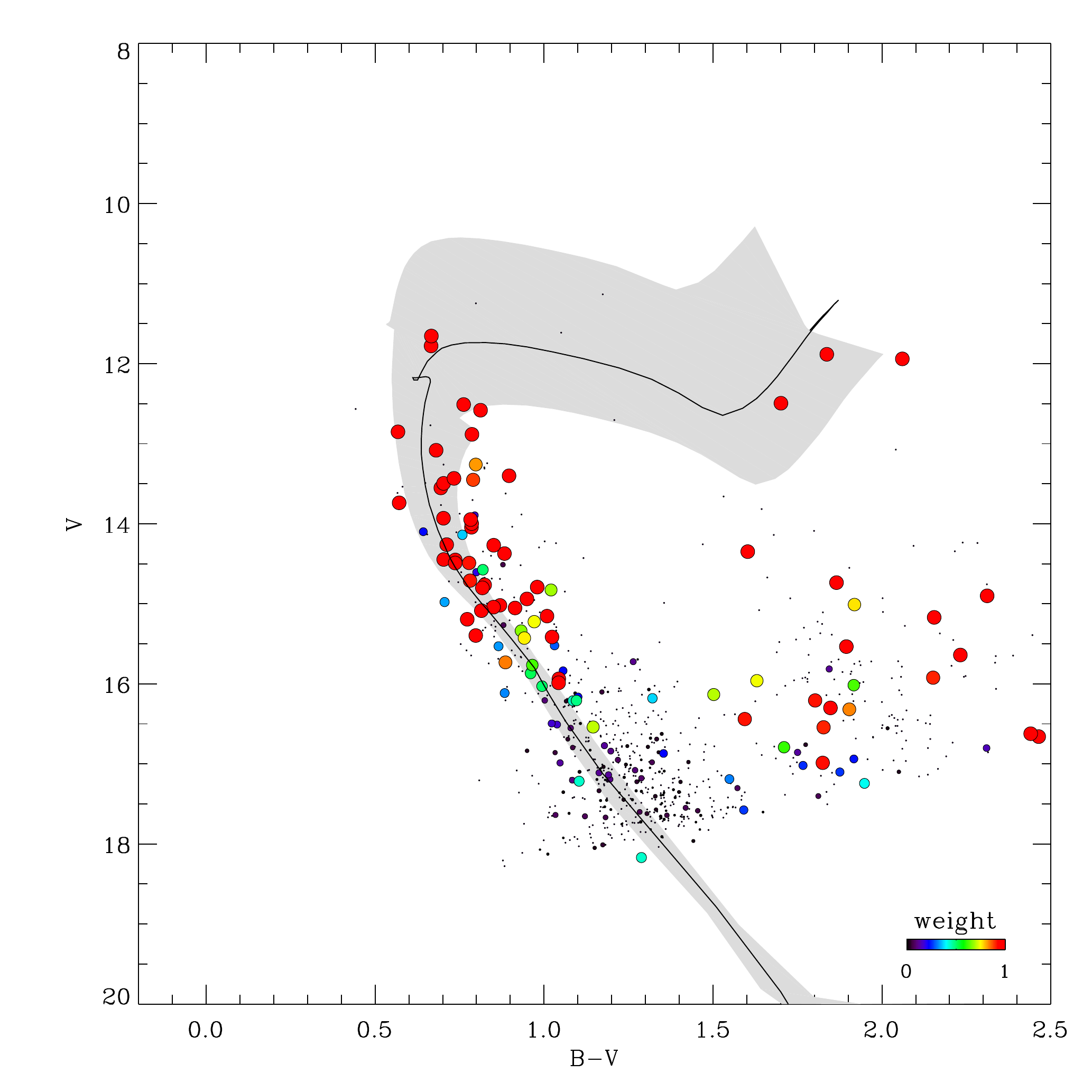}
\includegraphics[height = 6.0cm, width = 6.0cm]{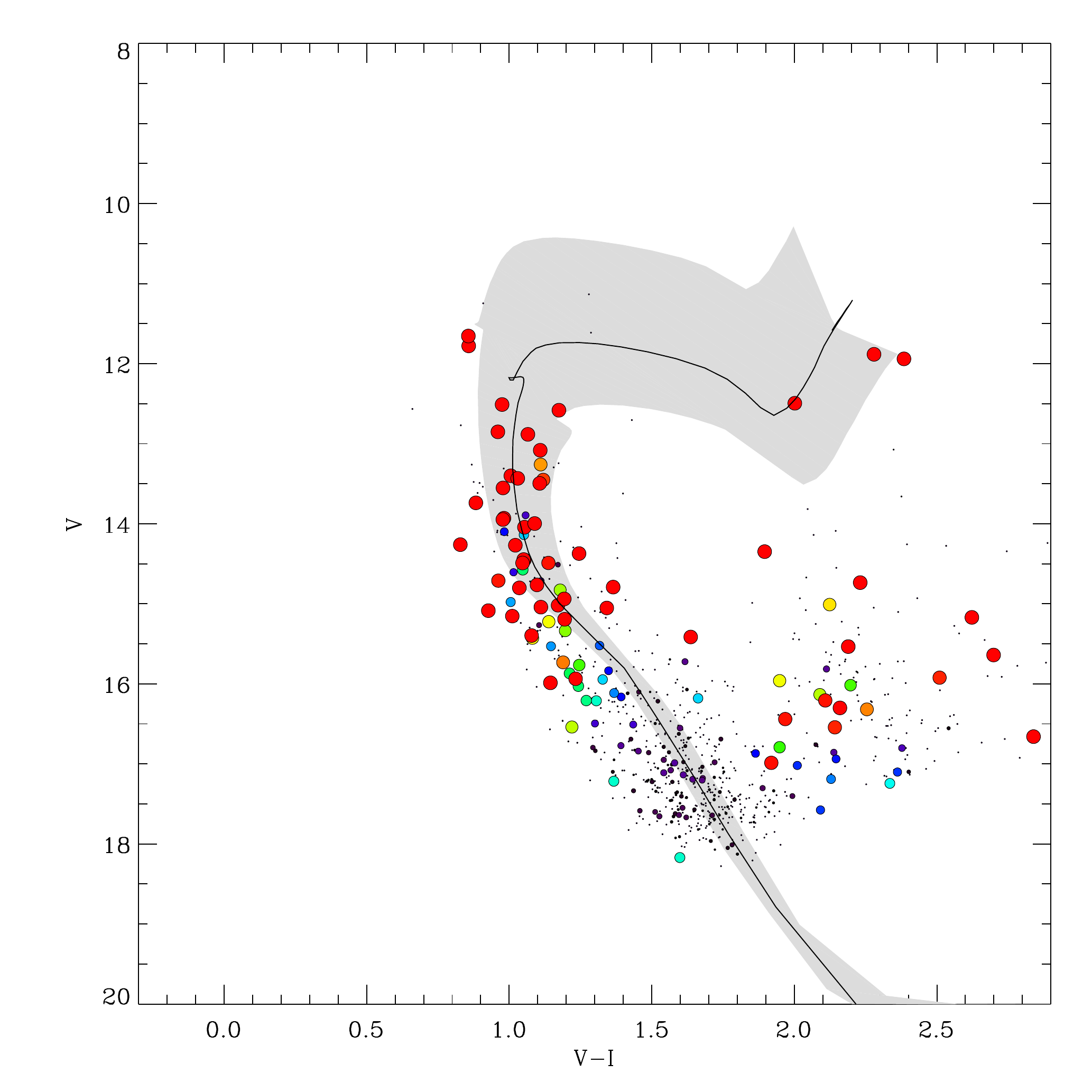}
   \caption{Same as \ref{bh200}, but for Collinder307.}
              \label{Collinder307}%
\end{figure*}

\begin{figure*}
\centering
\includegraphics[height = 6.0cm, width = 6.0cm]{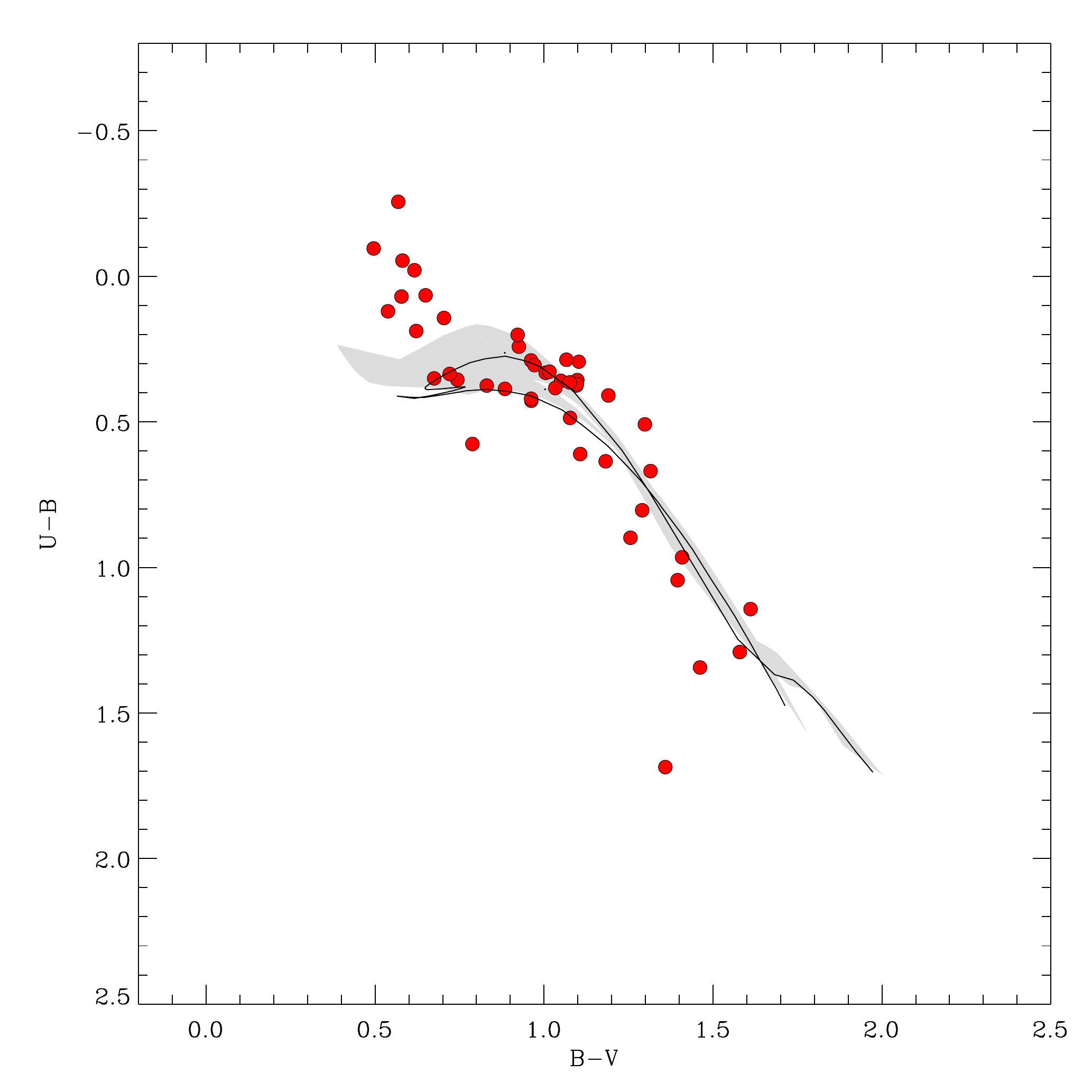}
\includegraphics[height = 6.0cm, width = 6.0cm]{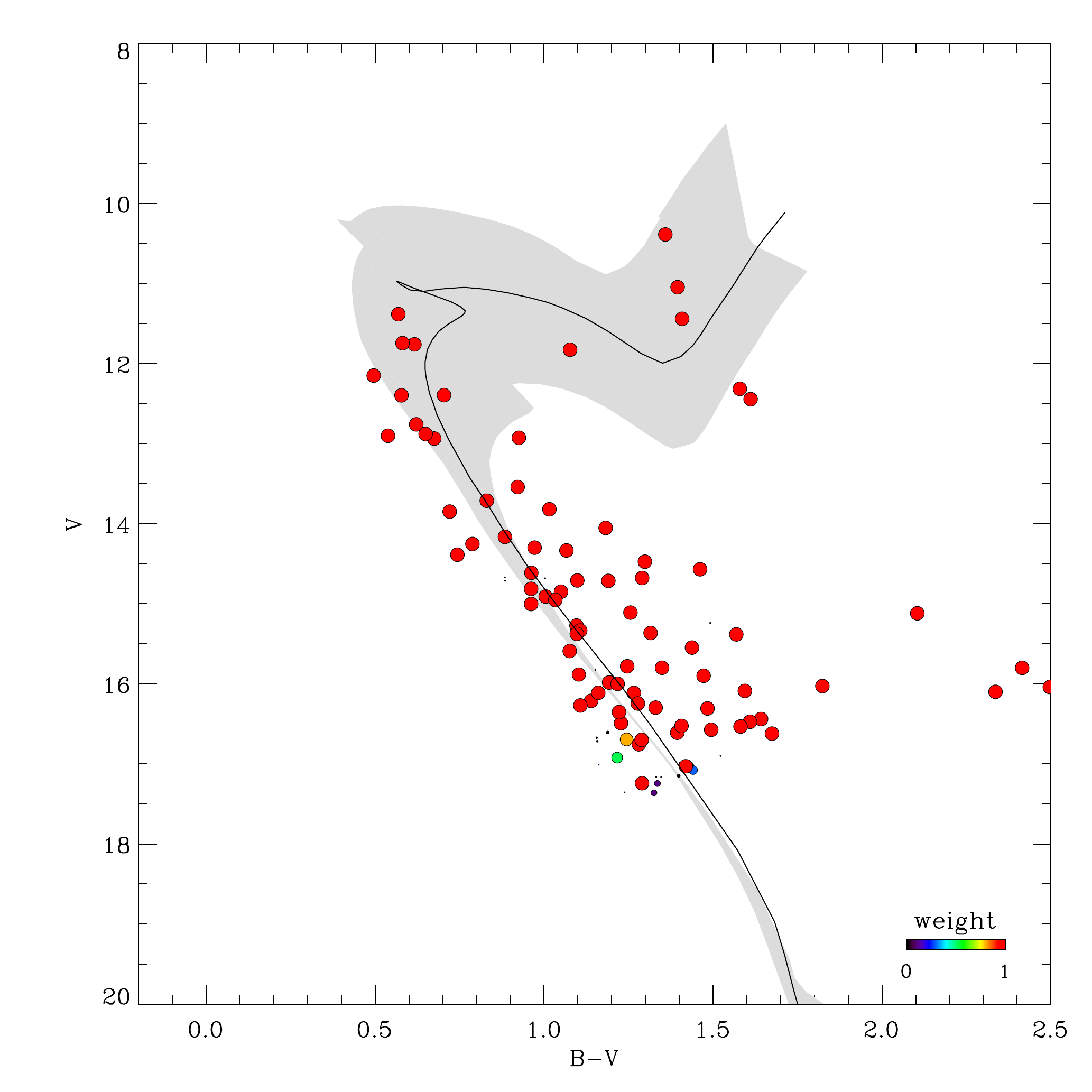}
\includegraphics[height = 6.0cm, width = 6.0cm]{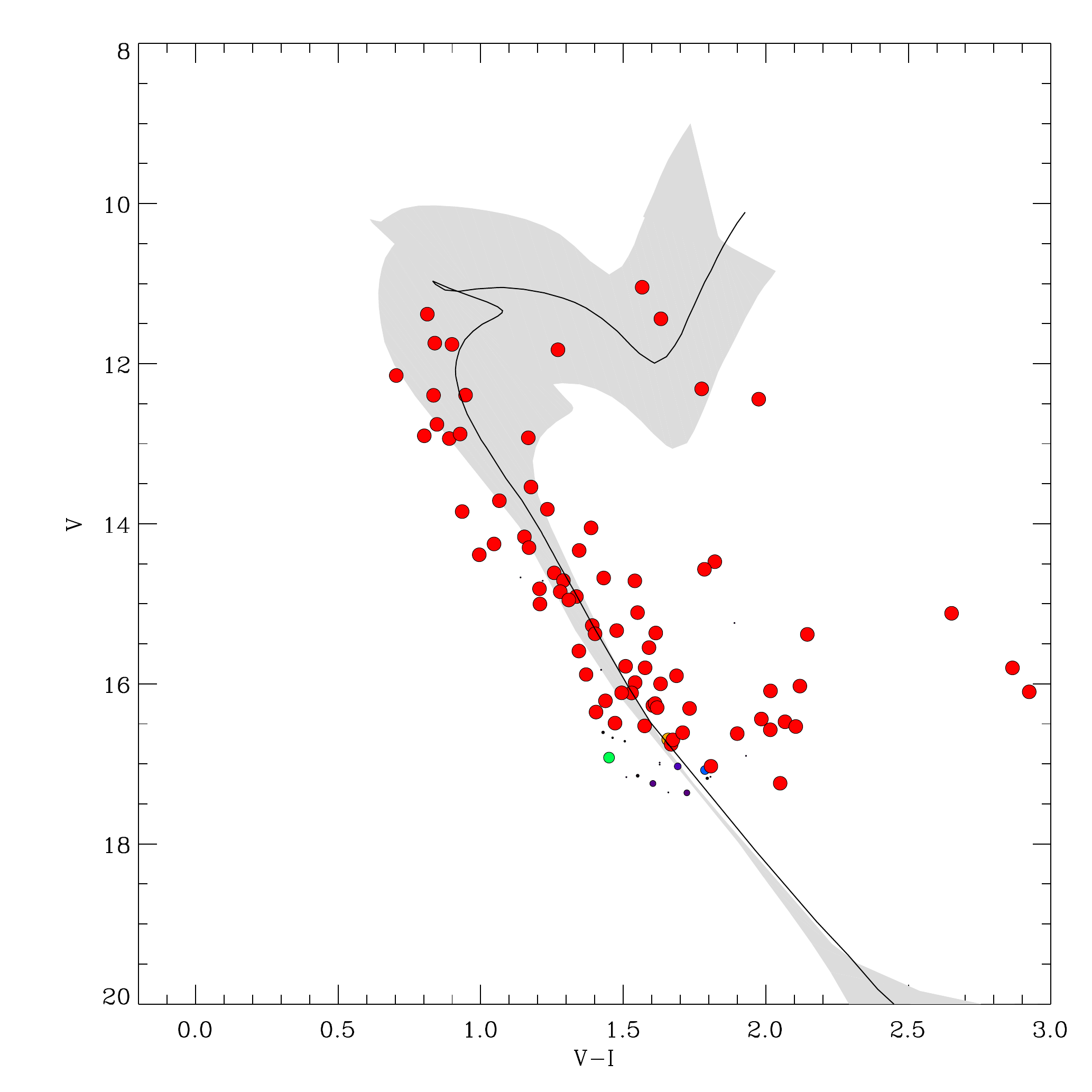}
   \caption{Same as Figure \ref{bh200}, but for ESO392-13.}
              \label{ESO392-13}%
\end{figure*}

\begin{figure*}
\centering
\includegraphics[height = 6.0cm, width = 6.0cm]{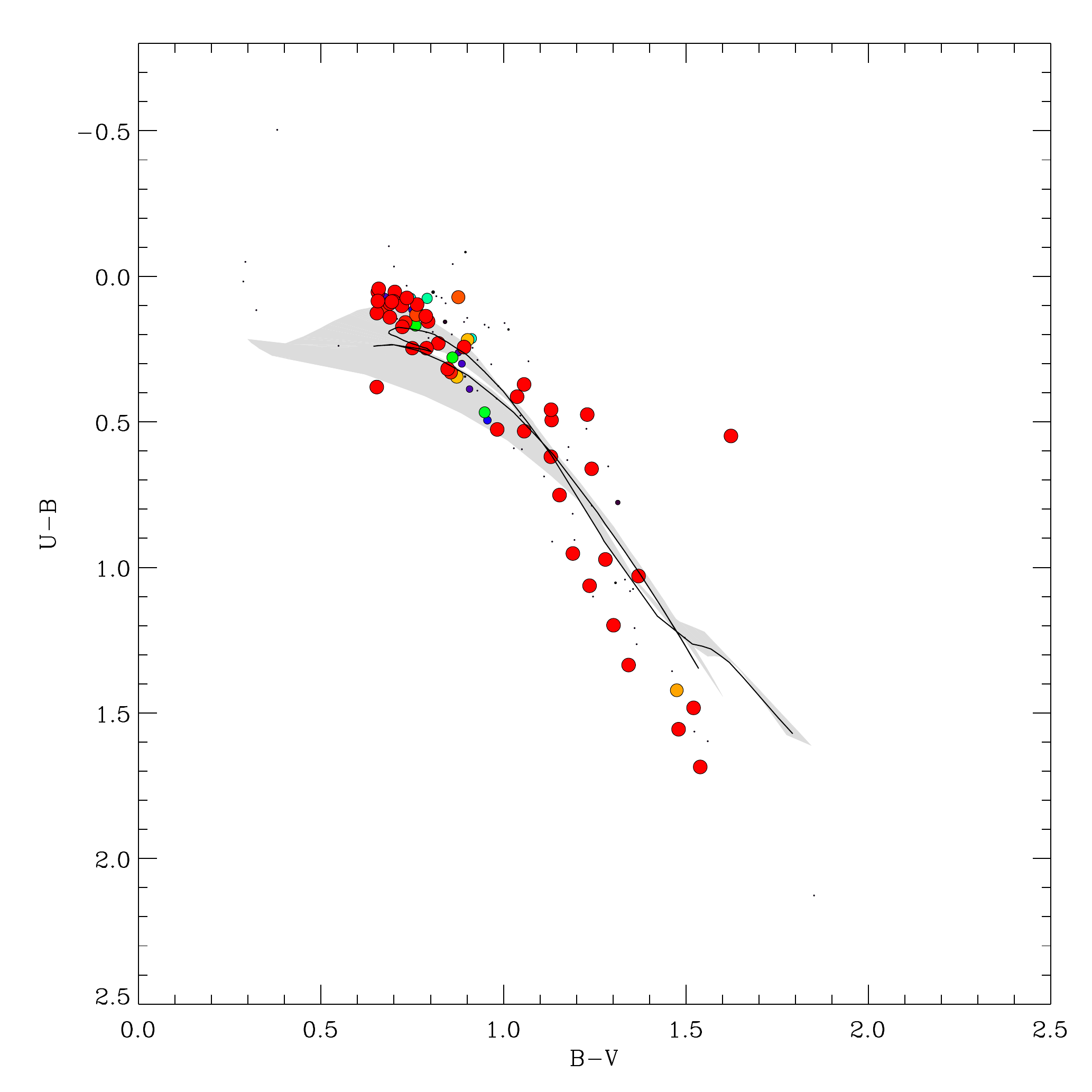}
\includegraphics[height = 6.0cm, width = 6.0cm]{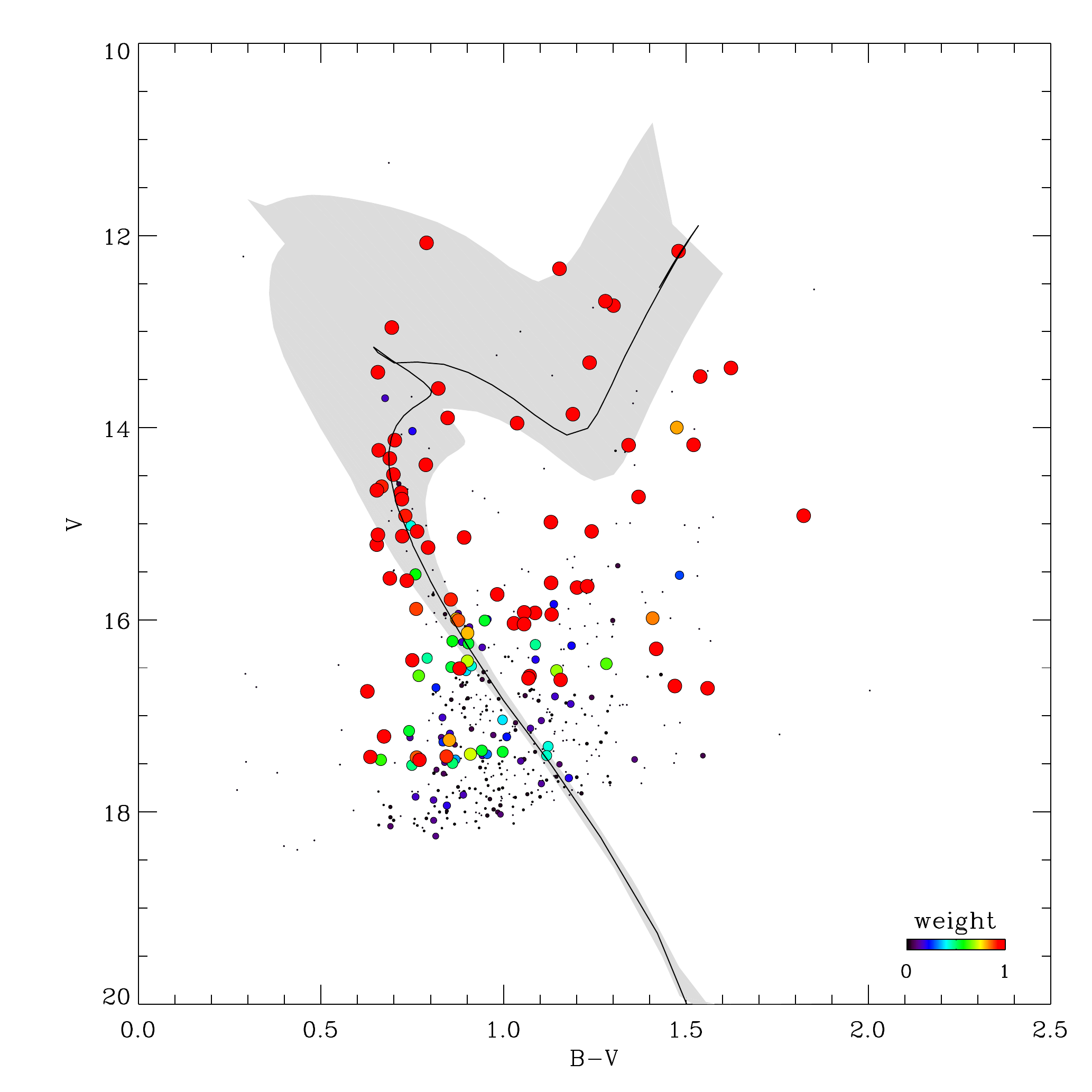}
\includegraphics[height = 6.0cm, width = 6.0cm]{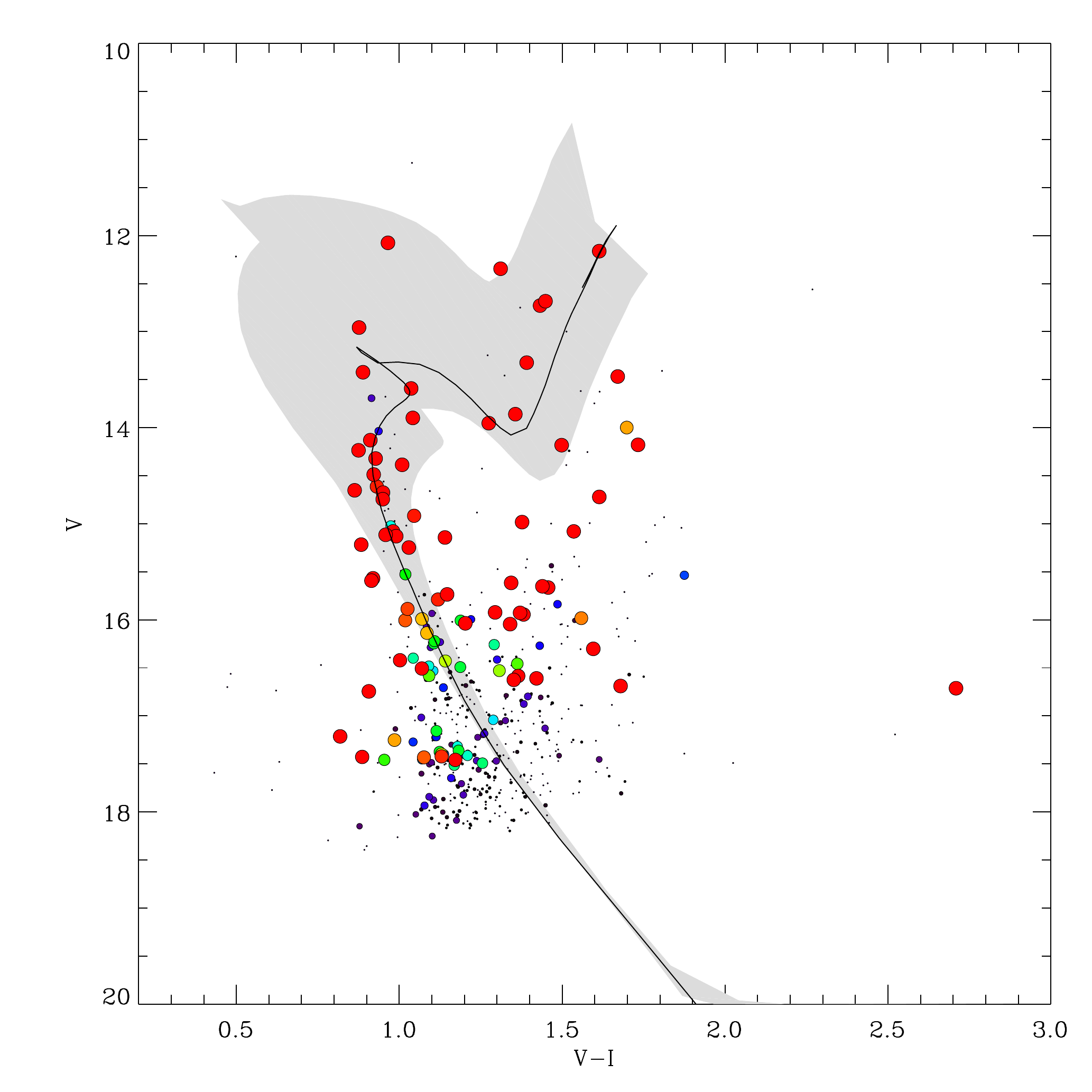}
   \caption{Same as Figure \ref{bh200}, but for ESO518-03.}
              \label{ESO518-03}%
\end{figure*}

\begin{figure*}
\centering
\includegraphics[height = 6.0cm, width = 6.0cm]{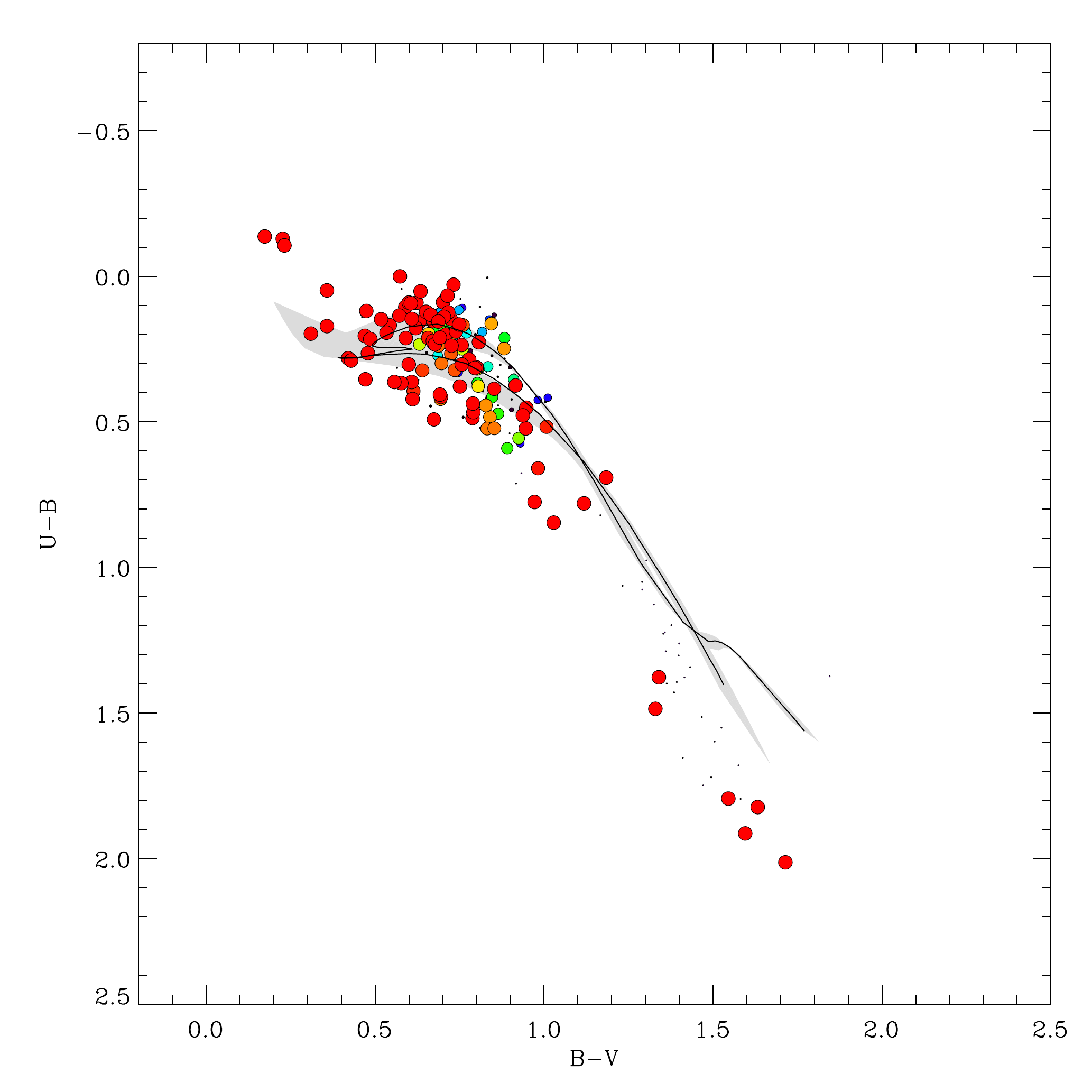}
\includegraphics[height = 6.0cm, width = 6.0cm]{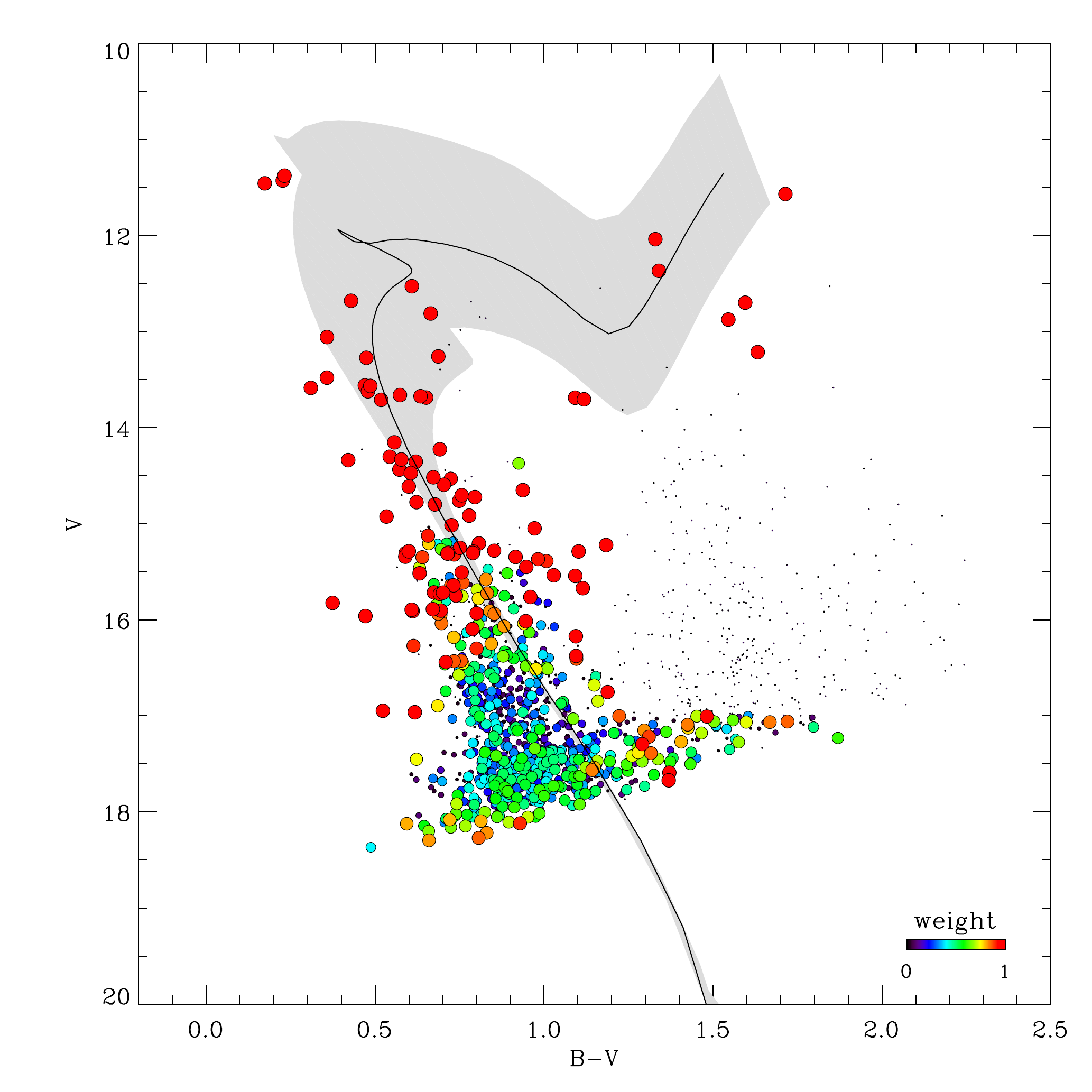}
\includegraphics[height = 6.0cm, width = 6.0cm]{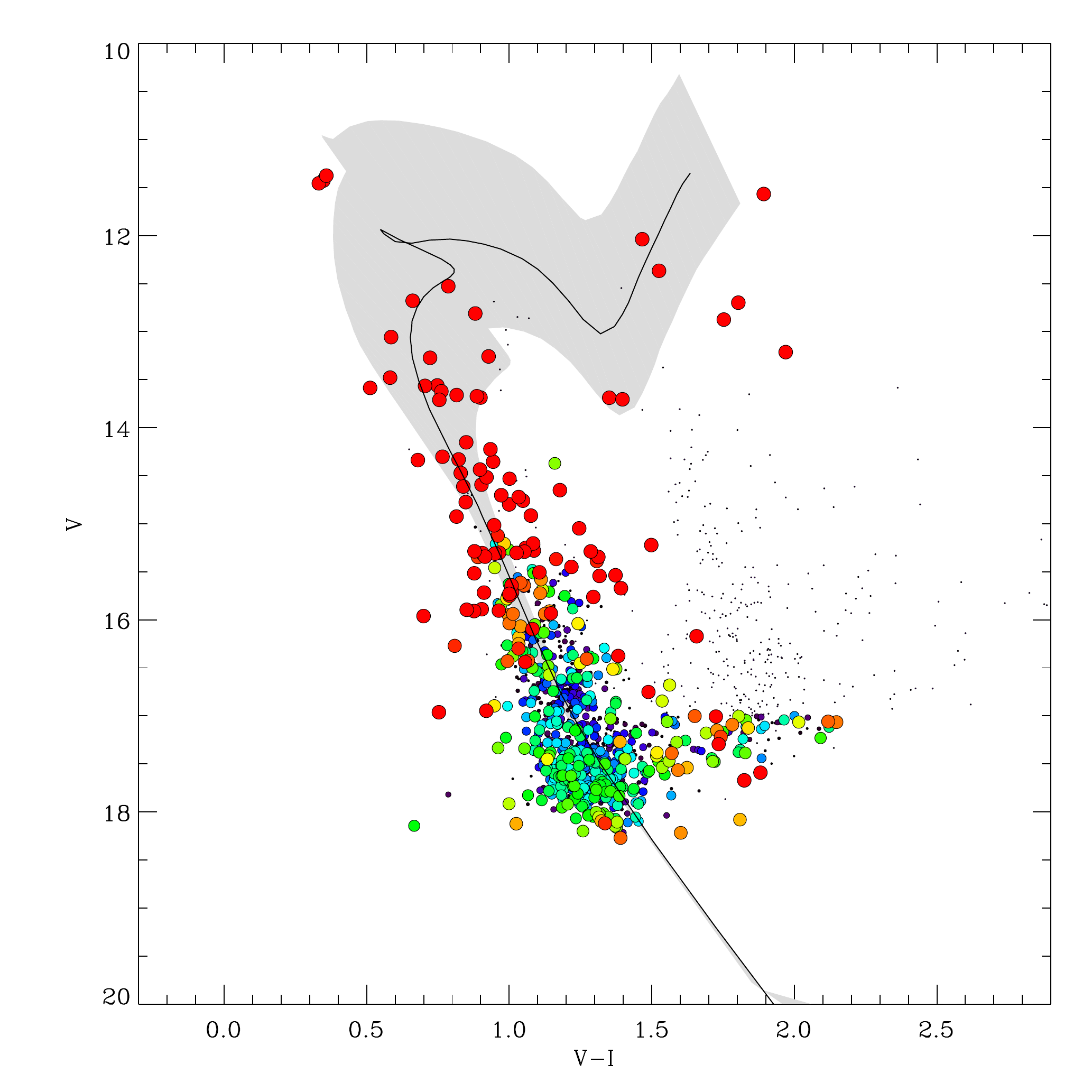}
   \caption{Same as Figure \ref{bh200}, but for Loden1002.}
              \label{Loden1002}%
\end{figure*}

\begin{figure*}
\centering
\includegraphics[height = 6.0cm, width = 6.0cm]{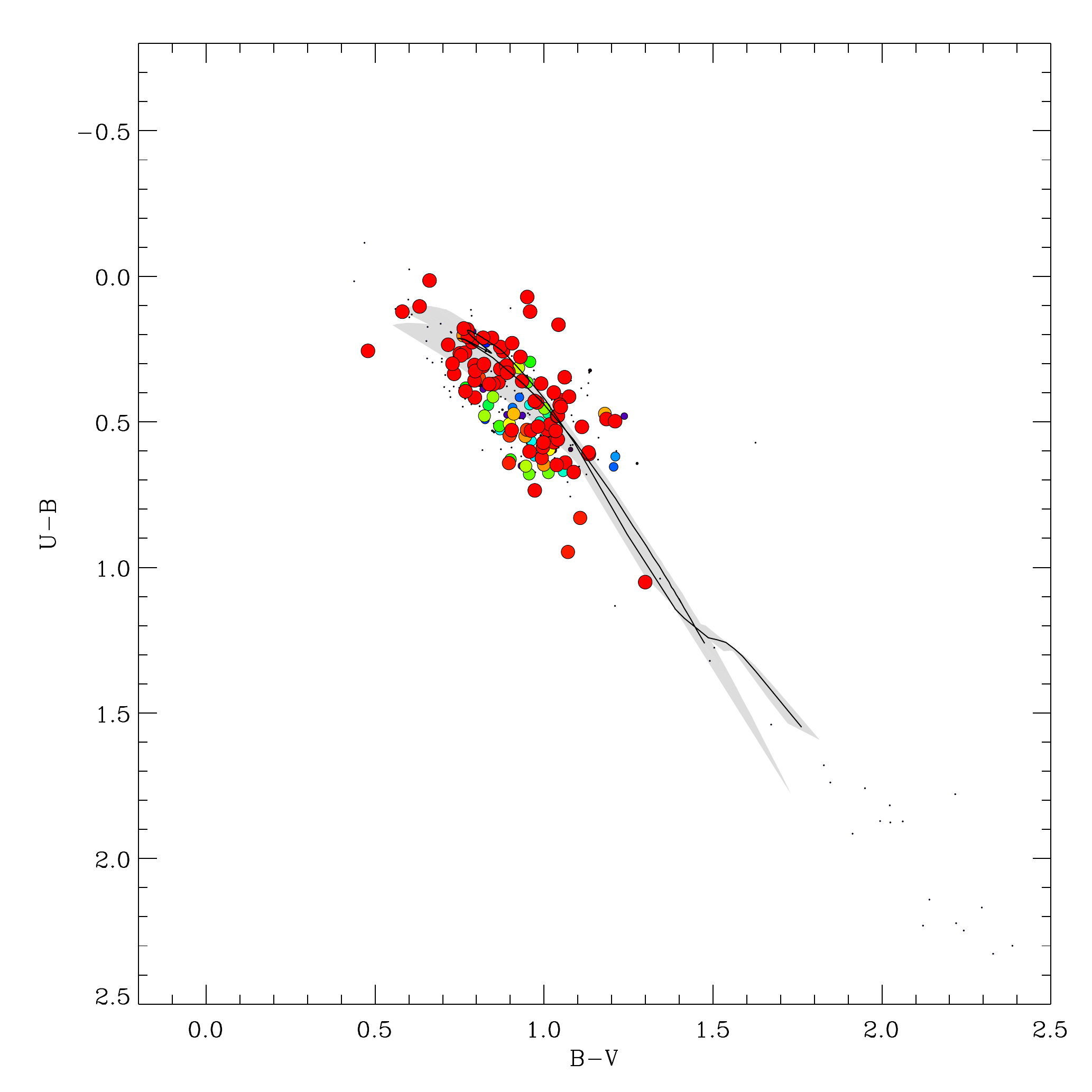}
\includegraphics[height = 6.0cm, width = 6.0cm]{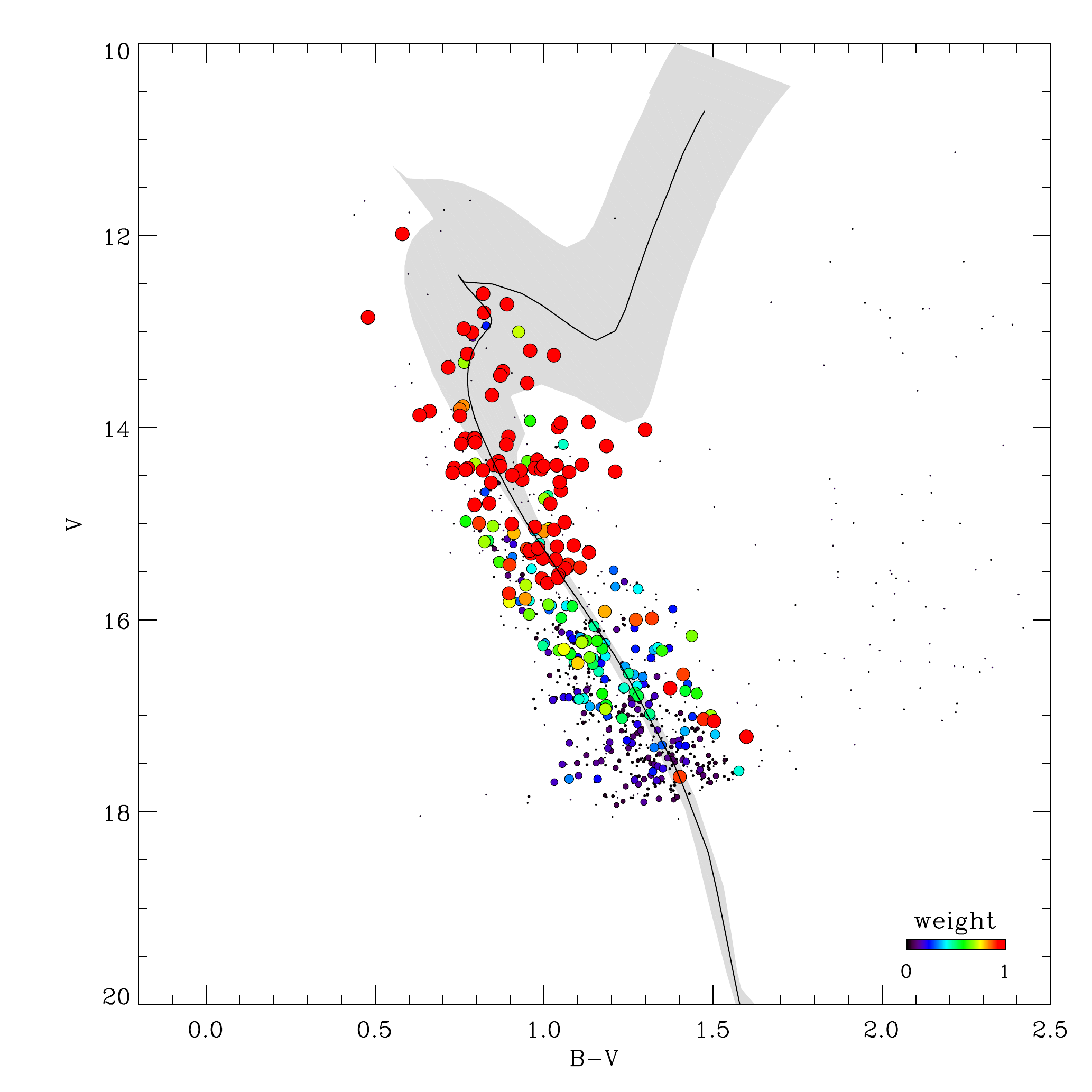}
\includegraphics[height = 6.0cm, width = 6.0cm]{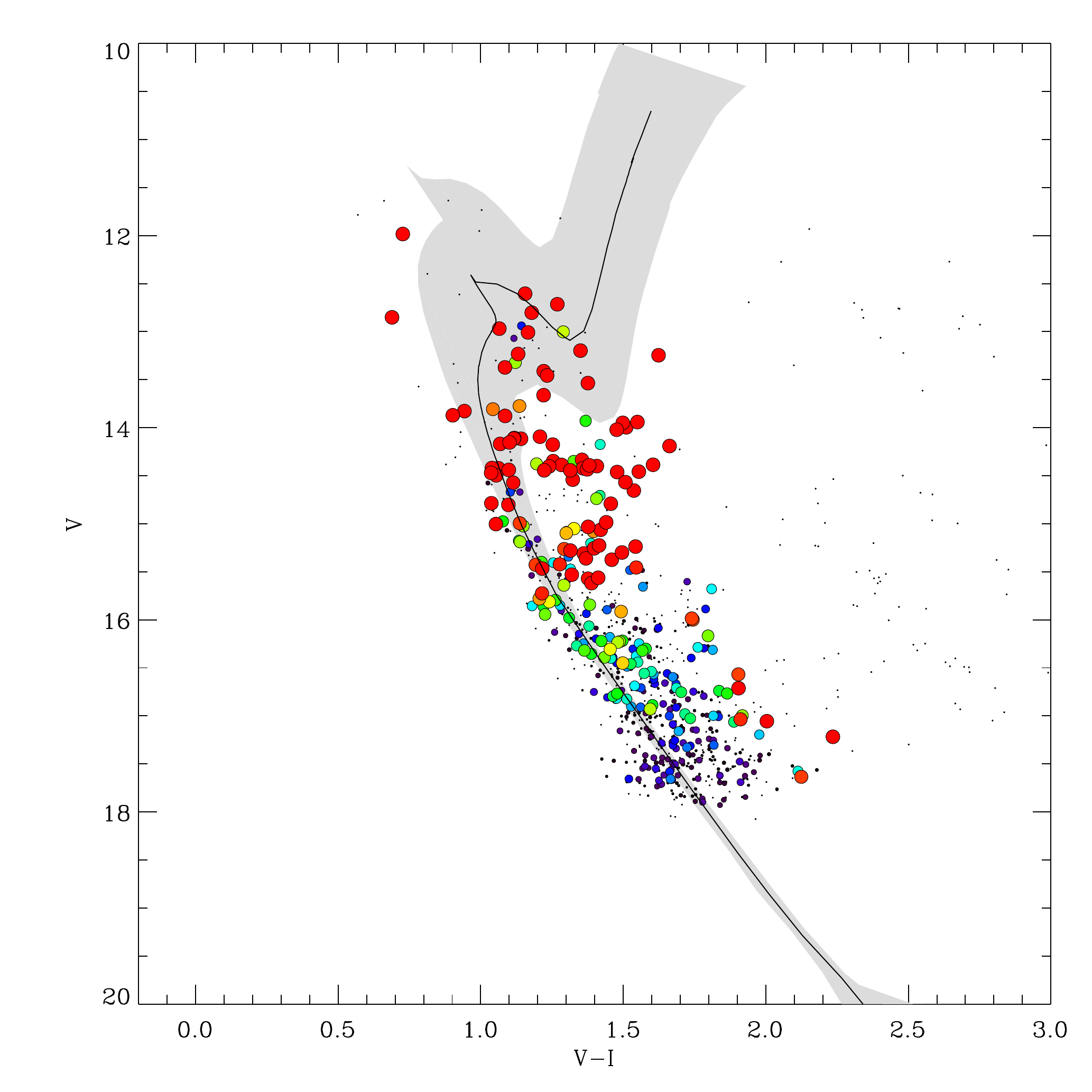}
   \caption{Same as Figure \ref{bh200}, but for Ruprecht121.}
              \label{Rup121}%
\end{figure*}

\begin{figure*}
\centering
\includegraphics[height = 6.0cm, width = 6.0cm]{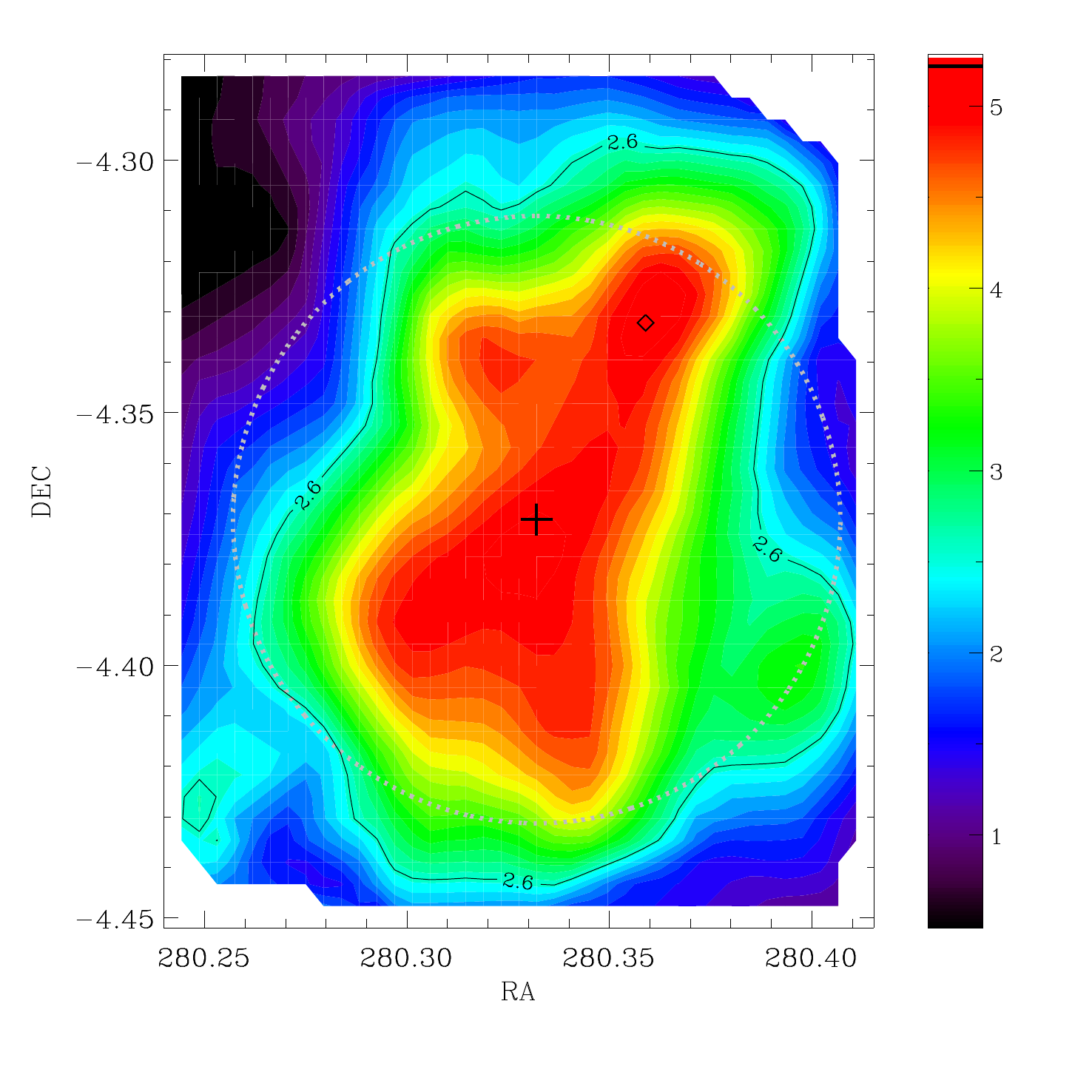}
\includegraphics[height = 6.0cm, width = 6.0cm]{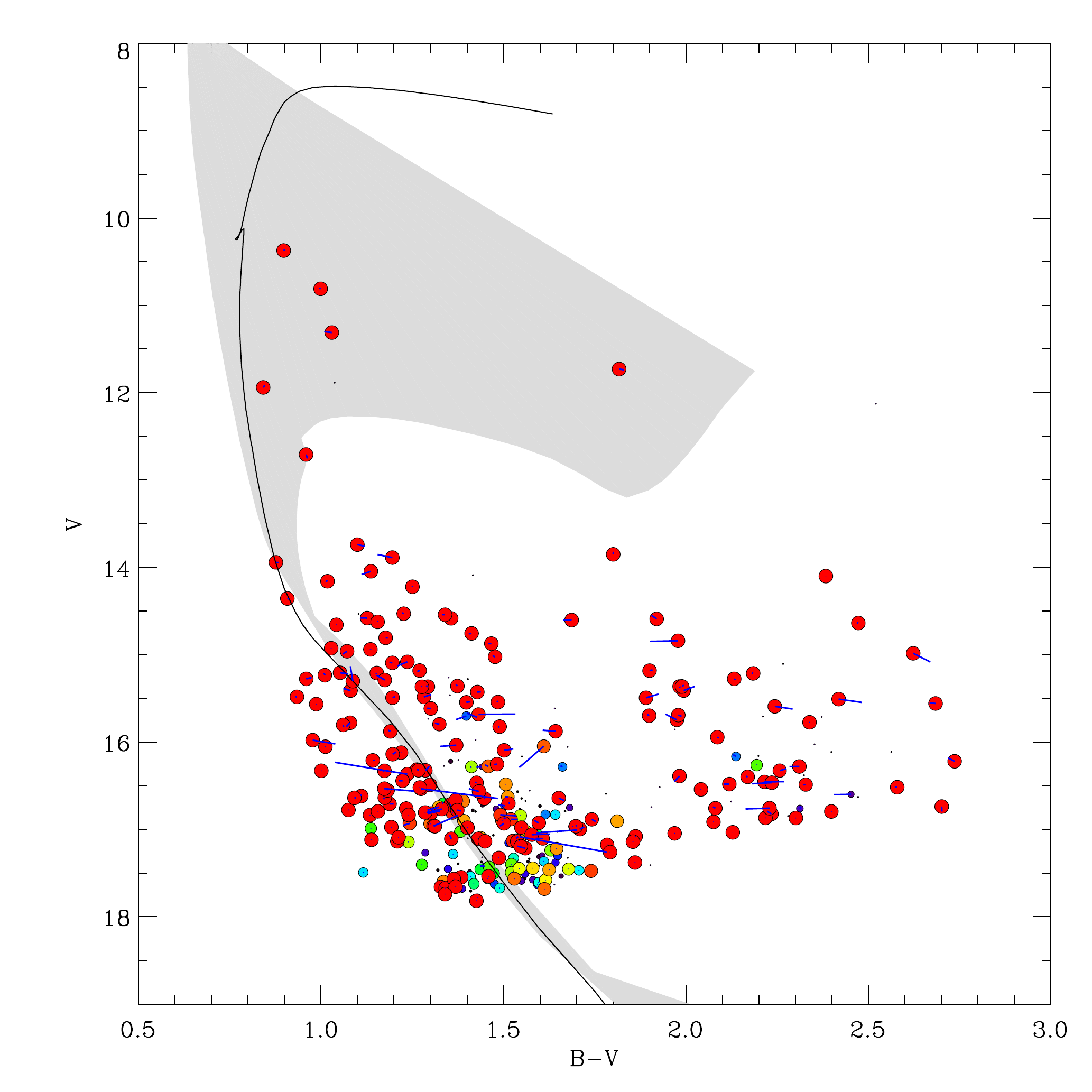}
   \caption{In the left panel are given the density map for the
     candidate Dolidze 33. Dotted line indicates the radius of the
     cluster and the solid line contour the density values above
     $1\sigma$ of the background level.  In the right panel are given
     the CMD with the stellar vector proper motion in arbitrary units
     overplotted.}
              \label{Dol33-den-vpd}
\end{figure*}

\begin{figure*}
\centering
\includegraphics[height = 6.0cm, width = 6.0cm]{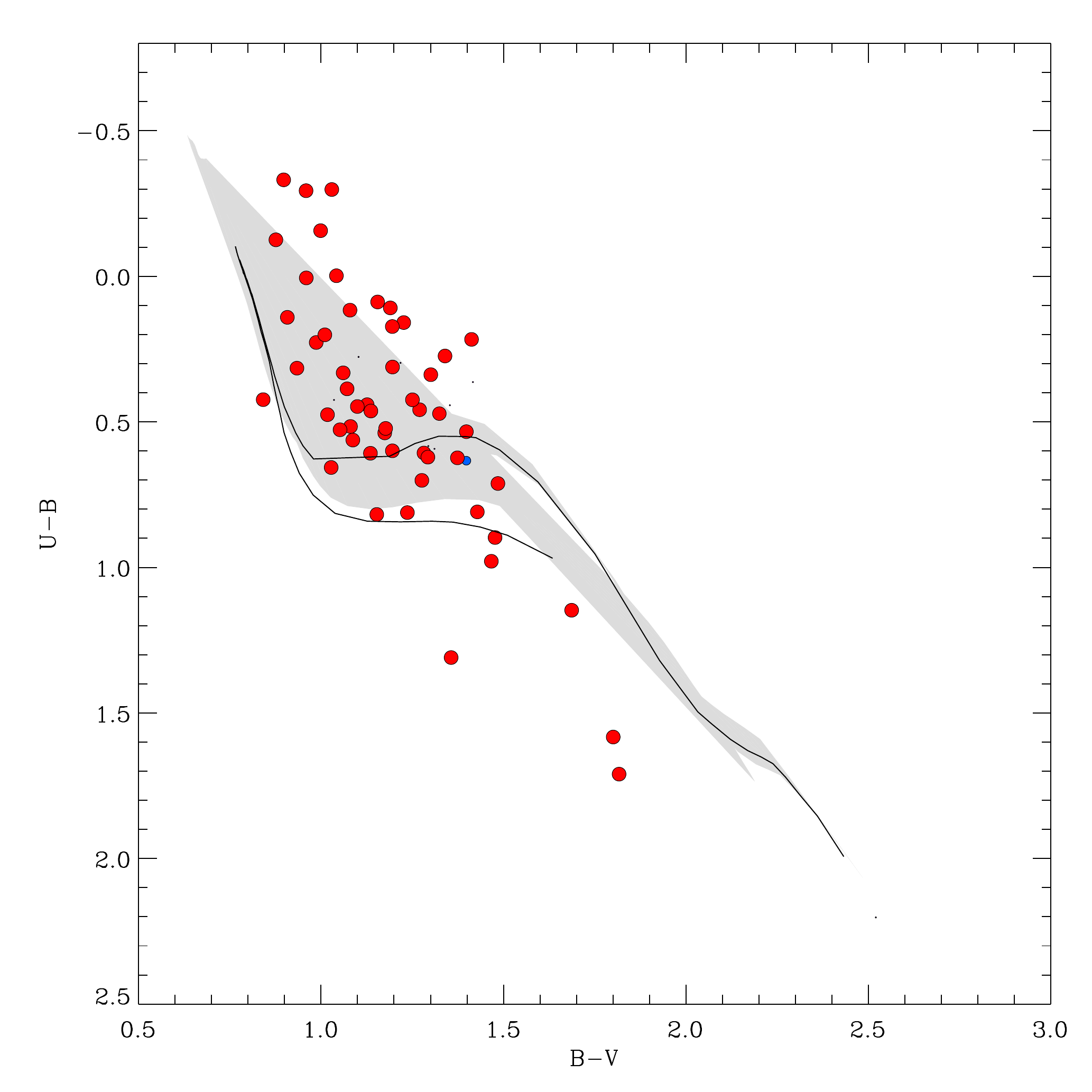}
\includegraphics[height = 6.0cm, width = 6.0cm]{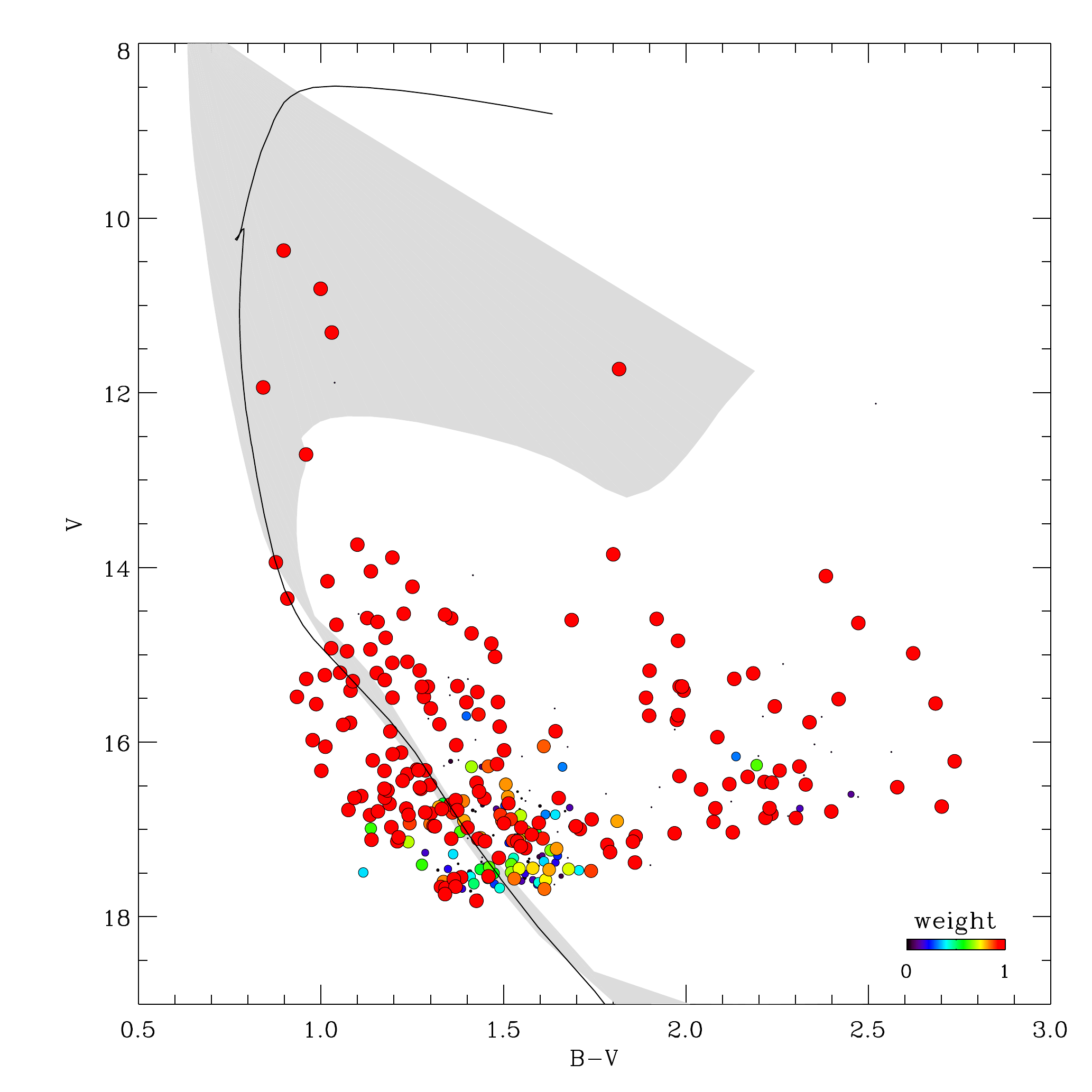}
\includegraphics[height = 6.0cm, width = 6.0cm]{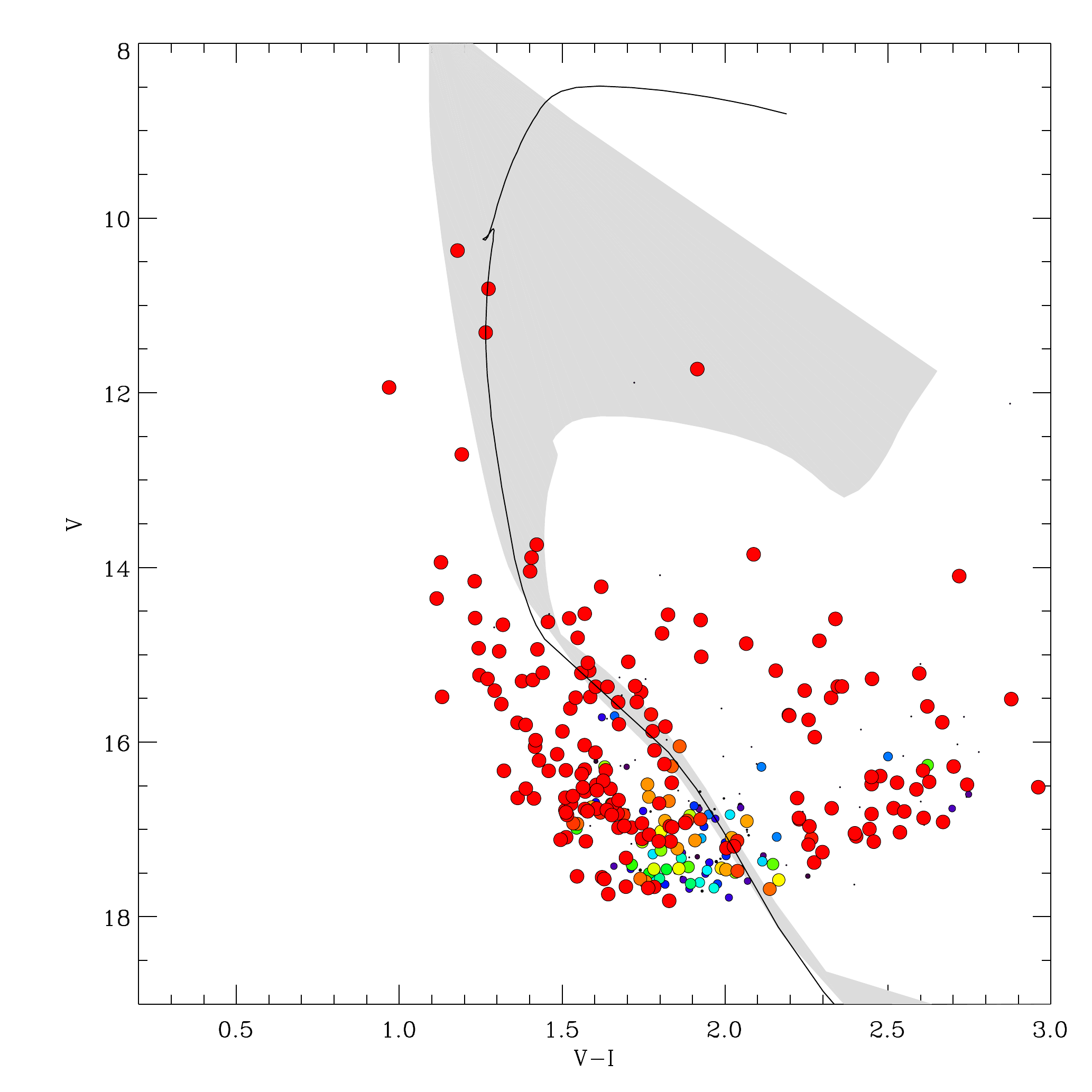}
   \caption{Same as Figure \ref{bh200}, but for Dolidze33.}
    \label{dol33}
\end{figure*}

\begin{figure*}
\centering
\includegraphics[height = 6.0cm, width = 6.0cm]{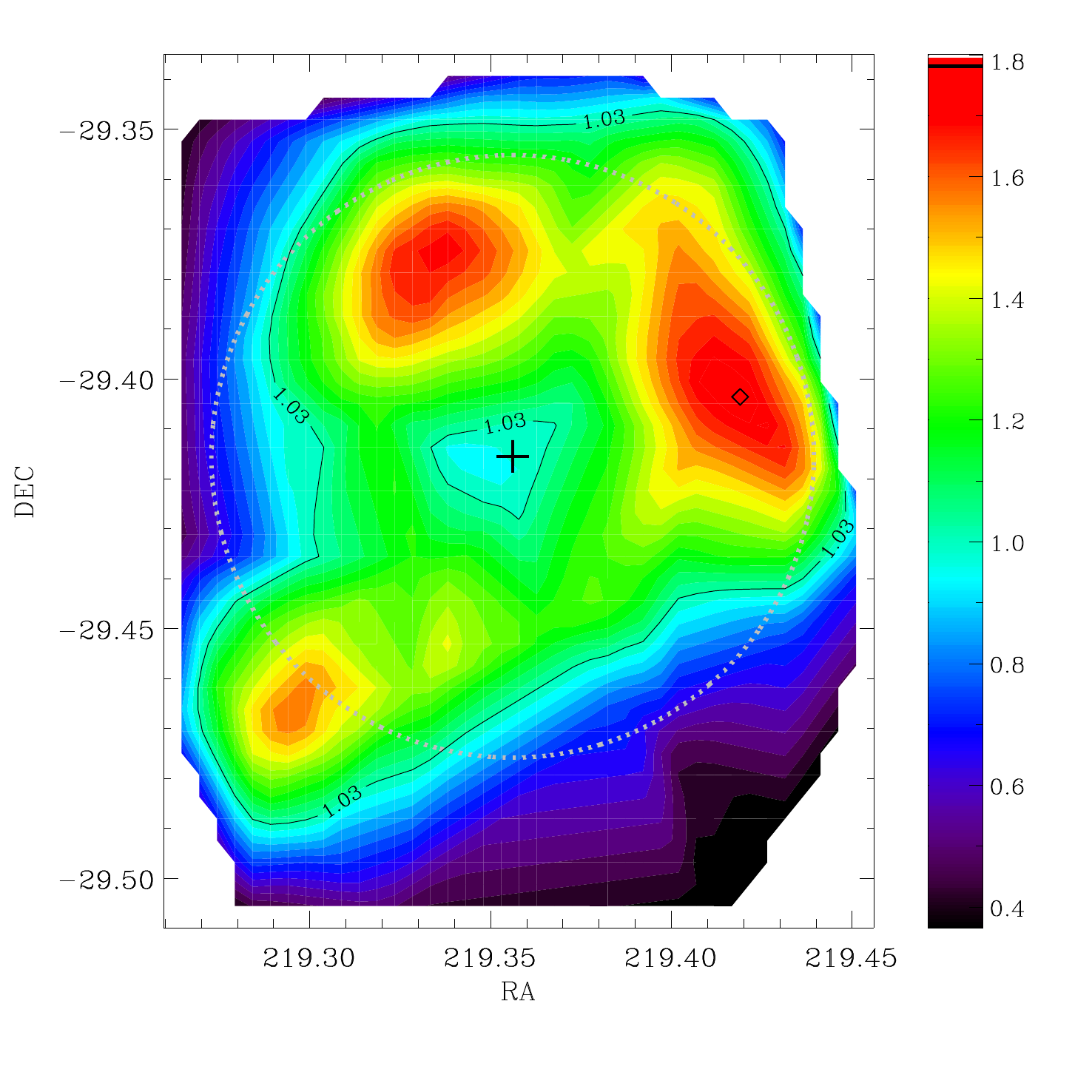}
\includegraphics[height = 6.0cm, width = 6.0cm]{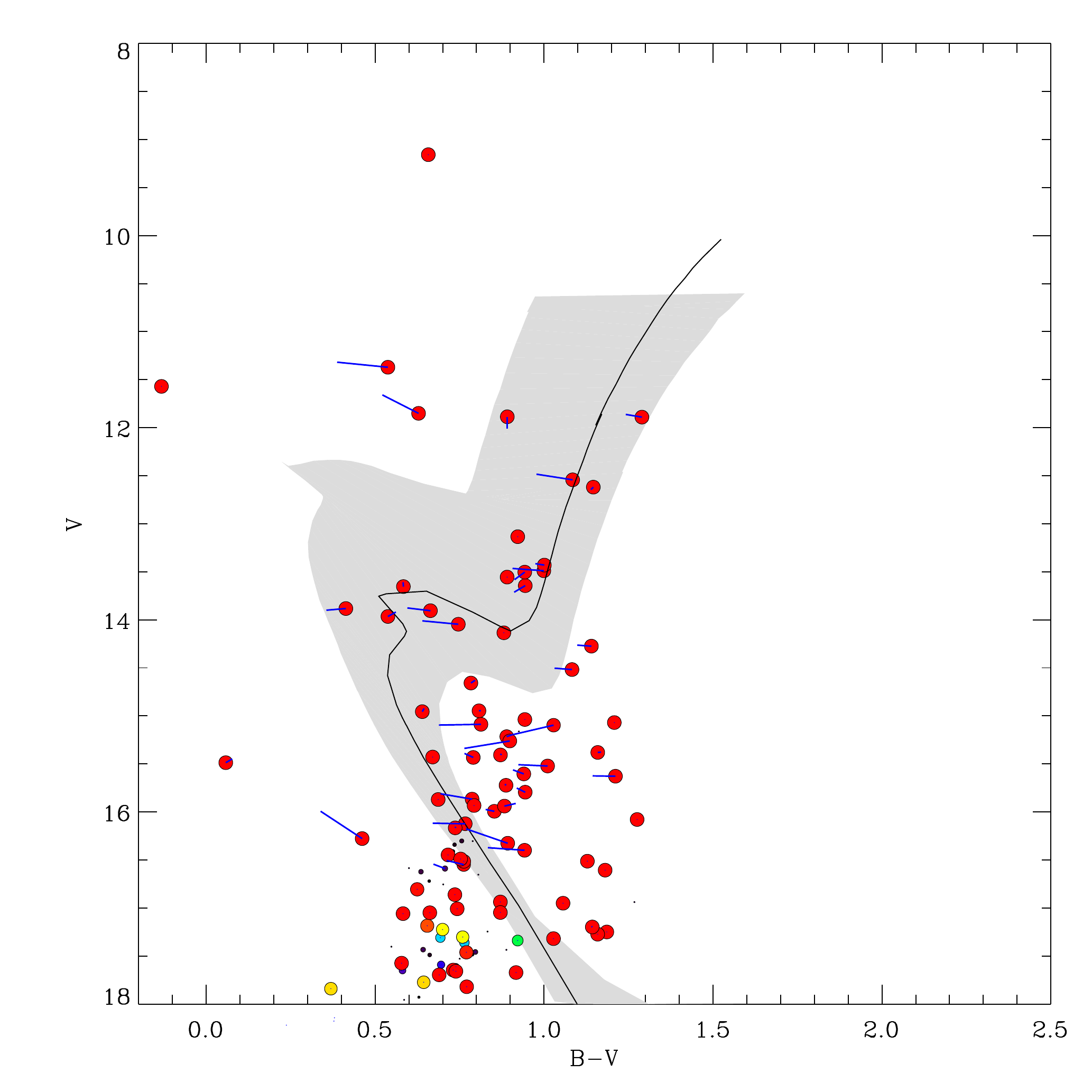}
   \caption{In the left panel are given the density map for the
     candidate ESO 447-29. Dotted line indicates the radius of the
     cluster and the solid line contour the density values above
     $1\sigma$ of the background level. In the right panel are given
     the CMD with the stellar vector proper motion in arbitrary units
     overplotted.}
              \label{dens-VPD-ESO447-29}
\end{figure*}

\begin{figure*}
\centering
\includegraphics[scale = 0.5]{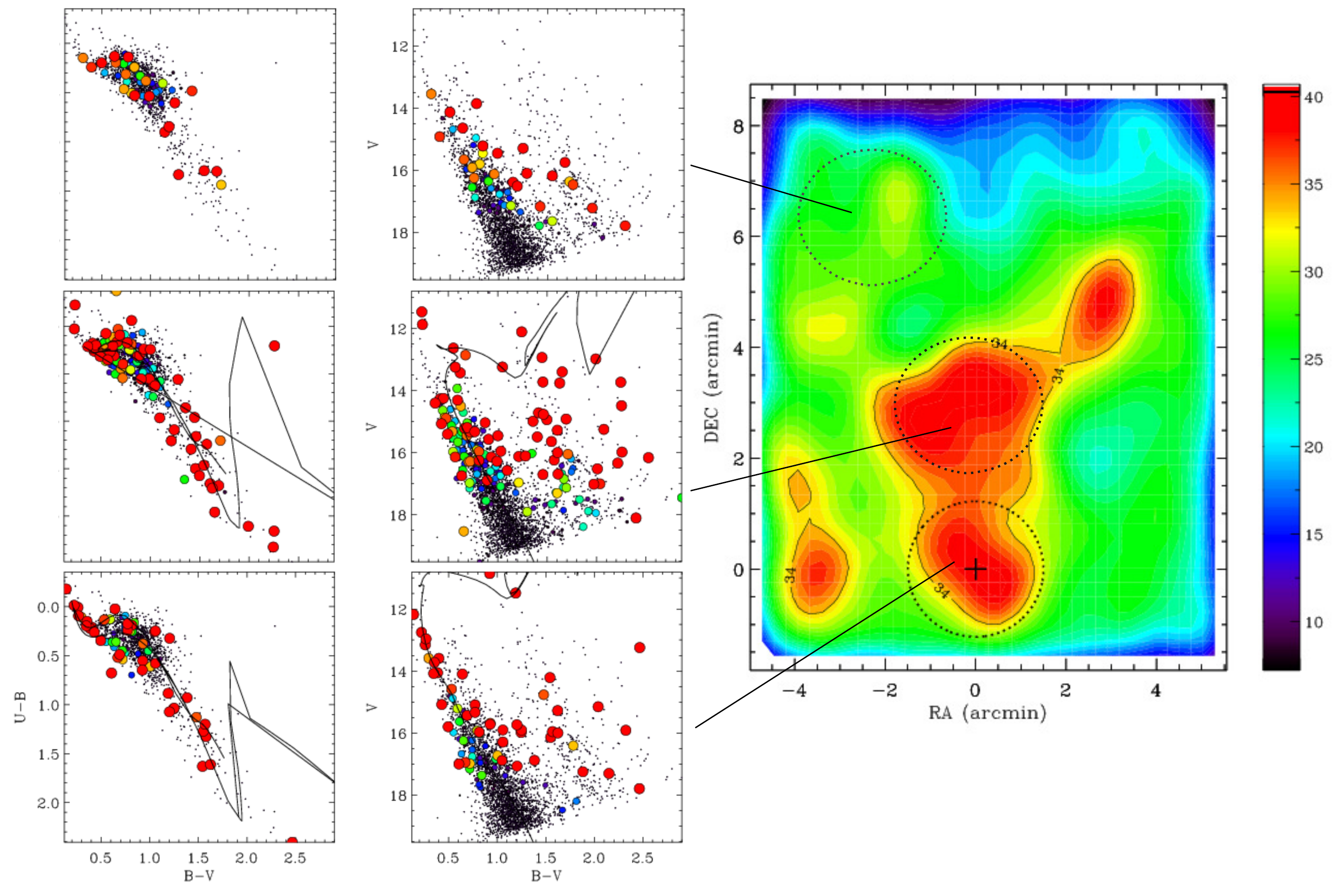}
\caption{In the right side of the panel is presented the density map
  of the observed region centered in the cluster Ruprecht 100.  In the
  left side of the panel are presented the CMDs related to the marked
  regions in the density map. The final fit for this cluster was then
  done with the cluster region being defined by an iso-density region
  taken at 32 $star/arcmin^2$}
\label{Rup100-panel}
\end{figure*}

\begin{figure*}
\centering
\includegraphics[height = 6.0cm, width = 6.0cm]{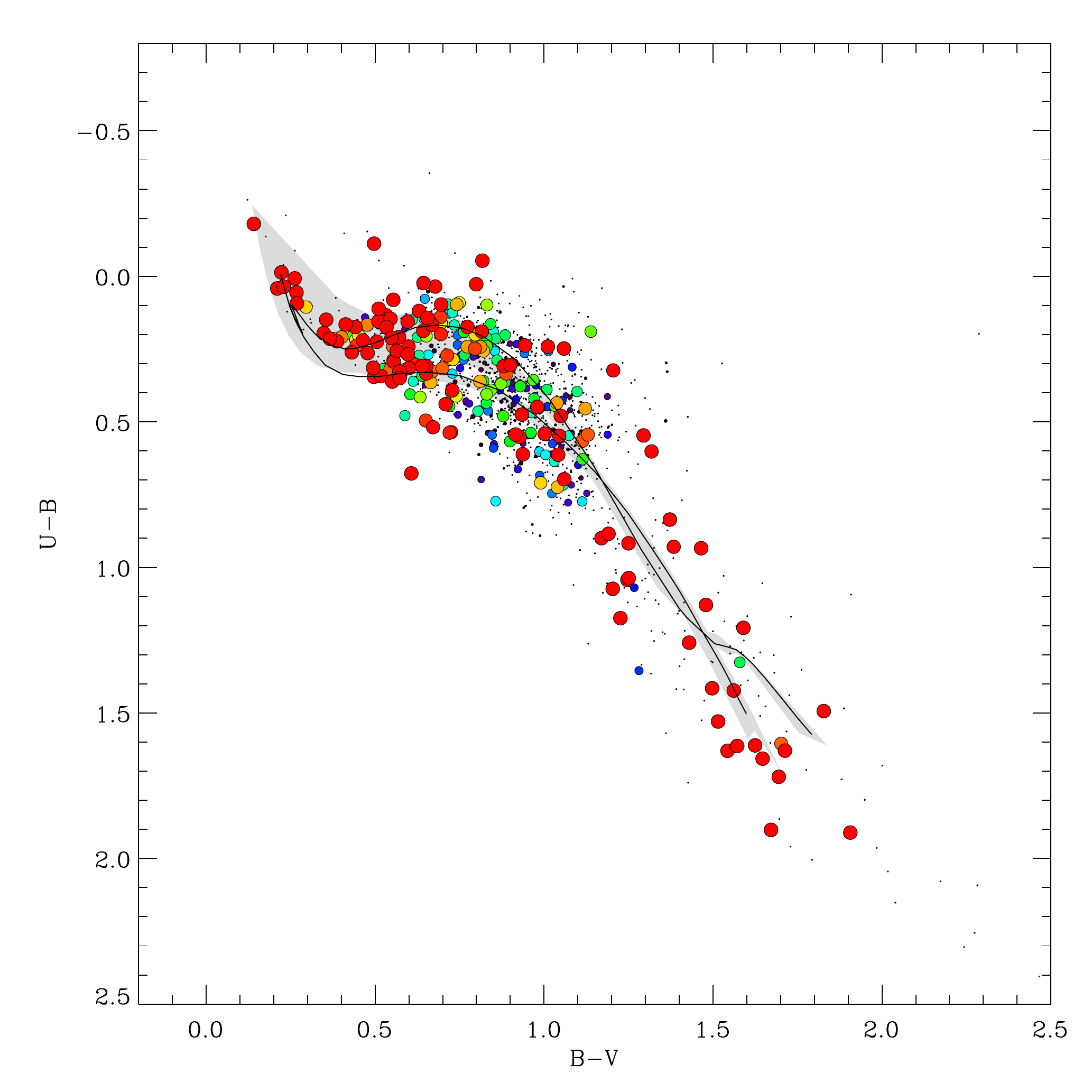}
\includegraphics[height = 6.0cm, width = 6.0cm]{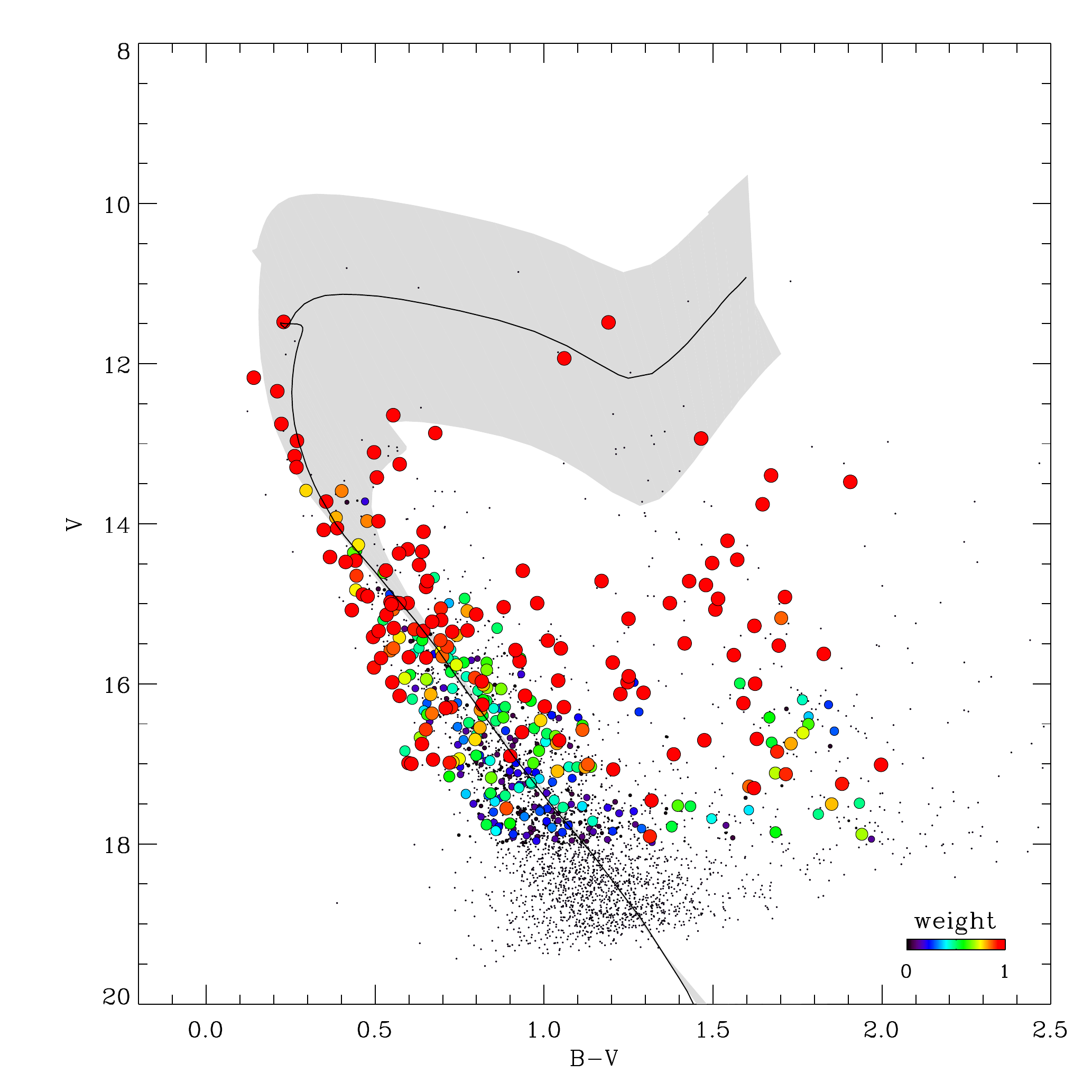}
\includegraphics[height = 6.0cm, width = 6.0cm]{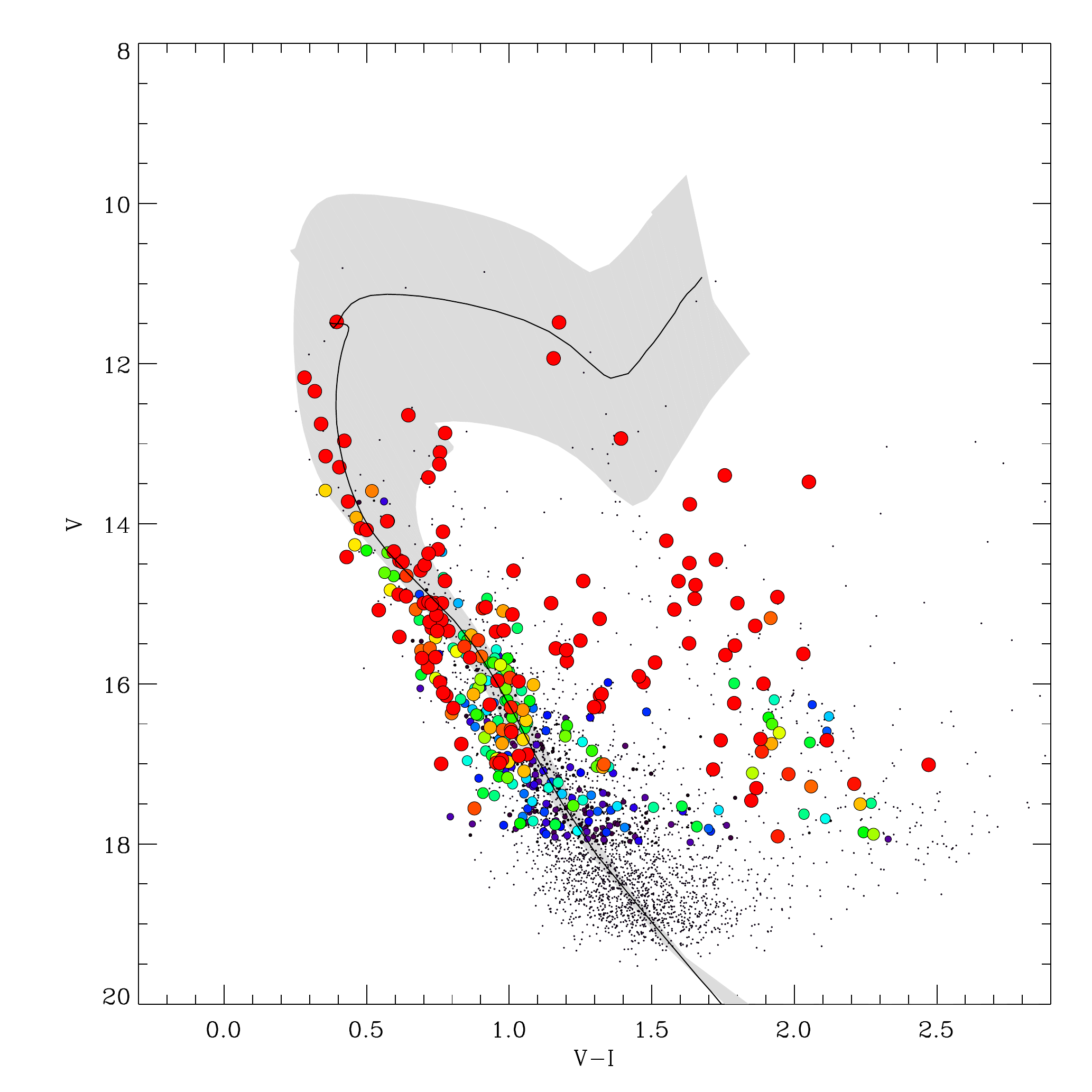}
   \caption{Same as Figure \ref{bh200}, but for Ruprecht 100.}
              \label{Rup100}%
\end{figure*}

The cluster Trumpler 25 is also in the Sagittarius-Carina arm direction
($l=349.1562^{\rm o}$ and $b = -01.7738^{\rm o}$) and the scenario is
similar to the one described for Ruprecht 100 (see CMD presented in
Figure \ref{Trumpler25}).  While the density map shows a peak density, the CMD
presented in the Figure \ref{Trumpler25} appears to have two distinct populations
with two possible turn-offs, the younger at $V\approx11$ and the older
one at $V\approx14$.

The analysis of the CMD with the stellar vector proper motion
over-plotted show that the five stars with $ V \leq 13$ and $(B-V)\leq
0.6$ that seem to represent the top of the main sequence of the
younger population, show proper motions that are inconsistent with a
group motion.  These stars received low membership when the cluster
region was defined by an iso-density region taken at about 8
$star/arcsec^2$.

For this case our interpretation is that the younger population is a
feature of the Sagittarius-Carina arm region.  It is also supported by
the analysis of Carraro (2006, c.f. their Figures 12 and 13).

\subsection{Comparison with the literature}
 
\begin{figure*}
\centering
\includegraphics[scale=0.4]{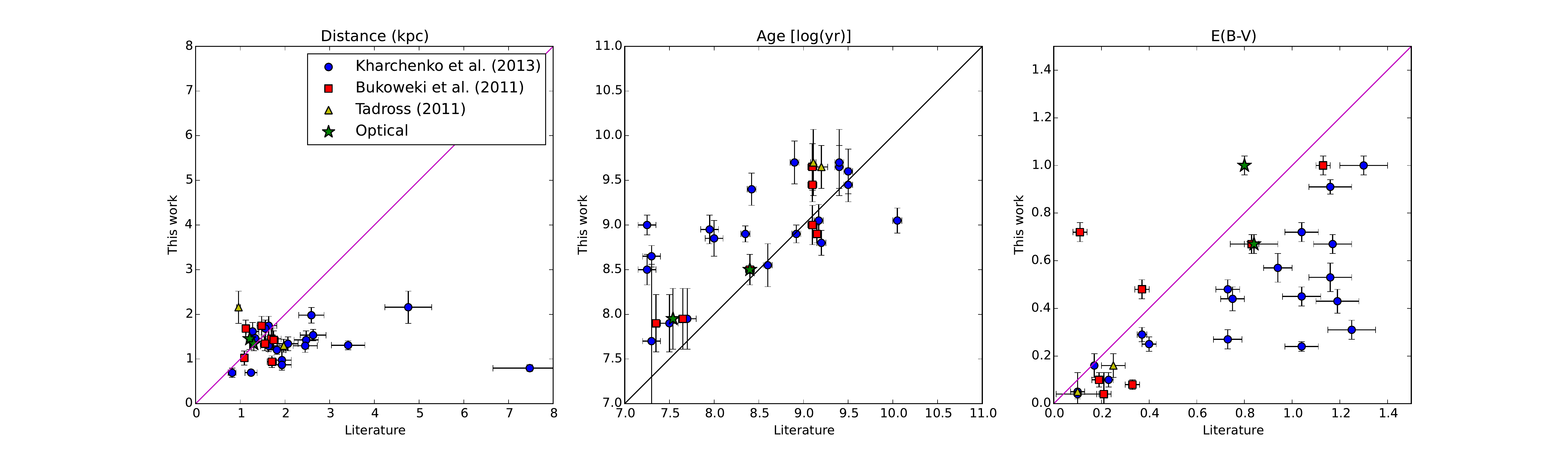}
\caption{Comparison of the results obtained in this work with other
  determinations from the literature as listed in Table 2. The
  identity loci is indicated by the solid line.}
              \label{comp-fig}%
\end{figure*}

The results we obtained in this work do agree in general with
  other results obtained in the literature. However important
  discrepancies emerge when we categorize determinations from
  different sources. In Figure \ref{comp-fig} we show the comparison
  of our results to those from the literature obtained by
  \citet{McSwain2005}, \citet{Kharchenko2013}, \citet{Bukowiecki2011},
  \citet{Tadross2011} and \citet{Carraro2006}. It is clear that the best
  agreement to our results are from those estimates based on optical
  data. This is to be expected since their data set retain the higher
  degree of similarity to our own data, despite different isochrone
  fitting procedures that migh have been used. The estimates of
  \citet{Bukowiecki2011} and \citet{Tadross2011} also agree well with
  our own even though they used 2MASS data to obtain their final
  parameter estimates. The results of \citet{Kharchenko2013} however
  show large discrepancies from our results. There are clear
  systematic differences in the age and E(B-V) obtained from their
  work. Curiously if we compare the common estimates of
  \citet{Bukowiecki2011} and \citet{Kharchenko2013} they do not agree
  among themselves, sometimes with large discrepancies beyond the
  uncertainties, despite the use of the same 2MASS data set. Although
  it is beyond the scope of this work these discrepancies warrant
  further investigation.

\section{Conclusions}

In this second paper of the series we present robust determination of
the fundamental parameters of 24 open clusters from UBVRI photometry.

The code was improved in many aspects and is now more consistent with
the formalism of probability distributions and a Bayesian approach to
account for prior information for the stars. It is also made available
for public use in a web-site developed to provide the results obtained
in this project.

We presented results of the estimation of fundamental parameters of
the 24 open clusters studied, 14 of which were objects for which we
obtained UBVRI photometry for the first time. Some clusters were
previously studied in Paper I but were re-evaluated in this work to
check for the consistency of the results. The results for these
clusters agree with those published in paper I, which validates the
new formalism and improvements made in the code.

The results for the other clusters studied, for which we used quality
UBVRI photometry obtained for the first time explore the current
possibilities and limitations of the code. In particular, we emphasize
that for some cases the code can work well with the automatically set
parameters such as radius, density cut-off and others, but for other
cases such as Ruprecht 100 or ESO 392-13 for example, a more careful
analysis is necessary.

On the other hand, for some of the objects classified as clusters for
which we obtained fundamental parameters, the results do not agree
with previously published values in the literature derived from 2MASS
data. We also found important discrepancies when comparing our results
to one obtained from 2MASS photometry. In particular we also encounter
important differences, not accommodated by the quoted uncertainties,
between estimates from distinct authors based on the same 2MASS
photometry.This subject is important and we plan to explore it further
in a future study with a greater sample of open clusters for which we
will have high quality UBVRI photometry.

Finally, given the fact that we have used quality UBVRI photometry,
with a robust photometric membership probability estimation procedure,
which takes into account the observational errors involved, we are
confident that we obtained non-subjective, reliable results with
realistic errors for the estimated parameters. These results will
contribute to improve the open cluster information published in the
DAML02 catalog available to all the community.

\section*{Acknowledgments}

The entire project was made possible by large amounts of observing
time and travel and other financial support from LNA/MCTI. We thank the
staff of the Pico dos Dias Observatory for the valuable support.
Special thanks to Rodrigo Prates.  W. S. Dias acknowledges the S\~ao
Paulo State Agency FAPESP (fellowship 2013/01115-6). H. Monteiro would
like to thank FAPEMIG grants APQ-02030-10 and CEX-PPM-00235-12 and
CAPES.  We employed catalogs from CDS/Simbad (Strasbourg) and
Digitized Sky Survey images from the Space Telescope Science Institute
(US Government grant NAG W-2166).

\clearpage

\appendix

\label{appendix}

 \section{Updates to the code}

One of the main modification in our code was a rewrite of the
algorithm that performed the non-parametric membership likelihood
estimation. We have done this to make the algorithm more consistent
with the formalism of probability distributions and a Bayesian
approach. The motivation for the change is the possibility of
incorporating new information as it becomes available in a more
transparent way through clearly stated prior probabilities.

In essence the changes introduced allowed us to take the information
from photometry, spatial density and distribution of stars as
individual probability density functions (PDFs). These PDFs were then
combined in a joint probability distribution that, after proper
normalization, was used in a Bayesian hypothesis test.

The hypothesis test is based on the assumption that any given observed
field of an open cluster will have stars that belong to the cluster
(defined here as $H_1$) and others that do not (defined here as
$H_2$). Throughout the text we also refer to stars that belong to
these two distinct populations as CLUSTER and FIELD stars respectively. We
also assume that the open cluster is always associated with a star
density enhancement observed in a field, from which the code can
determine a position which is defined as the cluster center. This
position can also be set by the user if it is needed. We also assume
that the further a star is from this defined center, the less likely
it is to be a CLUSTER star.

This approach is more interesting from a formal point of view because
it allows for missing data to be taken into account naturally.  In this
way if a star has no data for a given filter it will not have a PDF
for it and the final probabilities are estimated based only on the
available information for each star, after proper normalization.

Given the considerations above, for a given observed field the code
performs the following procedures:

\begin{itemize}

\item with the RA and DEC coordinates of the stars the code calculates
  the density map using a kernel density estimate with a bandwidth
  determined automatically or defined by the user. The code determines
  automatically the bandwidth to be used based on the well known
  Silverman's rule. The kernel bandwidth is then given by ${SIG\_PIX}
  \approx 1.06 \sigma N^{-1/5}$,where $\sigma$ is the standard
  deviation of the sample of distance between stars and N is the
  number of stars.

\item if not supplied by the user, from the density map the code
  determines the center of the cluster as the point where the density
  distribution peaks;

\item from the RA and DEC of the stars the code determines the radius
  of the cluster as the standard deviation of all the distances
  between stars in the sample;

\item with the determined radius the code separates the sample in two
  sub samples: CLUSTER and FIELD stars;

\item from the data for each filter we determine a probability density
  function (PDF) and define, as in previous versions of the code, a
  photometric PDF given by the joint probability of all filters as
  $PHOT_{PDF}=U_{PDF} \times B_{PDF} \times V_{PDF} \times R_{PDF} \times I_{PDF}$. The code obtains
  photometric PDFs for the CLUSTER and FIELD regions respectively;

\item from the density map the code obtains a PDF, $DENS_{PDF}$ using
  the kernel density estimate obtained in the first item;

\item since we take the distance of a given star to the cluster center
  as an information on the membership of that star, we also adopt a
  PDF that describes this assuming Gaussian distribution such that
  $DIST_{PDF}=\frac{1}{\sqrt{2\pi \sigma^2}} EXP[-0.5(D_S/RADIUS)^2]$
  where $D_S$ is the distance of star S to the center of the cluster;

\item from all the information listed above the code obtains final
  CLUSTER and FIELD joint distributions by calculating
  $CLUSTER_{PDF}=DIST_{PDF} \times PHOT_{PDF} \times DENS_{PDF}$ and normalizing;

\end{itemize}

After all the steps listed above are taken, the code performs a
Bayesian hypothesis test by calculating:

\begin{equation}
    P(H_1 \mid D)= \frac{P(D \mid H_1)\,P(H_1)}{P(D \mid
      H_1)\,P(H_1)\;+\;P(D \mid H_2)\,P(H_2)}
\end{equation}

where $H_1$ is the hypothesis that any given star of the data set $D$
is a cluster star and $H_2$ is the complementary that the star is a
field object. The prior probabilities $P(H_1)$ and $P(H_2)$ are
estimated from the data considering the two populations, CLUSTER and
FIELD, as determined by the code. From the star counts in the two
different populations we take $P(H_1)=n_C/n_T$ and $P(H_2)=n_F/n_T$,
where $n_C$ and $n_F$ are the number of stars in the CLUSTER and FIELD
populations.  $n_T$ is the total number of the stars in the sample.
We have tested other prior probabilities but they led to  minor
differences in the final results.

Here we also introduced the possibility of taking the actual spatial
structure of the cluster into account by allowing the CLUSTER region
to be defined by an iso-density limit obtained from the density
map. This tool is useful for clusters that show irregular stellar
densities for which the usual King profile assumption is clearly
inadequate. The procedure was critical in revising the results of the
previous paper for Ruprecht 100 as discussed in Section 4.

The final result of the procedure above is to provide a membership
probability for each star based on the data provided, which is then
used as a weight in the global optimization procedure.

Another important addition is the inclusion of $R_v$ as a new free
parameter. To accomplish this we implemented the detailed relations of
reddening developed by \citet{CCM89} which allowed us to establish the
extinction values in each filter used based on a given $R_v$ and we
refer the reader to that work for further details. The parameter space
limits in this work were used in accordance with the discussion of
\citet{T76} so that $2.0~<~R_V~<~4.0$.

Since we perform a bootstrap analysis of the fits we have access to an
estimate of the probability distributions of the solutions. This in
turn allows us to use the mode of the distributions to estimate final
solutions to the parameters. We have implemented this feature in the
current version of the code. The mode solution is now also taken to be
the adopted final solution.

\bibliographystyle{aa}
\bibliography{OPD-SURVEY-II-final}

\clearpage



\renewcommand*{\thesection}{\arabic{section}}
\setcounter{section}{0}

\begin{table} 
  \caption{The tuning parameters used in the cross-entropy fitting
    procedure. The first two columns (after the cluster identification)
    give the central coordinates. Column 4 gives the
    characteristic cluster radius ($R_{c}$), calculated based on the
    distribution of distances of each star of the sample to the center
    of the cluster.  Column 5 gives the width ($\sigma$) of the Gaussian
    kernel used in the density profile. Column 6 gives $P_{cut}$, the adopted normalized cut-off.  For
    all clusters $100\%$ of the stars were considered as binary. The
    values in the column radius marked with an asterisk corresponding to
    the density value used to select the stars.}
\label{table:1} 
\begin{tabular}{lccccc}
\hline
\hline
{Cluster} & {$\alpha_{c}$}  & {$\delta_{c}$} &  radius & sig-pix   &   {$P_{cut}$}\\
          &  J2000.0        &   J2000.0      &  arcmin &  arcmin   &        \\
\hline
\astrobj{BH~217}         & 259.071       &  -40.8222    &   1.0    & 1.0    & 0.51     \\ 
\astrobj{Dolidze~33}     & 280.328       &  -04.3797    &   2.5    & 1.0    & 0.95     \\ 
\astrobj{Dolidze~35}     & 291.36        &  +11.65      &   3.5    & 0.97   & 0.51      \\ 
\astrobj{ESO~021-06}     & 213.961       &  -78.5243    &   4.8    & 0.91   & 0.51     \\ 
\astrobj{ESO~099-06}     & 232.422       &  -64.8856    &   5.5    & 0.64   & 0.95      \\ 
\astrobj{ESO~456-09}     & 268.547       &  -32.4720    &  25.0*   & 0.55   & 0.90     \\  
\astrobj{Loden~991}      & 206.336       &  -62.0311    &   6.0    & 0.77   & 0.95      \\ 
\astrobj{Lynga~13}       & 252.240       &  -43.4276    &   2.0    & 0.69   & 0.51      \\ 
\astrobj{NGC~6588}       & 275.200       &  -63.8100    &   3.0    & 0.87   & 0.51      \\ 
\astrobj{Ruprecht~100}   & 181.450       &  -62.6500    &  32.0*   & 0.54   & 0.60     \\ 
\astrobj{BH~200}         & 252.484       &  -44.1799    &   5.5*   & 0.80   & 0.51     \\
\astrobj{ESO~008-06}     & 223.894       &  -83.4423    &   5.5    & 1.06   & 0.51     \\
\astrobj{ESO~099-06}     & 232.471       &  -64.8864    &   5.0    & 0.86   & 0.90     \\  
\astrobj{ESO~139-54}     & 271.170       &  -58.5388    &   5.0    & 0.85   & 0.90     \\  
\astrobj{ESO~397-01}     & 286.037       &  -33.3978    &   5.0    & 0.86   & 0.95     \\  
\astrobj{ESO~447-29}     & 219.356       &  -29.4155    &   5.5    & 1.00   & 0.90     \\  
\astrobj{NGC~6840}       & 298.813       &  +12.1261    &   5.0*   & 0.77   & 0.51     \\ 
\astrobj{Ruprecht~111}   & 218.942       &  -59.9390    &   2.1    & 0.74   & 0.51     \\ 
\astrobj{Trumpler~25}    & 218.942       &  -59.9390    &   8.2*   & 0.74   & 0.51     \\ 
\astrobj{Collinder~307}  & 248.784       &  -51.0006    &   1.9    & 0.71   & 0.51     \\ 
\astrobj{ESO~392-13}     & 261.726       &  -34.6952    &   5.5    & 1.04   & 0.51     \\  
\astrobj{ESO~518-03}     & 251.771       &  -25.8091    &   4.0    & 0.80   & 0.51     \\ 
\astrobj{Loden~1002}     & 208.562       &  -65.3297    &   5.0    & 0.65   & 0.95     \\ 
\astrobj{Ruprecht~121}   & 250.419       &  -46.1559    &   5.5    & 0.71   & 0.90     \\ 
\hline
\hline
\end{tabular}
\begin{flushleft} 
\tiny For Loden 991 we adopted a cut-off at $B-V\leq0.5$.  For NGC
6588 stars with $ V \leq 13.5$, for ESO139-54 stars with $ V \leq
16.0$.
For ESO 456-09 we adopted a cut-off at $B-V\geq1.8$ and stars with $ V
\leq 11.8$ received zero weight.  For Ruprecht 100 we adopted a
cut-off at $B-V\geq2.0$ and stars with $ V \leq 18.0$ received zero
weight.  For \astrobj{Loden~1002} stars with $B-V\geq 1.25$ and
$17.0 > V > 14.0$ were given a weight of zero.
\end{flushleft}
\end{table}


\begin{table*}
\label{res-table}
  \caption{Basic parameters obtained for the investigated  clusters.}
\small 
  \begin{tabular}{lccccccccc}
    \hline\hline
              & \multicolumn{4}{c}{This work}                      & &\multicolumn{3}{c}{Literature} \\
    {Cluster} &   {$E(B-V)$} & {Distance} & {Log(Age)} & {Z} & $R_v$   & {$E(B-V)$} &  {Distance} & {Log(Age)}   &  {Ref.} \\
              &      (mag)   &     (pc)   &     (yr)   &     &         &    (mag)        &      (pc)   &       (yr)   &          \\  
    \hline 
\astrobj{BH~217}       & $1.00\pm0.04$ & $1675\pm182$ & $7.75\pm0.35$  & $0.006\pm0.005$ & $3.02\pm0.01$  & $0.80          $    &  $ 1210            $    & $7.54               $ & 1   \\
\astrobj{BH~217}       &               &              &                &                 &                & $1.30\pm0.10   $    &  $ 1723\pm190      $    & $7.54\pm0.10        $ & 2   \\
\astrobj{BH~217}       &               &              &                &                 &                & $1.15\pm0.03   $    &  $ 1693\pm126      $    & $7.65\pm0.05        $ & 3   \\
\astrobj{Dolidze~33}   & $0.91\pm0.03$ & $ 975\pm226$ & $7.70\pm0.86$  & $0.006\pm0.005$ & $3.01\pm0.01$  & $1.16 \pm0.09  $    &  $ 1930\pm212      $    & $7.30\pm0.10        $ & 2   \\
\astrobj{Dolidze~35}   & $0.53\pm0.06$ & $ 869\pm89$  & $8.65\pm0.12$  & $0.027\pm0.007$ & $3.02\pm0.05$  & $1.25 \pm0.10  $    &  $ 1317\pm145      $    & $8.30\pm0.05        $ & 2   \\
\astrobj{ESO~021-06}   & $0.04\pm0.09$ & $1426\pm201$ & $9.45\pm0.19$  & $0.027\pm0.008$ & $3.13\pm0.04$  & $0.104\pm0.008$     &  $ 2473\pm272      $    & $9.50\pm0.05        $ & 2   \\
\astrobj{ESO~021-06}   &               &              &                &                 &                & $0.21\pm0.03   $    &  $ 1745\pm104      $    & $9.10\pm0.05        $ & 3   \\
\astrobj{ESO~099-06}   & $0.10\pm0.03$ & $1617\pm200$ & $9.70\pm0.24$  & $0.019\pm0.007$ & $3.07\pm0.04$  & $0.229\pm0.018  $   &  $ 1274\pm140      $    & $8.90\pm0.05        $ & 2   \\
\astrobj{ESO~456-09}   & $0.43\pm0.05$ & $ 795\pm76$  & $9.05\pm0.14$  & $0.019\pm0.007$ & $3.14\pm0.05$  & $1.187\pm0.094  $   &  $ 7475\pm822      $    & $10.05\pm0.05        $ & 2   \\
\astrobj{Loden~991}    & $0.31\pm0.04$ & $1534\pm127$ & $9.00\pm0.11$  & $0.027\pm0.005$ & $3.04\pm0.05$  & $1.249\pm0.100  $   &  $2628\pm289       $    & $ 7.25\pm0.10        $ & 2   \\
\astrobj{Lynga~13}     & $0.46\pm0.04$ & $1308\pm103$ & $8.80\pm0.14$  & $0.019\pm0.006$ & $3.04\pm0.05$  & $1.041\pm0.083  $   &  $3412\pm375       $    & $ 9.20\pm0.05        $ & 2   \\
\astrobj{NGC~6588}     & $0.05\pm0.01$ & $2159\pm357$ & $9.65\pm0.24$  & $0.027\pm0.008$ & $3.13\pm0.03$  & $0.042\pm0.003  $   &  $4757\pm523        $    & $ 9.40\pm0.05        $ & 2   \\
\astrobj{NGC~6588}     &               &              &                &                 &                & $0.10\pm0.03   $    &  $  960\pm 44      $    & $9.20\pm0.07        $ & 4   \\
\astrobj{Ruprecht~100} & $0.27\pm0.04$ & $1980\pm175$ & $8.55\pm0.24$  & $0.021\pm0.006$ & $2.78\pm0.38$  & $0.729\pm0.058  $   &  $2589\pm285        $    & $ 8.60\pm0.05        $ & 2   \\
\astrobj{BH~200}       & $0.72\pm0.04$ & $1746\pm208$ & $7.90\pm0.32$  & $0.005\pm0.004$ & $3.03\pm0.03$  &     $1.00\pm0.03$   &     $1475\pm106 $        &    $7.35\pm0.05$       & 3     \\
\astrobj{ESO~099-06}   & $0.09\pm0.02$ & $ 684\pm84 $ & $8.95\pm0.41$  & $0.021\pm0.007$ & $3.15\pm0.04$  &     $0.229\pm0.018$ &     $1274\pm140 $        &    $8.90\pm0.05$       & 2     \\ 
\astrobj{ESO~139-54}   & $0.08\pm0.02$ & $ 938\pm130$ & $9.00\pm0.22$  & $0.024\pm0.007$ & $3.08\pm0.05$  &     $0.33\pm0.03$   &     $1706\pm105 $        &    $9.10\pm0.05$       & 3     \\
\astrobj{ESO~397-01}   & $0.10\pm0.03$ & $1022\pm157$ & $9.65\pm0.26$  & $0.015\pm0.007$ & $3.11\pm0.04$  &     $0.19\pm0.03$   &     $1089\pm65  $        &    $9.10\pm0.05$       & 3     \\ 
\astrobj{ESO~447-29}   & $0.04\pm0.02$ & $1292\pm126$ & $9.60\pm0.25$  & $0.011\pm0.007$ & $3.09\pm0.04$  &     $0.104\pm0.008$  &     $1634\pm180 $        &    $9.50\pm0.05$       & 2     \\ 
\astrobj{NGC~6840}     & $0.16\pm0.05$ & $1296\pm147$ & $9.70\pm0.37$  & $0.015\pm0.007$ & $3.09\pm0.05$  &     $0.25\pm0.05$   &     $1970\pm90  $        &    $9.11\pm0.03$       & 4     \\ 
\astrobj{Ruprecht~111} & $0.48\pm0.04$ & $1681\pm189$ & $8.90\pm0.10$  & $0.014\pm0.007$ & $3.02\pm0.01$  &     $0.37\pm0.03$   &     $1121\pm70  $        &    $9.15\pm0.05$       & 3     \\ 
\astrobj{Trumpler~25}  & $0.57\pm0.06$ & $1203\pm 98$ & $8.90\pm0.09$  & $0.027\pm0.006$ & $3.04\pm0.03$  &     $0.937\pm0.100$ &     $1818       $        &    $8.350\pm0.05$       & 2     \\ 
\astrobj{Collinder~307}& $0.67\pm0.04$ & $1341\pm155$ & $8.50\pm0.17$  & $0.009\pm0.007$ & $3.06\pm0.06$  &     $0.84\pm0.10$   &     $1300       $        &    $8.40       $       & 5     \\ 
\astrobj{Collinder~307}&               &              &                &                 &                &     $0.83\pm0.03$   &     $1551\pm107 $        &    $8.40\pm0.05$       & 3     \\ 
\astrobj{ESO~392-13}   & $0.44\pm0.05$ & $ 691\pm104$ & $8.95\pm0.16$  & $0.014\pm0.006$ & $3.12\pm0.06$  &     $0.750\pm0.100$ &     $ 816       $        &    $7.950\pm0.10$       & 2     \\ 
\astrobj{ESO~518-03}   & $0.29\pm0.03$ & $1440\pm142$ & $9.05\pm0.18$  & $0.017\pm0.006$ & $3.14\pm0.04$  &     $0.369\pm0.100$ &     $1266       $        &    $9.170\pm0.05 $       & 2     \\ 
\astrobj{Loden~1002}   & $0.30\pm0.03$ & $1235\pm183$ & $9.05\pm0.81$  & $0.024\pm0.008$ & $3.19\pm0.02$  &     $0.396\pm0.100$  &     $1332       $        &    $8.000\pm0.10$       & 2     \\ 
\astrobj{Ruprecht~121} & $0.25\pm0.04$ & $ 698\pm 55$ & $9.45\pm0.22$  & $0.019\pm0.004$ & $3.19\pm0.01$  &     $1.041\pm0.100$ &     $1239       $        &    $8.425\pm0.05$       & 2     \\ 
   \hline      
  \end{tabular} 
  \begin{flushleft} 
\tiny
(1) \citet{McSwain2005}; (2) \citet{Kharchenko2013}; (3) \citet{Bukowiecki2011}; (4) \citet{Tadross2011}; (5) Carraro 2006

\end{flushleft}
\end{table*}


\end{document}